\documentclass[11pt]{article}
\usepackage{amsmath}   
\usepackage{amsfonts}    
\usepackage{mathrsfs} 
\usepackage{xfrac} 
\usepackage{tabu}
\usepackage{color}

\usepackage{tikz}
\usetikzlibrary{trees,snakes}
\usepackage{verbatim}

\usepackage{amsmath}
\usepackage{amsfonts}
\usepackage{amssymb}
\usepackage{graphicx}
\usepackage{tikz}
\usetikzlibrary{calc}
\usepackage{pgfplots}
\usepackage{cleveref}
\usepackage{graphicx}
\usepackage{subcaption}
\usetikzlibrary{trees,snakes}
\usepackage{verbatim}

\usepackage{mathrsfs} 
\usepackage{xfrac} 
\usepackage{tabu}
\usepackage{color}
\usepackage{tikz}
\usepackage{bbm}

\definecolor{myred}{RGB}{255, 0, 0}
\definecolor{myblue}{RGB}{0, 0, 255}
\newtheorem{theorem}{Theorem}
\newtheorem{lemma}{Lemma}
\newtheorem{proposition}{Proposition}

\newcommand{\OEP}{P_{\mbox{\tiny e}}^{\mbox{\tiny opt}}(\calC_{n})}
\newcommand{\Gset}{\calG_{n}(\delta,i,m,\bx_{m})}
\newcommand{\code}{{\mathscr C}}

\newcommand {\nn} {\nonumber}
\newcommand {\prob} {\mathbb{P}}
\newcommand{\IND}{\mathbbm{1}}
\newcommand{\DEF}{\overset{\Delta}{=}}

\newcommand{\dfn}{\stackrel{\triangle}{=}}

\newcommand {\lexe} {\stackrel{\cdot} {\le}}
\newcommand {\gexe} {\stackrel{\cdot} {\ge}}

\newcommand {\DEXE} {\stackrel{\circ}{=}}

\newcommand {\hQ} {\hat{Q}}

\newcommand {\bi} {\mbox{\boldmath $i$}}
\newcommand {\bj} {\mbox{\boldmath $j$}}

\newcommand {\bx} {\boldsymbol{x}}
\newcommand {\by} {\boldsymbol{y}}

\newcommand {\bX} {\boldsymbol{X}}
\newcommand{\calA}{{\cal A}}
\newcommand{\calB}{{\cal B}}
\newcommand{\calC}{{\cal C}}

\newcommand{\calE}{{\cal E}}

\newcommand{\calG}{{\cal G}}

\newcommand{\calJ}{{\cal J}}

\newcommand{\calQ}{{\cal Q}}

\newcommand{\calS}{{\cal S}}
\newcommand{\calT}{{\cal T}}

\newcommand{\calX}{{\cal X}}
\newcommand{\calY}{{\cal Y}}

\begin{document}
\thispagestyle{empty}
\title{Error Exponents in the Bee Identification Problem\footnote{
		This research was supported by the Israel Science Foundation (ISF) grant no.\ 137/18.}\\}
\author{\\ Ran Tamir (Averbuch) and Neri Merhav\\}
\maketitle
\begin{center}
The Andrew \& Erna Viterbi Faculty of Electrical Engineering \\
Technion - Israel Institute of Technology \\
Technion City, Haifa 3200003, ISRAEL \\
\{rans@campus, merhav@ee\}.technion.ac.il
\end{center}
\vspace{1.5\baselineskip}
\setlength{\baselineskip}{1.5\baselineskip}

\begin{abstract}
We derive various error exponents in the bee identification problem under two different decoding rules. Under na\"ive decoding, which decodes each bee independently of the others, we analyze a general discrete memoryless channel and a relatively wide family of stochastic decoders. Upper and lower bounds to the random coding error exponent are derived and proved to be equal at relatively high coding rates. Then, we propose a lower bound on the error exponent of the typical random code, which improves upon the random coding exponent at low coding rates. We also derive a third bound, which is related to expurgated codes, which turns out to be strictly higher than the other bounds, also at relatively low rates. We show that the universal maximum mutual information decoder is optimal with respect to the typical random code and the expurgated code. Moving further, we derive error exponents under optimal decoding, the relatively wide family of symmetric channels, and the maximum likelihood decoder. We first propose a random coding lower bound, and then, an improved bound which stems from an expurgation process. We show numerically that our second bound strictly improves upon the random coding bound at an intermediate range of coding rates, where a bound derived in a previous work no longer holds.   \\

\noindent
{\bf Index Terms:}  Bee identification problem, error exponent, expurgated exponent, typical random code, permutation recovery.
\end{abstract}

\clearpage
\section{Introduction}

The bee identification problem is a problem of correctly identifying a massive amount of data which have been shuffled and corrupted by noise. 
Specifically, consider the following problem. Let $\calC_{n}$ be a codebook composed by $e^{nR}$ codewords. Assume that these codewords are randomly permuted and afterwards, each one of them is fed into a discrete memoryless channel (DMC). Based on a set of channel outputs, one has to correctly decode the underlying permutation.
   
While originally motivated in a study on the social interactions between bees in a beehive \cite{OriginalBees}, the bee identification problem (to be defined formally later on) and its variants already found its way to information theory in a few different research areas. We mention here just a few.  
In \cite{Shahi1}, a strongly asynchronous massive access channel was investigated. In this model, $e^{n \nu}$ different users transmit a randomly selected message among $e^{nR}$ ones. The decoder has to correctly decode all messages, and furthermore, to correctly identify the users' identities.  
In a different topic, the problem of identifying the underlying probability distributions of a set of a massive number of observed sequences under the constraint that each sequence is generated i.i.d.\ by a distinct distribution has been considered in \cite{Shahi2}. 
Fundamental limits of data storage via unordered DNA molecules was studied in \cite{Heckel1}, and it noisy version was analyzed in \cite{Heckel2}.
Other aspects of the permutation recovery problem have been investigated in \cite{PWC}.

Recently, the bee identification problem has been studied from the viewpoint of its exponential error bounds. In \cite{BEES}, the codebook is composed by binary codewords, which are permuted and fed into a binary symmetric channel (BSC). In that work, two different decoding techniques have been considered; independent decoding and joint decoding. In independent decoding, each channel output is decoded separately, and in joint decoding, one uses all channel output sequences together in order to recover the underlying permutation. Under any of these decoders, the authors derive two kinds of bounds on the optimal error exponent: (i) random coding error exponent, and, (ii) error exponent which relies on characteristics of typical random binary codes \cite{BargForney}. They show that for any of the two decoders, the error exponent of the typical random code (TRC) is strictly higher than the random coding error exponent at relatively low coding rates, as is already known to happen in ordinary channel coding over a general DMC \cite{MERHAV_TYPICAL}, \cite{PRAD2014}. In \cite{BEES}, a converse bound is also derived, which is proved to have the same value as the value of the TRC exponent under joint decoding at rate zero. In a different work \cite{BEES2}, the same authors of \cite{BEES} study the capacity and the error exponent of the bee identification problem, but when some fraction of the bees are assumed to be outside the beehive. The authors provide an exact characterization of the error exponent and they prove that independent decoding is optimal.           

The focus of this work is on extensions and refinements of the error exponent analysis of the same decoding rules studied in \cite{BEES}.  
In particular, the main contributions of this work are the following. 

\begin{enumerate}
	\item In na\"ive (independent) decoding, we adopt a slightly relaxed definition for the probability of error; while in \cite{BEES}, error counts even if a single bee is incorrectly decoded, here, we refer to an error event only when at least $L$ bees are erroneously decoded. We believe that such a relaxed definition may be more suitable in this kind of problem (and others as well), which accounts for a massive amount of data.
	\item For the ensemble of uniformly randomly drawn constant composition codes, we provide different exponential error bounds for a general DMC and a wide class of stochastic decoders, collectively referred to as the generalized likelihood decoder (GLD). We provide the following results:
	\begin{enumerate}
		\item Both upper and lower bounds on the random coding error exponent, which turn to match each other at relatively high coding rates, at least for some specific DMCs.
		\item A lower bound on the error exponent of the TRC. We show on a numerical example that it strictly improves upon the random coding exponent at low coding rates.
		\item An error exponent which stems from expurgated codes in ordinary channel coding. This exponent is strictly higher at low coding rates relative to the TRC exponent.
	\end{enumerate}
	\item We show that the universal maximum mutual information (MMI) decoder is optimal with respect to the TRC and the expurgated code, a fact that was recently asserted in ordinary channel coding \cite{TM_UNIVERSAL}.
	\item We provide exponential error bounds under optimal (joint) decoding, but under a slightly less general model: (i) the general DMC is replaced by the family of symmetric channels, which includes the BSC as a special case. (ii) The wide family of GLDs is confined only to the (optimal) maximum likelihood (ML) decoder. (iii) The ensemble of constant composition codes is switched to the i.i.d.\ random coding ensemble. Under this setting, we provide two different lower bounds to the optimal error exponent:
	\begin{enumerate}
		\item The first is a lower bound on the random coding error exponent, which is given by a relatively simple expression, that does not include any optimization problems.
		\item The second is derived by code expurgation, and it improves upon the previous one at low coding rates. Our second bound matches the bound in \cite{BEES} that relies on characteristics of typical random binary codes, but it holds for a wider set of coding rates. Specifically, it still improves upon the random coding lower bound at rates where the bound in \cite{BEES} no longer holds.
	\end{enumerate}
\end{enumerate}            

The remaining part of the paper is organized as follows. 
In Section 2, we establish notation conventions. 
In Section 3, we formalize the models and the main objectives of this work. 
In Section 4, we provide and discuss the main results, and in the Appendixes, we prove them.

\section{Notation Conventions}

Throughout the paper, random variables will be denoted by capital letters, realizations will be denoted by the corresponding lower case letters, 
and their alphabets in calligraphic font.
Random vectors and their realizations will be denoted, 
respectively, by boldfaced capital and lower case letters. 
Their alphabets will be superscripted by their dimensions. 
%For example, the random vector $\bX = (X_{1}, \dotsc , X_{n})$, 
%($n$ -- positive integer) 
%may take a specific vector value $\bx = (x_{1}, \dotsc , x_{n})$ 
%in $\calX^{n}$, 
%the $n$-th order Cartesian power of $\cal{X}$, 
%which is the alphabet of each component of this vector. 
%Sources and channels will be subscripted by the names of the relevant random variables/vectors and their conditionings, whenever applicable, 
%following the standard notation conventions, e.g., $Q_{X}$, $Q_{Y|X}$, and so on. 
%When there is no room for ambiguity, these subscripts will be omitted. 
For a generic joint distribution $Q_{XY} = \{Q_{XY}(x,y), x \in \mathcal{X}, y \in \mathcal{Y} \}$, which will often be abbreviated by $Q$, information measures will be denoted in 
the conventional manner, but with a subscript $Q$, that is, $I_{Q}(X;Y)$ is the mutual information between $X$ and $Y$, and similarly for other quantities. 
The weighted divergence between 
two conditional distributions (channels), say, $Q_{Y|X}$ and $W = \{W(y|x), 
x \in \calX, y \in \calY \}$, with weighting $Q_{X}$ is defined as
\begin{align}
D(Q_{Y|X} || W | Q_{X}) 
= \sum_{x \in \calX} Q_{X}(x) 
\sum_{y \in \calY} Q_{Y|X}(y|x) \log \frac{Q_{Y|X}(y|x)}{W(y|x)},
\end{align}
where logarithms, here and throughout the sequel, are taken to the natural base.
%Logarithms are taken to the natural base.
The probability of an event $\mathcal{E}$ will be denoted by 
$\prob \{\cal{E}\}$, and the expectation operator will be denoted by $\mathbb{E}[\cdot]$. 
The indicator function of an event $\calE$ 
will be denoted by $\IND\{\calE\}$. 
The notation $[t]_{+}$ will stand for $\max \{0,t\}$.

For two positive sequences, $\{a_{n}\}$ and $\{b_{n}\}$, the notation $a_{n} \doteq b_{n}$ will stand for equality in the exponential scale, that is, $\lim_{n \to \infty} (1/n) \log \left(a_{n}/b_{n}\right) = 0$. Similarly, $a_{n} \lexe b_{n}$ means that $\limsup_{n \to \infty} (1/n) \log \left(a_{n}/b_{n}\right) \leq 0$, and so on.
Accordingly, the notation $a_{n} \doteq e^{-n \infty}$ means that $a_{n}$ decays at a super--exponential rate (e.g.\ double--exponentially).

By the same token, for two positive sequences, $\{a_{n}\}$ and $\{b_{n}\}$, whose elements are both smaller than one (for all large enough $n$), the notation $a_{n} \DEXE b_{n}$ will stand for equality in the double--exponential scale, that is, 
\begin{align}
\lim_{n \to \infty} \frac{1}{n} \log \left(\frac{\log b_{n}}{\log a_{n}} \right) = 0.
\end{align}
%Similarly, $a_{n} \DLEXE b_{n}$ means that 
%\begin{align}
%\limsup_{n \to \infty} \frac{1}{n} \log \left(\frac{\log b_{n}}{\log a_{n}} \right) \leq 0,
%\end{align}
%and $a_{n} \DGEXE b_{n}$ stands for 
%\begin{align}
%\liminf_{n \to \infty} \frac{1}{n} \log \left(\frac{\log b_{n}}{\log a_{n}} \right) \geq 0.
%\end{align}

The empirical distribution of a sequence $\boldsymbol{x} \in \mathcal{X}^{n}$, which will 
be denoted by $\hat{P}_{\boldsymbol{x}}$, is the vector of 
relative frequencies, $\hat{P}_{\bx}(x)$, 
of each symbol $x \in \mathcal{X}$ in $\bx$.
The joint empirical distribution of a pair of sequences, denoted by $\hat{P}_{\bx \by}$, is similarly defined.
%Likewise, the joint empirical distribution of a pair of sequences $\bx \in \mathcal{X}^{n}$ and $\by \in \mathcal{Y}^{n}$, which will 
%be denoted by $\hat{P}_{\bx \by}$, is the matrix of 
%relative frequencies, $\hat{P}_{\bx \by}(x,y)$, 
%of each pair of symbols $(x,y) \in \mathcal{X} \times \mathcal{Y}$ in $(\bx,\by)$. 
The type class of $Q_{X}$, denoted $\mathcal{T}(Q_{X})$, is the set of all vectors $\bx \in \calX^{n}$ with $\hat{P}_{\bx} = Q_{X}$. 
In the same spirit, the joint type class of $Q_{XY}$, denoted $\calT(Q_{XY})$, is the set of all pairs of sequences $(\bx,\by) \in \calX^{n} \times \calY^{n}$ with $\hat{P}_{\bx\by} = Q_{XY}$.

Throughout the paper, we will make a frequent use of the fact that 
\begin{align}
\label{SME}
\sum_{i=1}^{k_{n}} a_{n}(i) \doteq \max_{1 \leq i \leq k_{n}} a_{n}(i)
\end{align}
as long as $\{a_{n}(i)\}$ are nonnegative exponential functions of an integer $n$ and $k_{n} \doteq 1$. This exponential equivalence will be termed henceforth the {\it summation--maximization equivalence} (SME). The sequence $k_{n}$ will represent the number of type classes possible for a given block length $n$, which is polynomial in $n$.

\section{Problem Setting and Objectives}

Consider a DMC, $W = \{W(y|x):~ x \in \calX,~ y \in \calY\}$, where $\calX$ is a finite input alphabet, $\calY$ is a finite output alphabet, and $W(y|x)$ is the channel input-output single--letter transition probability from $x$ to $y$. When fed by a vector $\bx=(x_{1},x_{2},\ldots,x_{n}) \in \calX^{n}$, the channel responds by producing an output vector $\by=(y_{1},y_{2},\ldots,y_{n}) \in \calY^{n}$, according to 
\begin{align}
W(\by|\bx) = \prod_{i=1}^{n} W(y_{i}|x_{i}).
\end{align}
Let $\calC_{n} = \{\bx_{1},\bx_{2}, \dotsc, \bx_{M}\}$ be a set of $M=e^{nR}$ codewords, $R$ being the coding rate in nats per channel use.
Let $\tilde{\calC}_{n} = \{\tilde{\bx}_{1},\tilde{\bx}_{2}, \dotsc, \tilde{\bx}_{M}\}$ be some random permutation of $\calC_{n}$, drawn by the channel from the set of all possible permutations of $\{1,2,\ldots,M\}$, according to the uniform distribution.
Let $\{\tilde{\by}_{1},\tilde{\by}_{2}, \dotsc, \tilde{\by}_{M}\}$, where $\tilde{\by}_{i}$, $i \in \{1,2,\dotsc,M\}$, is the channel output when the channel is fed by $\tilde{\bx}_{i}$.    
Based on the set $\{\tilde{\by}_{1},\tilde{\by}_{2}, \dotsc, \tilde{\by}_{M}\}$, we would like to decode and find out which codeword in $\calC_{n}$ is the source for each of these channel outputs. 

At this point, we distinguish between two different decoders.

%\clearpage
\subsection{The Na\"ive Decoder} 

We consider the ensemble of constant composition codes: for a given distribution $Q_{X}$ over $\calX$, all vectors in $\calC_{n}$ are uniformly and independently drawn from the type class $\calT(Q_{X})$.

In na\"ive decoding, one takes each channel output sequence $\tilde{\by}_{i}$ and decodes for one codeword from $\calC_{n}$ using the GLD.
The GLD is a stochastic decoder, that chooses the estimated message $\hat{m}$ according to the following posterior probability mass function, induced by $\tilde{\by}_{i}$:
\begin{align}
\label{StrongGLD}
\prob \left\{ \hat{M}=m \middle| \tilde{\by}_{i} \right\}  =
\frac{\exp \{n g( \hat{P}_{\bx_{m} \tilde{\by}_{i} } ) \}} {\sum_{m'=1}^{M}  
	\exp \{n g( \hat{P}_{\bx_{m'} \tilde{\by}_{i} } ) \} } ,
\end{align}
where $\hat{P}_{\bx_{m} \tilde{\by}_{i} }$ is the empirical distribution of $(\bx_{m}, \tilde{\by}_{i})$, and $g(\cdot)$ is a given continuous, real--valued functional of this empirical distribution.
The GLD provides a unified framework which covers several important special cases, e.g., matched likelihood decoding, mismatched decoding, ML decoding, and universal decoding.

For a given codebook, define the following enumerator, which counts the total number of incorrect decodings:
\begin{align}
N_{\mbox{\tiny e}}(\calC_{n}) 
= \sum_{m=1}^{M} \IND \left\{\text{decoding of $\bx_{m}$ has failed} \right\}.
\end{align}
%Referring to the overall error event, we have different options. 
%In \cite{BEES}, the authors decided on a rather strict definition for the overall error event as the one that counts for at least one erroneous decoding, such that the probability of error in their work is given by $\prob \{N_{\mbox{\tiny e}}(\calC_{n}) \geq 1\}$.
%\begin{align}
%P_{\mbox{\tiny e}} = \prob \{N_{\mbox{\tiny e}}(\calC_{n}) \geq 1\}. 
%\end{align}  
%We believe that a more tolerant option makes more sense, since a possible incorrect decoding of very few bees may not be catastrophic. 
In this work, we allow for at most $L \in \mathbb{N}$ incorrect decodings, such that the probability of error is defined by  
\begin{align}
P_{\mbox{\tiny e}}(\calC_{n}) = \prob \{N_{\mbox{\tiny e}}(\calC_{n}) \geq L\}. 
\end{align}
The random coding error exponent is defined in the usual manner as
\begin{align} \label{DEF_RCE}
\mathsf{E}_{\mbox{\tiny r}}(R)
=\lim_{n \to \infty} -\frac{1}{n} \log \mathbb{E} \left[ P_{\mbox{\tiny e}}(\calC_{n}) \right],  
\end{align}
while the error exponent of the TRC is defined by
\begin{align} \label{DEF_TRC}
\mathsf{E}_{\mbox{\tiny trc}}(R)
=\lim_{n \to \infty} -\frac{1}{n} \mathbb{E} \left[\log P_{\mbox{\tiny e}}(\calC_{n}) \right].  
\end{align}
Finding exact expressions for \eqref{DEF_RCE} and \eqref{DEF_TRC} appears to be difficult. We derive lower and upper bounds on  \eqref{DEF_RCE} and a lower bound on \eqref{DEF_TRC}.

Another objective is to prove the existence of a sequence of codes %$\mathscr{C} = \{\calC(n)\}_{n=1}^{\infty}$
$\code = \{\calC_{n}\}_{n=1}^{\infty}$, whose error exponent is strictly higher than $\mathsf{E}_{\mbox{\tiny r}}(R)$ and $\mathsf{E}_{\mbox{\tiny trc}}(R)$, at least at low coding rates, and obtain a single--letter expression that lower bounds the following limit
\begin{align}
\mathsf{E}(\code) 
= \liminf_{n \to \infty} - \frac{1}{n} \log P_{\mbox{\tiny e}}(\calC_{n}). 
\end{align}

\subsection{The Optimal Decoder}

Under optimal decoding, the constant composition ensemble is much more complicated to analyze, since ordinary analysis tools, like the method of types, are no longer applicable.   
Hence, the constant composition ensemble is now replaced by the i.i.d.\ ensemble, where the $M$ codewords are drawn independently, and each one is drawn under the product distribution
\begin{align}
P(\bx) = \prod_{i=1}^{n} P_{X}(x_{i}),
\end{align}
where $P_{X}$ is some probability mass function on $\calX$. 
Let $\Pi(M)$ be the set of all possible permutations of $\{1,2,\ldots,M\}$. The maximum likelihood decoder is given by 
\begin{align}
\hat{\pi}(\by_{1},\ldots,\by_{M}) = \operatorname*{arg\,max}_{\pi \in \Pi(M)} \prod_{m=1}^{M} W(\by_{m}|\bx_{\pi(m)}). 
\end{align}
The probability of error is defined as
\begin{align}
\OEP = \frac{1}{|\Pi(M)|} \sum_{\pi \in \Pi(M)} \sum_{\by_{1} \in \calY^{n}} \cdots \sum_{\by_{M} \in \calY^{n}} \prod_{m=1}^{M} W(\by_{m}|\bx_{\pi(m)}) \IND \{\hat{\pi}(\by_{1},\ldots,\by_{M}) \neq \pi \}.
\end{align}
Under optimal decoding, we have two objectives. First, to obtain a lower bound on the random coding error exponent
\begin{align} \label{DEF_RCE_OPT}
\mathsf{E}_{\mbox{\tiny r}}^{\mbox{\tiny opt}}(R)
=\lim_{n \to \infty} -\frac{1}{n} \log \mathbb{E} \left[ \OEP \right],  
\end{align}
and second, to prove the existence of a sequence of codes $\code = \{\calC_{n}\}_{n=1}^{\infty}$, whose error probability decays exponentially at a strictly higher rate than $\mathsf{E}_{\mbox{\tiny r}}^{\mbox{\tiny opt}}(R)$, and obtain the tightest possible single--letter expression that lower bounds the following limit
\begin{align}
\mathsf{E}^{\mbox{\tiny opt}}(\code) 
= \liminf_{n \to \infty} - \frac{1}{n} \log \OEP. 
\end{align}

\section{Main Results}
\subsection{Na\"ive Decoding}

In order to present upper and lower bounds on the random coding error exponent, we first provide some definitions.  
Define the set $\calQ(Q_{X}) = \{Q_{X'|X}: Q_{X'}=Q_{X}\}$ and 
\begin{align}
\label{ALPHA_DEF}
\alpha(R,Q_{Y}) &= \max_{Q_{\tilde{X}|Y} \in \calS(Q_{X},Q_{Y})} \{g(Q_{\tilde{X}Y}) + R - I_{Q}(\tilde{X};Y)\}, \\
%%%%%%%%%%%%%%%%%%%%%%%%%%%%%%%%%%%%%%%%%%%%%
\label{Beta_DEF}
\beta(R,Q_{Y}) &= \max_{\{Q_{\tilde{X}|Y}:~ Q_{\tilde{X}}=Q_{X}\}} \{g(Q_{\tilde{X}Y}) + [R - I_{Q}(\tilde{X};Y)]_{+}\},
\end{align}
where $\calS(Q_{X},Q_{Y})=\{Q_{\tilde{X}|Y}:~I_{Q}(\tilde{X};Y) \leq R,~ Q_{\tilde{X}}=Q_{X}\}$, as well as 
\begin{align}
\label{LAMBDA_DEF}
\Lambda(Q_{XX'},R) &= \min_{Q_{Y|XX'}} \{ D(Q_{Y|X} \| W |Q_{X}) + I_{Q}(X';Y|X) + \beta(R,Q_{Y}) - g(Q_{X'Y}) \}, \\ 
%%%%%%%%%%%%%%%%%%%%%%%%%%%%%%%%%%%%%%%%%%%%%%%
\label{Gamma_DEF}
\Gamma(Q_{XX'},R) &= \min_{Q_{Y|XX'}} \{ D(Q_{Y|X} \| W |Q_{X}) + I_{Q}(X';Y|X) \nn \\
&~~~~~~~~+ [\max\{g(Q_{XY}), \alpha(R,Q_{Y})\} - g(Q_{X'Y})]_{+} \}.
\end{align}
%Define the set
%\begin{align} \label{DEF_J}
%\calJ(R,s) = \left\{Q_{X'|X} \in \calQ(Q_{X}):~ 
%\left[R-I_{Q}(X;X')\right]_{+} \geq \Lambda(Q_{XX'},R) - s \right\} .
%\end{align}
Finally, define the exponent functions
\begin{align} \label{RC_UB_expression}
E_{\mbox{\tiny r}}^{\mbox{\tiny ub}}(R,L) 
= \min_{Q_{X'|X} \in \calQ(Q_{X})}
\left[L \cdot \Gamma(Q_{XX'},R) - L \cdot [2R - I_{Q}(X;X')]_{+} + [I_{Q}(X;X')-2R]_{+}\right]_{+}
\end{align}
and
\begin{align} \label{RC_LB_expression}
E_{\mbox{\tiny r}}^{\mbox{\tiny lb}}(R,L) = \min_{Q_{X'|X} \in \calQ(Q_{X})} 
L \cdot \max\left\{[I_{Q}(X;X')-R]_{+}, \Lambda(Q_{XX'},R) + I_{Q}(X;X') - 2R \right\}.
\end{align}

Our first result in this section is the following theorem, which is proved in appendices A and B.

\begin{theorem} \label{THEOREM_Naive}
	Consider the ensemble of random constant composition codes $\calC_{n}$ of rate $R$ and composition $Q_{X}$. Then,
	\begin{align} \label{THM1_UB}
	\lim_{n \to \infty} -\frac{1}{n} \log \mathbb{E} \left[ P_{\mbox{\tiny e}}(\calC_{n}) \right] \geq E_{\mbox{\tiny r}}^{\mbox{\tiny ub}}(R,L).
	\end{align}
	Also,
	\begin{align} \label{THM1_LB}
	\lim_{n \to \infty} -\frac{1}{n} \log \mathbb{E} \left[ P_{\mbox{\tiny e}}(\calC_{n}) \right] \leq E_{\mbox{\tiny r}}^{\mbox{\tiny lb}}(R,L).
	\end{align}
\end{theorem}

\subsubsection*{Discussion}

For $L=1$, the exponent function \eqref{RC_UB_expression} is at least as tight as in \cite[Eq.\ (14)]{BEES}. To see why this is true, consider a GLD with $g(Q_{XY})=I_{Q}(X;Y)$. In this case, $\alpha(R,Q_{Y})=R$ and we get that
\begin{align} 
E_{\mbox{\tiny r}}^{\mbox{\tiny ub}}(R,1)
%%%%%%%%%%%%%%%%%%%%%%%%%%%%%%%%%%%%%%%%%%%%%%%%%%%%%%%%%%%%%%%
&\geq \min_{Q_{X'|X} \in \calQ(Q_{X})}
 [\Gamma(Q_{XX'},R) + I_{Q}(X;X') - 2R ]_{+} \\
%%%%%%%%%%%%%%%%%%%%%%%%%%%%%%%%%%%%%%%%%%%%%%%%%%%%%%%%%%%%%%% 
&= \min_{\{Q_{X'Y|X},~Q_{X'}=Q_{X}\}} [ D(Q_{Y|X} \| W |Q_{X}) + I_{Q}(X';Y|X) \nn \\
&~~~~~~~~+ [\max\{I_{Q}(X;Y), R\} - I_{Q}(X';Y)]_{+} +I_{Q}(X;X') - 2R ]_{+} \\
%%%%%%%%%%%%%%%%%%%%%%%%%%%%%%%%%%%%%%%%%%%%%%%%%%%%%%%%%%%%%%% 
&= \min_{\{Q_{X'Y|X},~Q_{X'}=Q_{X}\}} [ D(Q_{Y|X} \| W |Q_{X}) + I_{Q}(X;X'|Y) \nn \\
&~~~~~~~~+ [\max\{I_{Q}(X;Y), R\} - I_{Q}(X';Y)]_{+} +I_{Q}(X';Y) - 2R ]_{+} \\
%%%%%%%%%%%%%%%%%%%%%%%%%%%%%%%%%%%%%%%%%%%%%%%%%%%%%%%%%%%%%%% 
&= \min_{\{Q_{X'Y|X},~Q_{X'}=Q_{X}\}} [ D(Q_{Y|X} \| W |Q_{X}) + I_{Q}(X;X'|Y) \nn \\
&~~~~~~~~+ \max\{I_{Q}(X;Y), I_{Q}(X';Y), R\} - 2R ]_{+} \\
%%%%%%%%%%%%%%%%%%%%%%%%%%%%%%%%%%%%%%%%%%%%%%%%%%%%%%%%%%%%%%% 
&= \min_{Q_{Y|X}} [ D(Q_{Y|X} \| W |Q_{X}) + \max\{I_{Q}(X;Y), R\} - 2R ]_{+} \\
%%%%%%%%%%%%%%%%%%%%%%%%%%%%%%%%%%%%%%%%%%%%%%%%%%%%%%%%%%%%%%% 
&= \min_{Q_{Y|X}} [ D(Q_{Y|X} \| W |Q_{X}) + [I_{Q}(X;Y) - R]_{+} - R ]_{+} \\
\label{BottomLine}
&= [E_{\mbox{\tiny r}}(R) - R]_{+},
\end{align}
where $E_{\mbox{\tiny r}}(R)$ is the random coding error exponent in ordinary channel coding. 
The expression in \eqref{BottomLine} is the same as in \cite[Eq.\ (14)]{BEES}, but for a general DMC, which proves our claim.

On the one hand,
for any $L \geq 2$, $E_{\mbox{\tiny r}}^{\mbox{\tiny lb}}(R,L)$ is larger than $E_{\mbox{\tiny r}}^{\mbox{\tiny ub}}(R,L)$, at least at low coding rates, since at rate zero,
\begin{align}
E_{\mbox{\tiny r}}^{\mbox{\tiny ub}}(0,L) 
&= \min_{Q_{X'|X} \in \calQ(Q_{X})}
\left\{L \cdot \Gamma(Q_{XX'},0) + I_{Q}(X;X') \right\} \\
E_{\mbox{\tiny r}}^{\mbox{\tiny lb}}(0,L) 
&= \min_{Q_{X'|X} \in \calQ(Q_{X})} 
L \cdot \left\{\Lambda(Q_{XX'},0) + I_{Q}(X;X') \right\}
\end{align} 
and $\Lambda(Q_{XX'},R) \geq \Gamma(Q_{XX'},R)$. Moreover, we note the following fact: when $L$ grows, the exponent function $E_{\mbox{\tiny r}}^{\mbox{\tiny lb}}(R,L)$ grows without bound, while the exponent function $E_{\mbox{\tiny r}}^{\mbox{\tiny ub}}(R,L)$ converges to the finite function
\begin{align}
\tilde{E}_{\mbox{\tiny r}}(R) = \min_{\{Q_{X'|X} \in \calQ(Q_{X}):~ [2R-I_{Q}(X;X')]_{+} \geq \Gamma(Q_{XX'},R) \}}
[I_{Q}(X;X')-2R]_{+}.
\end{align}
Since we expect the exponential rate of decay of the probability of error to increase without bound as the number of incorrectly decoded bees grows, we believe that the true exponential rate of decay of $\mathbb{E} \left[ P_{\mbox{\tiny e}}(\calC_{n}) \right]$ is closer to $E_{\mbox{\tiny r}}^{\mbox{\tiny lb}}(R,L)$ at relatively low coding rates, rather than to $E_{\mbox{\tiny r}}^{\mbox{\tiny ub}}(R,L)$. Unfortunately, we were not able to further tighten the exponential rate of decay of the upper bound on $\mathbb{E} \left[ P_{\mbox{\tiny e}}(\calC_{n}) \right]$. 
%the expected error probability.

On the other hand, we argue that $E_{\mbox{\tiny r}}^{\mbox{\tiny lb}}(R,L) = E_{\mbox{\tiny r}}^{\mbox{\tiny ub}}(R,L)$ at relatively high coding rates, at least for some DMCs. As for $E_{\mbox{\tiny r}}^{\mbox{\tiny ub}}(R,L)$, we claim that there exists some rate $R^{*}(L)$, such that for all $R \ge R^{*}(L)$, the clipping operator around $I_{Q}(X;X')-2R$ in \eqref{RC_UB_expression} is active. 
To see why this is true, assume conversely, that is, 
there exist arbitrarily high rates, such that the clipping operator around $I_{Q}(X;X')-2R$ is inactive, while the clipping operator around $2R-I_{Q}(X;X')$ is active. Since $\Gamma(Q_{XX'},R)$ increases linearly with a slope of one at high rates, due to the behavior of $\alpha(R,Q_{Y})$, $E_{\mbox{\tiny r}}^{\mbox{\tiny ub}}(R,L)$ increases without bound, which is a contradiction. Hence, at relatively high rates,
\begin{align} \label{UB_high_rates}
E_{\mbox{\tiny r}}^{\mbox{\tiny ub}}(R,L)
= \min_{Q_{X'|X} \in \calQ(Q_{X})}
L \cdot \left[\Gamma(Q_{XX'},R) + I_{Q}(X;X') - 2R \right]_{+}.
\end{align}  
For the exponent function $E_{\mbox{\tiny r}}^{\mbox{\tiny lb}}(R,L)$, note that for sufficiently high rates, the clipping operator around $I_{Q}(X;X')-R$ in \eqref{RC_LB_expression} is active, such that, 
\begin{align}
E_{\mbox{\tiny r}}^{\mbox{\tiny lb}}(R,L)
= \min_{Q_{X'|X} \in \calQ(Q_{X})}
L \cdot \left[\Lambda(Q_{XX'},R) + I_{Q}(X;X') - 2R \right]_{+}.
\end{align} 
Finally, it can be easily proved, using similar techniques as in \cite[Section 5]{MERHAV2017}, that for some specific channels, like the $z$-channel or the binary erasure channel, an equality between $\Lambda(Q_{XX'},R)$ and $\Gamma(Q_{XX'},R)$ holds, which asserts that $E_{\mbox{\tiny r}}^{\mbox{\tiny lb}}(R,L) = E_{\mbox{\tiny r}}^{\mbox{\tiny ub}}(R,L)$ at relatively high rates. 

We conclude from \eqref{UB_high_rates} that for any $L$, there exists $R_{\mbox{\tiny max}}$, such that $E_{\mbox{\tiny r}}^{\mbox{\tiny ub}}(R,L) > 0$ if and only if $R < R_{\mbox{\tiny max}}$. An explicit lower bound on $R_{\mbox{\tiny max}}$ can be derived as follows using the lower bound in \eqref{BottomLine}. The requirement $[E_{\mbox{\tiny r}}(R) - R]_{+}>0$ is equivalent to      
\begin{align}
R 
&< \min_{Q_{Y|X}} \{ D(Q_{Y|X} \| W |Q_{X}) + [I_{Q}(X;Y) - R]_{+} \} \\
&= \min_{Q_{Y|X}} \max_{t \in [0,1]} \{ D(Q_{Y|X} \| W |Q_{X}) + t(I_{Q}(X;Y) - R) \}, 
\end{align}
which, in turn, is equivalent to
\begin{align}
\forall Q_{Y|X},~~\exists t \in [0,1],~~ R<D(Q_{Y|X} \| W |Q_{X}) + t(I_{Q}(X;Y) - R),
\end{align}
or, to
\begin{align}
\forall Q_{Y|X},~~\exists t \in [0,1],~~ 
R< \frac{D(Q_{Y|X} \| W |Q_{X}) + tI_{Q}(X;Y)}{1+t}.
\end{align}
Hence, we conclude that
\begin{align}
R_{\mbox{\tiny max}}
&\geq \min_{Q_{Y|X}} \max_{t \in [0,1]} \left\{ \frac{D(Q_{Y|X} \| W |Q_{X}) + tI_{Q}(X;Y)}{1+t} \right\} \\
&= \min_{Q_{Y|X}} \max \left\{D(Q_{Y|X} \| W |Q_{X}), \frac{D(Q_{Y|X} \| W |Q_{X}) + I_{Q}(X;Y)}{2} \right\} \\
\label{Maximal_Rate}
&= \min_{Q_{Y|X}} \left\{D(Q_{Y|X} \| W |Q_{X}) +\frac{1}{2} \cdot [I_{Q}(X;Y) - D(Q_{Y|X} \| W |Q_{X}) ]_{+}  \right\}.
\end{align}

Following the studies in \cite{BargForney}, \cite{MERHAV_TYPICAL}, and \cite{PRAD2014} on TRCs in ordinary channel coding, we claim that also in the bee identification problem, the random coding error exponent, which is bounded from above and below in Theorem \ref{THEOREM_Naive}, does not yield the true exponential behavior of the error probability of a randomly chosen code, since it is dominated by the relatively bad codes in the ensemble, rather than the channel noise, at least at low coding rates.
Due to the definition of the TRC exponent, the derivation of a single-letter expression is not as easy as in ordinary random coding (for example, see the proof in \cite[Section 5]{MERHAV_TYPICAL}), since the expectations over the randomness of the ensemble and over the randomness of the channel cannot be switched, which is one of the first steps in random coding analysis. 
We next present a lower bound on the error exponent of the TRC. Define the exponent function
\begin{align} \label{TRC_naive}
E_{\mbox{\tiny trc}}(R,L) = \min_{\{Q_{X'|X} \in \calQ(Q_{X}):~ I_{Q}(X;X') \leq 2R\}}
L \cdot \left[\Gamma(Q_{XX'},R) + I_{Q}(X;X') - 2R \right]_{+}.
\end{align}
Then, our second result is the following theorem, which is proved in Appendix D.
\begin{theorem} \label{THEOREM_Naive_TRC}
	Consider the ensemble of random constant composition codes $\calC_{n}$ of rate $R$ and composition $Q_{X}$. Then,
	\begin{align} \label{THM2_UB}
	\lim_{n \to \infty} -\frac{1}{n} \mathbb{E} \left[\log P_{\mbox{\tiny e}}(\calC_{n}) \right] \geq E_{\mbox{\tiny trc}}(R,L).
	\end{align}
\end{theorem}
Several comments are now in order.
\begin{itemize}
	\item Since each bee is decoded independently, the error probability depends heavily on the statistical characteristics of the type class enumerators,
	\begin{align} \label{NQ_def}
	N(Q_{XX'}) \dfn \sum_{m=0}^{M-1} \sum_{m' \neq m} \IND \left\{(\bX_{m},\bX_{m'}) \in \calT(Q_{XX'}) \right\},
	\end{align}
	which also play a pivotal role in the proofs of the main results in \cite{MERHAV_TYPICAL} and \cite{TMWG}. Specifically, the result in Theorem \ref{THEOREM_Naive_TRC} is related to the values of $\{N(Q_{XX'})\}$ in a TRC, which is $\exp\{n (2R-I_{Q}(X;X'))\}$ if $2R \geq I_{Q}(X;X')$ and zero otherwise. This fact was already asserted in \cite{MERHAV_TYPICAL} and it explains the constraint in the minimization problem in \eqref{TRC_naive}.   
	\item By applying \eqref{TRC_naive} to the BSC, a symmetric input assignment, the ML decoder, and $L=1$, one arrive to a similar result as in \cite[Theorem 3]{BEES}. Nevertheless, we mention a relatively significant difference between the two derivations. On the one hand, the bound in \cite{BEES} is heavily based on the behavior of typical random binary codes \cite{BargForney}, and thus, it cannot be directly generalized to larger alphabets. On the other hand, in this work, we directly derive (a lower bound on) the error exponent of the TRC, which holds for any DMC.     
	\item Although we only propose here a lower bound on the TRC exponent, we conjecture that a matching upper bound also holds, and leave it to future work. Furthermore, we believe that a concentration property holds, i.e., that the exponential rate of decay of the error probability of a randomly chosen code is close to $E_{\mbox{\tiny trc}}(R,L)$ with a very high probability. A similar property in ordinary channel coding was already proved in \cite{TMWG}.      
\end{itemize}

In ordinary channel coding, the random coding error exponent, as well as the error exponent of the TRC are improved at relatively low coding rates by code expurgation. Upon using the result in \cite[Section 5]{MERHAV2017}, which is an error exponent under the assumption of a GLD, we are able to derive a bound which is tighter than $E_{\mbox{\tiny r}}^{\mbox{\tiny ub}}(R,L)$ and $E_{\mbox{\tiny trc}}(R,L)$, at least at low coding rates. 
Let us define the exponent function
\begin{align} \label{EX_naive}
E_{\mbox{\tiny ex}}(R,L) = \min_{\{Q_{X'|X} \in \calQ(Q_{X}):~ I_{Q}(X;X') \leq R\}}
L \cdot \left[\Gamma(Q_{XX'},R) + I_{Q}(X;X') - 2R \right]_{+}.
\end{align}
Then, our third result is the following theorem, which is proved in Appendix E.
\begin{theorem} \label{THEOREM_Naive_EX}
	There exists a sequence of constant composition codes, $\{\calC_{n},~n=1,2,\dotsc\}$, with composition $Q_{X}$, such that
	\begin{align}
	\liminf_{n \to \infty} - \frac{1}{n} \log P_{\mbox{\tiny e}}(\calC_{n}) \geq E_{\mbox{\tiny ex}}(R,L).
	\end{align}
\end{theorem}

The qualitative behavior of $E_{\mbox{\tiny trc}}(R,L)$ and $E_{\mbox{\tiny ex}}(R,L)$ is similar to the behavior of the TRC exponent and the expurgated exponent in ordinary channel coding. At rate zero, they are equal, but at positive low rates, $E_{\mbox{\tiny trc}}(R,L) < E_{\mbox{\tiny ex}}(R,L)$. At relatively high coding rates, the minimization constraints in \eqref{TRC_naive} and \eqref{EX_naive} become inactive and these exponent functions, as well as the lower bound on the random coding error exponent given in \eqref{UB_high_rates} are all equal.   

In ordinary channel coding, it has been lately proved in \cite{TM_UNIVERSAL} that the MMI decoder is optimal with respect to the TRC and with respect to the expurgated code. One may wonder whether a similar phenomenon also holds in the bee identification problem.     
Note that the exponent functions in \eqref{TRC_naive} and \eqref{EX_naive} strongly resembles the error exponent of the TRC \cite[Eq.\ (18)]{MERHAV_TYPICAL} and the expurgated exponent \cite[Eq.\ (42)]{MERHAV2017} in ordinary channel coding. Since the proof in \cite{TM_UNIVERSAL} exclusively relies on upper and lower-bounding the term $\Gamma(Q_{XX'},R)$, we conclude that in the current setting, the MMI-based na\"ive decoder is optimal with respect to both the TRC and the expurgated code, i.e., it performs as good as the ML-based na\"ive decoder. This fact may be quite important from the practical point of view, since the effective channel that reads the bee bar-codes may vary with time, due to thermal effects in electro-optical detectors and more. 
%except for three differences: (i) the factor $L$, which obviously equals one in ordinary channel coding 

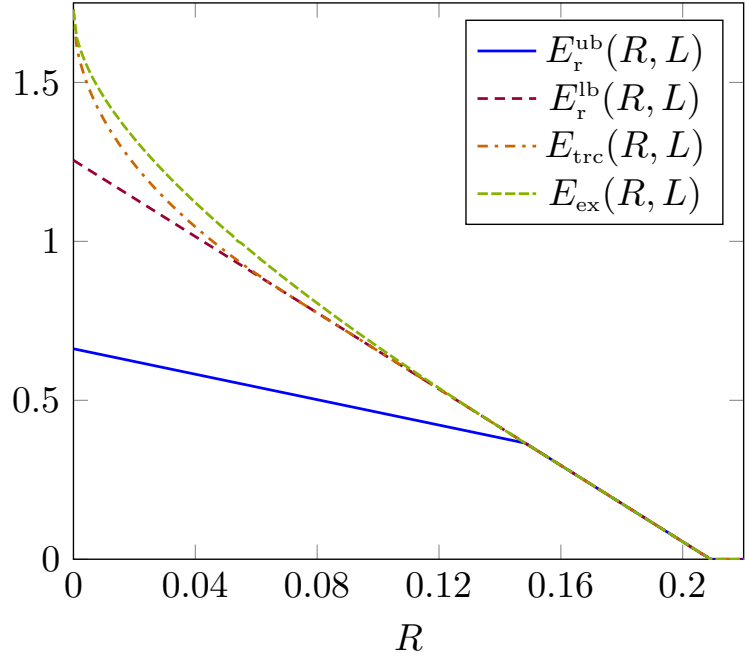
\begin{figure}[ht!]
	\centering
	\begin{tikzpicture}[scale=1.3]
	\begin{axis}[
	disabledatascaling,
	%x=25cm,
	%y=30cm,
	scaled x ticks=false,
	xticklabel style={/pgf/number format/fixed,
		/pgf/number format/precision=3},
	scaled y ticks=false,
	yticklabel style={/pgf/number format/fixed,
		/pgf/number format/precision=3},
	xlabel={$R$},
	xmin=0, xmax=0.22,
	ymin=0, ymax=1.75,
	xtick={0,0.04,0.08,0.12,0.16,0.20},
	legend pos=north east,
	%    ymajorgrids=true,
	%    grid style=dashed,
	]
	
	\addplot[smooth,color=blue,thick]
	table[row sep=crcr] 
	{
0	0.662014186	\\
0.001	0.660014186	\\
0.002	0.658014186	\\
0.003	0.656014186	\\
0.004	0.654014186	\\
0.005	0.652014186	\\
0.006	0.650014186	\\
0.007	0.648014186	\\
0.008	0.646014186	\\
0.009	0.644014186	\\
0.01	0.642014186	\\
0.011	0.640014186	\\
0.012	0.638014186	\\
0.013	0.636014186	\\
0.014	0.634014186	\\
0.015	0.632014186	\\
0.016	0.630014186	\\
0.017	0.628014186	\\
0.018	0.626014186	\\
0.019	0.624014186	\\
0.02	0.622014186	\\
0.021	0.620014186	\\
0.022	0.618014186	\\
0.023	0.616014186	\\
0.024	0.614014186	\\
0.025	0.612014186	\\
0.026	0.610014186	\\
0.027	0.608014186	\\
0.028	0.606014186	\\
0.029	0.604014186	\\
0.03	0.602014186	\\
0.031	0.600014186	\\
0.032	0.598014186	\\
0.033	0.596014186	\\
0.034	0.594014186	\\
0.035	0.592014186	\\
0.036	0.590014186	\\
0.037	0.588014186	\\
0.038	0.586014186	\\
0.039	0.584014186	\\
0.04	0.582014186	\\
0.041	0.580014186	\\
0.042	0.578014186	\\
0.043	0.576014186	\\
0.044	0.574014186	\\
0.045	0.572014186	\\
0.046	0.570014186	\\
0.047	0.568014186	\\
0.048	0.566014186	\\
0.049	0.564014186	\\
0.05	0.562014186	\\
0.051	0.560014186	\\
0.052	5.58E-01	\\
0.053	5.56E-01	\\
0.054	5.54E-01	\\
0.055	5.52E-01	\\
0.056	5.50E-01	\\
0.057	5.48E-01	\\
0.058	5.46E-01	\\
0.059	5.44E-01	\\
0.06	5.42E-01	\\
0.061	5.40E-01	\\
0.062	5.38E-01	\\
0.063	5.36E-01	\\
0.064	5.34E-01	\\
0.065	5.32E-01	\\
0.066	5.30E-01	\\
0.067	5.28E-01	\\
0.068	5.26E-01	\\
0.069	5.24E-01	\\
0.07	0.522014186	\\
0.071	0.520014186	\\
0.072	0.518014186	\\
0.073	0.516014186	\\
0.074	0.514014186	\\
0.075	0.512014186	\\
0.076	0.510014186	\\
0.077	0.508014186	\\
0.078	0.506014186	\\
0.079	0.504014186	\\
0.08	0.502014186	\\
0.081	0.500014186	\\
0.082	0.498014186	\\
0.083	0.496014186	\\
0.084	0.494014186	\\
0.085	0.492014186	\\
0.086	0.490014186	\\
0.087	0.488014186	\\
0.088	0.486014186	\\
0.089	0.484014186	\\
0.09	0.482014186	\\
0.091	0.480014186	\\
0.092	0.478014186	\\
0.093	0.476014186	\\
0.094	0.474014186	\\
0.095	0.472014186	\\
0.096	0.470014186	\\
0.097	0.468014186	\\
0.098	0.466014186	\\
0.099	0.464014186	\\
0.1	0.462014186	\\
0.101	0.460014186	\\
0.102	0.458014186	\\
0.103	0.456014186	\\
0.104	0.454014186	\\
0.105	0.452014186	\\
0.106	0.450014186	\\
0.107	0.448014186	\\
0.108	0.446014186	\\
0.109	0.444014186	\\
0.11	0.442014186	\\
0.111	0.440014186	\\
0.112	0.438014186	\\
0.113	0.436014186	\\
0.114	0.434014186	\\
0.115	0.432014186	\\
0.116	0.430014186	\\
0.117	0.428014186	\\
0.118	0.426014186	\\
0.119	0.424014186	\\
0.12	0.422014186	\\
0.121	0.420014186	\\
0.122	0.418014186	\\
0.123	0.416014186	\\
0.124	0.414014186	\\
0.125	0.412014186	\\
0.126	0.410014186	\\
0.127	0.408014186	\\
0.128	0.406014186	\\
0.129	0.404014186	\\
0.13	0.402014186	\\
0.131	0.400014186	\\
0.132	0.398014186	\\
0.133	0.396014186	\\
0.134	0.394014186	\\
0.135	0.392014186	\\
0.136	0.390014186	\\
0.137	0.388014186	\\
0.138	0.386014186	\\
0.139	0.384014186	\\
0.14	0.382014186	\\
0.141	0.380014186	\\
0.142	0.378014186	\\
0.143	0.376014186	\\
0.144	0.374014186	\\
0.145	0.372014186	\\
0.146	0.370014186	\\
0.147	0.368014186	\\
0.148	0.366014186	\\
0.149	0.361131892	\\
0.15	0.355131892	\\
0.151	0.349131892	\\
0.152	0.343131892	\\
0.153	0.337131892	\\
0.154	0.331131892	\\
0.155	0.325131892	\\
0.156	0.319131892	\\
0.157	0.313131892	\\
0.158	0.307131892	\\
0.159	0.301131892	\\
0.16	0.295131892	\\
0.161	0.289131892	\\
0.162	0.283131892	\\
0.163	0.277131892	\\
0.164	0.271131892	\\
0.165	0.265131892	\\
0.166	0.259131892	\\
0.167	0.253131892	\\
0.168	0.247131892	\\
0.169	0.241131892	\\
0.17	0.235131892	\\
0.171	0.229131892	\\
0.172	0.223131892	\\
0.173	0.217131892	\\
0.174	0.211131892	\\
0.175	0.205131892	\\
0.176	0.199131892	\\
0.177	0.193131892	\\
0.178	0.187131892	\\
0.179	0.181131892	\\
0.18	0.175131892	\\
0.181	0.169131892	\\
0.182	0.163131892	\\
0.183	0.157131892	\\
0.184	0.151131892	\\
0.185	0.145131892	\\
0.186	0.139131892	\\
0.187	0.133131892	\\
0.188	0.127131892	\\
0.189	0.121131892	\\
0.19	0.115131892	\\
0.191	0.109131892	\\
0.192	0.103131892	\\
0.193	0.097131892	\\
0.194	0.091131892	\\
0.195	0.085131892	\\
0.196	0.079131892	\\
0.197	0.073131892	\\
0.198	0.067131892	\\
0.199	0.061131892	\\
0.2	0.055131892	\\
0.201	0.049131892	\\
0.202	0.043131892	\\
0.203	0.037131892	\\
0.204	0.031131892	\\
0.205	0.025131892	\\
0.206	0.019131892	\\
0.207	0.013131892	\\
0.208	0.007131892	\\
0.209	0.001131892	\\
0.21	0	\\
0.211	0	\\
0.212	0	\\
0.213	0	\\
0.214	0	\\
0.215	0	\\
0.216	0	\\
0.217	0	\\
0.218	0	\\
0.219	0	\\
0.22	0	\\
	};
	\legend{}
	\addlegendentry{$E_{\mbox{\tiny r}}^{\mbox{\tiny ub}}(R,L)$}

		\addplot[smooth,color=black!20!purple,thick,dash pattern={on 3pt off 2pt}]
	table[row sep=crcr]
	{
0	1.255133694	\\
0.01	1.195133694	\\
0.02	1.135133694	\\
0.03	1.075133694	\\
0.04	1.015133694	\\
0.05	0.955133694	\\
0.06	0.895133694	\\
0.07	0.835133694	\\
0.08	0.775133694	\\
0.09	0.715133694	\\
0.1	0.655133694	\\
0.11	0.595133694	\\
0.12	0.535133694	\\
0.13	0.475133694	\\
0.14	0.415133694	\\
0.15	0.35514	\\
0.16	0.29514	\\
0.17	0.23514	\\
0.18	0.17514	\\
0.19	0.11514	\\
0.2	0.05514	\\
0.21	0	\\
0.22	0	\\
0.23	0	\\
0.24	0	\\
0.25	0	\\
0.26	0	\\
0.27	0	\\
0.28	0	\\
	};
	\addlegendentry{$E_{\mbox{\tiny r}}^{\mbox{\tiny lb}}(R,L)$}

	\addplot[smooth,color=black!20!orange,thick,dash pattern={on 3pt off 2pt on 1pt off 2pt}]
	table[row sep=crcr]
	{
0	1.73E+00	\\
0.001	1.617791666	\\
0.002	1.57284677	\\
0.003	1.538279835	\\
0.004	1.509222092	\\
0.005	1.483621859	\\
0.006	1.46020205	\\
0.007	1.438814542	\\
0.008	1.418849377	\\
0.009	1.40068322	\\
0.01	1.382853611	\\
0.011	1.366233846	\\
0.012	1.350327942	\\
0.013	1.335101975	\\
0.014	1.320523888	\\
0.015	1.306169892	\\
0.016	1.292422924	\\
0.017	1.279253168	\\
0.018	1.266632157	\\
0.019	1.254178485	\\
0.02	1.241892856	\\
0.021	1.230111485	\\
0.022	1.218481438	\\
0.023	1.20732117	\\
0.024	1.19629607	\\
0.025	1.185708283	\\
0.026	1.174946577	\\
0.027	1.164606877	\\
0.028	1.154665344	\\
0.029	1.14455964	\\
0.03	1.134837556	\\
0.031	1.125222041	\\
0.032	1.115960756	\\
0.033	1.106552161	\\
0.034	1.097483786	\\
0.035	1.088509041	\\
0.036	1.079628144	\\
0.037	1.071052606	\\
0.038	1.062353639	\\
0.039	1.053946834	\\
0.04	1.04562146	\\
0.041	1.037377883	\\
0.042	1.029216567	\\
0.043	1.021307938	\\
0.044	1.013305341	\\
0.045	1.005542837	\\
0.046	0.997850691	\\
0.047	0.990229125	\\
0.048	0.982678556	\\
0.049	0.975199312	\\
0.05	0.967916734	\\
0.051	0.960574585	\\
0.052	9.53E-01	\\
0.053	9.46E-01	\\
0.054	9.39E-01	\\
0.055	9.32E-01	\\
0.056	9.25E-01	\\
0.057	9.18E-01	\\
0.058	9.12E-01	\\
0.059	9.05E-01	\\
0.06	8.98E-01	\\
0.061	8.92E-01	\\
0.062	8.85E-01	\\
0.063	8.79E-01	\\
0.064	8.72E-01	\\
0.065	8.66E-01	\\
0.066	8.60E-01	\\
0.067	8.54E-01	\\
0.068	8.47E-01	\\
0.069	8.41E-01	\\
0.07	0.835156902	\\
0.071	0.829131892	\\
0.072	0.823131892	\\
0.073	0.817131892	\\
0.074	0.811131892	\\
0.075	0.805131892	\\
0.076	0.799131892	\\
0.077	0.793131892	\\
0.078	0.787131892	\\
0.079	0.781131892	\\
0.08	0.775131892	\\
0.081	0.769131892	\\
0.082	0.763131892	\\
0.083	0.757131892	\\
0.084	0.751131892	\\
0.085	0.745131892	\\
0.086	0.739131892	\\
0.087	0.733131892	\\
0.088	0.727131892	\\
0.089	0.721131892	\\
0.09	0.715131892	\\
0.091	0.709131892	\\
0.092	0.703131892	\\
0.093	0.697131892	\\
0.094	0.691131892	\\
0.095	0.685131892	\\
0.096	0.679131892	\\
0.097	0.673131892	\\
0.098	0.667131892	\\
0.099	0.661131892	\\
0.1	0.655131892	\\
0.101	0.649131892	\\
0.102	0.643131892	\\
0.103	0.637131892	\\
0.104	0.631131892	\\
0.105	0.625131892	\\
0.106	0.619131892	\\
0.107	0.613131892	\\
0.108	0.607131892	\\
0.109	0.601131892	\\
0.11	0.595131892	\\
0.111	0.589131892	\\
0.112	0.583131892	\\
0.113	0.577131892	\\
0.114	0.571131892	\\
0.115	0.565131892	\\
0.116	0.559131892	\\
0.117	0.553131892	\\
0.118	0.547131892	\\
0.119	0.541131892	\\
0.12	0.535131892	\\
0.121	0.529131892	\\
0.122	0.523131892	\\
0.123	0.517131892	\\
0.124	0.511131892	\\
0.125	0.505131892	\\
0.126	0.499131892	\\
0.127	0.493131892	\\
0.128	0.487131892	\\
0.129	0.481131892	\\
0.13	0.475131892	\\
0.131	0.469131892	\\
0.132	0.463131892	\\
0.133	0.457131892	\\
0.134	0.451131892	\\
0.135	0.445131892	\\
0.136	0.439131892	\\
0.137	0.433131892	\\
0.138	0.427131892	\\
0.139	0.421131892	\\
0.14	0.415131892	\\
0.141	0.409131892	\\
0.142	0.403131892	\\
0.143	0.397131892	\\
0.144	0.391131892	\\
0.145	0.385131892	\\
0.146	0.379131892	\\
0.147	0.373131892	\\
0.148	0.367131892	\\
0.149	0.361131892	\\
0.15	0.355131892	\\
0.151	0.349131892	\\
0.152	0.343131892	\\
0.153	0.337131892	\\
0.154	0.331131892	\\
0.155	0.325131892	\\
0.156	0.319131892	\\
0.157	0.313131892	\\
0.158	0.307131892	\\
0.159	0.301131892	\\
0.16	0.295131892	\\
0.161	0.289131892	\\
0.162	0.283131892	\\
0.163	0.277131892	\\
0.164	0.271131892	\\
0.165	0.265131892	\\
0.166	0.259131892	\\
0.167	0.253131892	\\
0.168	0.247131892	\\
0.169	0.241131892	\\
0.17	0.235131892	\\
0.171	0.229131892	\\
0.172	0.223131892	\\
0.173	0.217131892	\\
0.174	0.211131892	\\
0.175	0.205131892	\\
0.176	0.199131892	\\
0.177	0.193131892	\\
0.178	0.187131892	\\
0.179	0.181131892	\\
0.18	0.175131892	\\
0.181	0.169131892	\\
0.182	0.163131892	\\
0.183	0.157131892	\\
0.184	0.151131892	\\
0.185	0.145131892	\\
0.186	0.139131892	\\
0.187	0.133131892	\\
0.188	0.127131892	\\
0.189	0.121131892	\\
0.19	0.115131892	\\
0.191	0.109131892	\\
0.192	0.103131892	\\
0.193	0.097131892	\\
0.194	0.091131892	\\
0.195	0.085131892	\\
0.196	0.079131892	\\
0.197	0.073131892	\\
0.198	0.067131892	\\
0.199	0.061131892	\\
0.2	0.055131892	\\
0.201	0.049131892	\\
0.202	0.043131892	\\
0.203	0.037131892	\\
0.204	0.031131892	\\
0.205	0.025131892	\\
0.206	0.019131892	\\
0.207	0.013131892	\\
0.208	0.007131892	\\
0.209	0.001131892	\\
0.21	0	\\
0.211	0	\\
0.212	0	\\
0.213	0	\\
0.214	0	\\
0.215	0	\\
0.216	0	\\
0.217	0	\\
0.218	0	\\
0.219	0	\\
0.22	0	\\
	};
	\addlegendentry{$E_{\mbox{\tiny trc}}(R,L)$}

	\addplot[smooth,color=black!30!lime,thick,dash pattern={on 3pt off 1pt}]
table[row sep=crcr]
{
0	1.72693882	\\
0.001	1.647220755	\\
0.002	1.611791666	\\
0.003	1.58456786	\\
0.004	1.56084677	\\
0.005	1.53984066	\\
0.006	1.520279835	\\
0.007	1.502094929	\\
0.008	1.485222092	\\
0.009	1.469073548	\\
0.01	1.453621859	\\
0.011	1.438328884	\\
0.012	1.42420205	\\
0.013	1.410200935	\\
0.014	1.396814542	\\
0.015	1.384020727	\\
0.016	1.370849377	\\
0.017	1.358724273	\\
0.018	1.34668322	\\
0.019	1.334726386	\\
0.02	1.322853611	\\
0.021	1.311507646	\\
0.022	1.300233846	\\
0.023	1.289032361	\\
0.024	1.278327942	\\
0.025	1.267684401	\\
0.026	1.257101975	\\
0.027	1.246580789	\\
0.028	1.236523888	\\
0.029	1.226119972	\\
0.03	1.216169892	\\
0.031	1.20627084	\\
0.032	1.196422924	\\
0.033	1.187003626	\\
0.034	1.177253168	\\
0.035	1.167921775	\\
0.036	1.158632157	\\
0.037	1.149384399	\\
0.038	1.140178485	\\
0.039	1.131014607	\\
0.04	1.121892856	\\
0.041	1.1131534	\\
0.042	1.104111485	\\
0.043	1.095443178	\\
0.044	1.086481438	\\
0.045	1.077884482	\\
0.046	1.06932117	\\
0.047	1.060791675	\\
0.048	1.05229607	\\
0.049	1.043834426	\\
0.05	1.035708283	\\
0.051	1.027310341	\\
0.052	1.02E+00	\\
0.053	1.01E+00	\\
0.054	1.00E+00	\\
0.055	9.95E-01	\\
0.056	9.87E-01	\\
0.057	9.79E-01	\\
0.058	9.71E-01	\\
0.059	9.63E-01	\\
0.06	9.55E-01	\\
0.061	9.47E-01	\\
0.062	9.39E-01	\\
0.063	9.32E-01	\\
0.064	9.24E-01	\\
0.065	9.16E-01	\\
0.066	9.09E-01	\\
0.067	9.01E-01	\\
0.068	8.93E-01	\\
0.069	8.86E-01	\\
0.07	0.878509041	\\
0.071	0.871167634	\\
0.072	0.863628144	\\
0.073	0.856330321	\\
0.074	0.849052606	\\
0.075	0.841795234	\\
0.076	0.834353639	\\
0.077	0.827140044	\\
0.078	0.819946834	\\
0.079	0.812773862	\\
0.08	0.80562146	\\
0.081	0.798489388	\\
0.082	0.791377883	\\
0.083	0.78428699	\\
0.084	0.777216567	\\
0.085	0.770340503	\\
0.086	0.763307938	\\
0.087	0.756296264	\\
0.088	0.749305341	\\
0.089	0.742495463	\\
0.09	0.735542837	\\
0.091	0.728611009	\\
0.092	0.721850691	\\
0.093	0.714957331	\\
0.094	0.708229125	\\
0.095	0.701374501	\\
0.096	0.694678556	\\
0.097	0.687997299	\\
0.098	0.681199312	\\
0.099	0.674550629	\\
0.1	0.667916734	\\
0.101	0.661175773	\\
0.102	0.654574585	\\
0.103	0.647988475	\\
0.104	0.6414172	\\
0.105	0.634751675	\\
0.106	0.628213414	\\
0.107	0.621690463	\\
0.108	0.615182398	\\
0.109	0.608689713	\\
0.11	0.602211984	\\
0.111	0.595749704	\\
0.112	0.589302453	\\
0.113	0.582870722	\\
0.114	0.576454092	\\
0.115	0.570053055	\\
0.116	0.563667193	\\
0.117	0.557296998	\\
0.118	0.551011785	\\
0.119	0.544669431	\\
0.12	0.538342487	\\
0.121	0.532031264	\\
0.122	0.525735529	\\
0.123	0.519510307	\\
0.124	0.51324284	\\
0.125	0.506991152	\\
0.126	0.500801141	\\
0.127	0.494577833	\\
0.128	0.48837042	\\
0.129	0.482215874	\\
0.13	0.476037138	\\
0.131	0.469905518	\\
0.132	0.463755595	\\
0.133	0.45764701	\\
0.134	0.451525952	\\
0.135	0.445440603	\\
0.136	0.43934855	\\
0.137	0.433286551	\\
0.138	0.427223647	\\
0.139	0.421185114	\\
0.14	0.415156902	\\
0.141	0.409136556	\\
0.142	0.403131892	\\
0.143	0.397131892	\\
0.144	0.391131892	\\
0.145	0.385131892	\\
0.146	0.379131892	\\
0.147	0.373131892	\\
0.148	0.367131892	\\
0.149	0.361131892	\\
0.15	0.355131892	\\
0.151	0.349131892	\\
0.152	0.343131892	\\
0.153	0.337131892	\\
0.154	0.331131892	\\
0.155	0.325131892	\\
0.156	0.319131892	\\
0.157	0.313131892	\\
0.158	0.307131892	\\
0.159	0.301131892	\\
0.16	0.295131892	\\
0.161	0.289131892	\\
0.162	0.283131892	\\
0.163	0.277131892	\\
0.164	0.271131892	\\
0.165	0.265131892	\\
0.166	0.259131892	\\
0.167	0.253131892	\\
0.168	0.247131892	\\
0.169	0.241131892	\\
0.17	0.235131892	\\
0.171	0.229131892	\\
0.172	0.223131892	\\
0.173	0.217131892	\\
0.174	0.211131892	\\
0.175	0.205131892	\\
0.176	0.199131892	\\
0.177	0.193131892	\\
0.178	0.187131892	\\
0.179	0.181131892	\\
0.18	0.175131892	\\
0.181	0.169131892	\\
0.182	0.163131892	\\
0.183	0.157131892	\\
0.184	0.151131892	\\
0.185	0.145131892	\\
0.186	0.139131892	\\
0.187	0.133131892	\\
0.188	0.127131892	\\
0.189	0.121131892	\\
0.19	0.115131892	\\
0.191	0.109131892	\\
0.192	0.103131892	\\
0.193	0.097131892	\\
0.194	0.091131892	\\
0.195	0.085131892	\\
0.196	0.079131892	\\
0.197	0.073131892	\\
0.198	0.067131892	\\
0.199	0.061131892	\\
0.2	0.055131892	\\
0.201	0.049131892	\\
0.202	0.043131892	\\
0.203	0.037131892	\\
0.204	0.031131892	\\
0.205	0.025131892	\\
0.206	0.019131892	\\
0.207	0.013131892	\\
0.208	0.007131892	\\
0.209	0.001131892	\\
0.21	0	\\
0.211	0	\\
0.212	0	\\
0.213	0	\\
0.214	0	\\
0.215	0	\\
0.216	0	\\
0.217	0	\\
0.218	0	\\
0.219	0	\\
0.22	0	\\
};
\addlegendentry{$E_{\mbox{\tiny ex}}(R,L)$}

	\end{axis}
	
	\end{tikzpicture}
	\caption{Error exponents for the $z$--channel ($w=0.9$ and $L=3$).}\label{Z-Channel-Numeric}
\end{figure}

We demonstrate some of the above discussed properties of the different error exponents in a specific numerical example. Consider the $z$-channel with alphabets $\calX=\calY=\{0,1\}$, conditional probabilities of $W(0|0)=1-W(1|0)=0.9$, and let the input assignment be $Q_{X}(0)=Q_{X}(1)=1/2$. Also, we use the decoding metric $g(Q) = \mathbb{E}_{Q} \log W(Y|X)$, which is equivalent to ML decoding. In Figure \ref{Z-Channel-Numeric}, all four error exponents are plotted for the choice $L=3$. As discussed earlier, at low coding rates, $E_{\mbox{\tiny r}}^{\mbox{\tiny lb}}(R,L) > E_{\mbox{\tiny r}}^{\mbox{\tiny ub}}(R,L)$, but for any $R \geq 0.1483$, $E_{\mbox{\tiny r}}^{\mbox{\tiny lb}}(R,L) = E_{\mbox{\tiny r}}^{\mbox{\tiny ub}}(R,L)$, i.e., we have an exact random coding error exponent. Although not shown here, this tightness holds for any coding rate for $L=1$. 
At low coding rates, indeed $E_{\mbox{\tiny ex}}(R,L) > E_{\mbox{\tiny trc}}(R,L)$, and both of these exponent functions strictly improve upon the random coding error exponent, similarly as in ordinary channel coding. At high coding rates, all the exponent functions coincide. 
As for the maximal attainable coding rate, all exponent functions are strictly positive as long as $R < 0.2092$. This maximal rate is also predicted by the lower bound in \eqref{Maximal_Rate}, which is relatively surprising, since the bound in \eqref{Maximal_Rate} was derived from an exponent function which is related to a GLD with decoding metric $g(Q)=I_{Q}(X;Y)$, not the matched decoder.

\subsection{Optimal Decoding}

In order to present our first result in this section, which is a lower bound to the random coding error exponent, we first make a few definitions. 
A DMC $W$ is called symmetric if its probability transition matrix is doubly stochastic, i.e., every row is given by a permutation of any other row, and the same for its columns.
For $x,x' \in \calX$, define 
\begin{align}
B(x,x') = \sum_{y \in \calY} 
\sqrt{W(y|x) W(y|x')}.
\end{align}
%as well as
%\begin{align}
%\Phi = \sum_{x \in \calX} \sum_{x' \in \calX} P_{X}(x) P_{X}(x') [B(x,x')]^{2},
%\end{align}
%and 
%\begin{align}
%\Psi = \sum_{x \in \calX}  \sum_{x' \in \calX} P_{X}(x) P_{X}(x') B(x,x').
%\end{align}
For $\sigma \geq 1$, define
\begin{align}
\Xi(\sigma) = \sum_{x \in \calX} \sum_{x' \in \calX} P_{X}(x) P_{X}(x') [B(x,x')]^{2/\sigma} ,
\end{align}
and 
\begin{align}
\Omega(\sigma) = \sum_{x \in \calX}  \sum_{x' \in \calX} P_{X}(x) P_{X}(x') [B(x,x')]^{1/\sigma}.
\end{align} 
Also, define the exponent function
\begin{align} \label{OPT_RC_exponent}
E_{\mbox{\tiny r}}^{\mbox{\tiny opt}}(R) =  \left[ \min \left\{ -\log \Xi(1) - 2R, -2 \log \Omega(1) - 3R \right\} \right]_{+}.
\end{align}
The proof of the following result is very similar to the proof of Theorem \ref{Theorem_EX} below, and hence omitted.
\begin{theorem} \label{Theorem_OPT}
	Assume that $W$ is a symmetric channel and that $P_{X}$ is the uniform distribution. Then, under optimal decoding,
	\begin{align}
	\lim_{n \to \infty} -\frac{1}{n} \log \mathbb{E} \left[\OEP\right] \geq E_{\mbox{\tiny r}}^{\mbox{\tiny opt}}(R).
	\end{align}
\end{theorem}

\subsubsection*{Discussion}
As can be seen in \eqref{OPT_RC_exponent}, the overall error event may be dominated by two different error events, depending on the quality of the channel and on the coding rate. 
This fact has already been asserted in \cite{BEES}, but here, we elaborate more on it. 
On the one hand, for relatively good channels, and for any coding rate, the dominating error event is when two bees are switched. On the other hand, for relatively bad channels, it depends on the coding rate; at relative low coding rates, two bees are incorrectly decoded, but at relatively high rates, three bees are erroneously identified. In order to demonstrate these issues more quantitatively, we now refer to the BSC.    
For a BSC with crossover probability $p \in (0,1/2)$, one easily finds that
\begin{align}
\Xi(1) &= \frac{1}{2} + 2p(1-p), \\
\Omega(1) &= \frac{1}{2} + \sqrt{p(1-p)} .
\end{align}
Then, the critical channel parameter in this case is the one that solves the equation:
\begin{align}
\left[\frac{1}{2} + 2p(1-p)\right]^{3} = \left[\frac{1}{2} + \sqrt{p(1-p)}\right]^{4}, 
\end{align}
which can be found numerically as $p^{*} \approx 0.01466$. Furthermore, for BSCs with a crossover parameter in the range $(p^{*},1/2)$, the phase transition in the rate axis occurs at 
\begin{align}
R^{*}(p) = \log \frac{\Xi(1)}{\Omega^{2}(1)}.
\end{align}

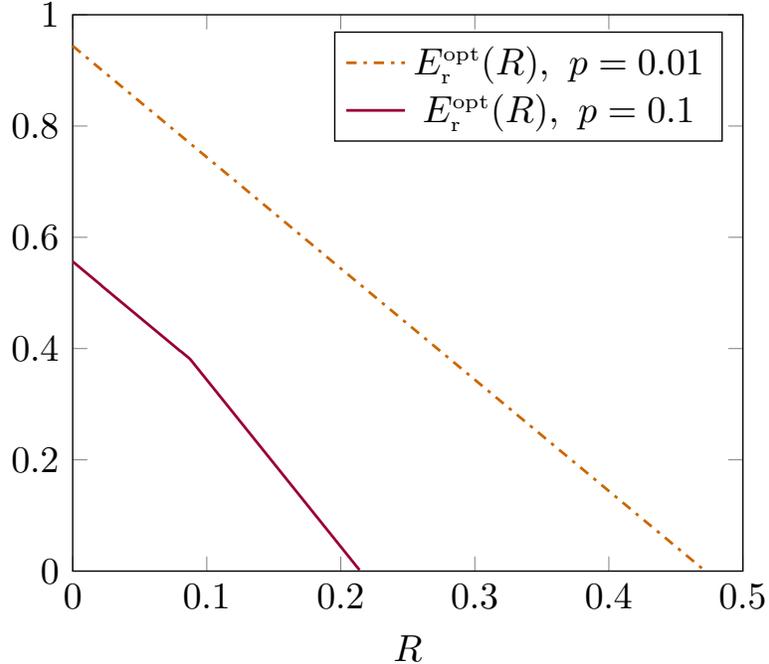
\begin{figure}[ht!]
	\centering
	\begin{tikzpicture}[scale=1.3]
	\begin{axis}[
	disabledatascaling,
	%x=25cm,
	%y=30cm,
	scaled x ticks=false,
	xticklabel style={/pgf/number format/fixed,
		/pgf/number format/precision=3},
	scaled y ticks=false,
	yticklabel style={/pgf/number format/fixed,
		/pgf/number format/precision=3},
	xlabel={$R$},
	xmin=0, xmax=0.5,
	ymin=0, ymax=1.0,
	xtick={0,0.1,0.2,0.3,0.4,0.5},
	legend pos=north east,
	%    ymajorgrids=true,
	%    grid style=dashed,
	]
	
	\addplot[smooth,color=black!20!orange,thick,dash pattern={on 3pt off 2pt on 1pt off 2pt}]
	table[row sep=crcr] 
	{
		0	0.943971461	\\
		0.01	0.923971461	\\
		0.02	0.903971461	\\
		0.03	0.883971461	\\
		0.04	0.863971461	\\
		0.05	0.843971461	\\
		0.06	0.823971461	\\
		0.07	0.803971461	\\
		0.08	0.783971461	\\
		0.09	0.763971461	\\
		0.1	0.743971461	\\
		0.11	0.723971461	\\
		0.12	0.703971461	\\
		0.13	0.683971461	\\
		0.14	0.663971461	\\
		0.15	0.643971461	\\
		0.16	0.623971461	\\
		0.17	0.603971461	\\
		0.18	0.583971461	\\
		0.19	0.563971461	\\
		0.2	0.543971461	\\
		0.21	0.523971461	\\
		0.22	0.503971461	\\
		0.23	0.483971461	\\
		0.24	0.463971461	\\
		0.25	0.443971461	\\
		0.26	0.423971461	\\
		0.27	0.403971461	\\
		0.28	0.383971461	\\
		0.29	0.363971461	\\
		0.3	0.343971461	\\
		0.31	0.323971461	\\
		0.32	0.303971461	\\
		0.33	0.283971461	\\
		0.34	0.263971461	\\
		0.35	0.243971461	\\
		0.36	0.223971461	\\
		0.37	0.203971461	\\
		0.38	0.183971461	\\
		0.39	0.163971461	\\
		0.4	0.143971461	\\
		0.41	0.123971461	\\
		0.42	0.103971461	\\
		0.43	0.083971461	\\
		0.44	0.063971461	\\
		0.45	0.043971461	\\
		0.46	0.023971461	\\
		0.47	0.003971461	\\
	};
	\legend{}
	\addlegendentry{$E_{\mbox{\tiny r}}^{\mbox{\tiny opt}}(R),~p=0.01$}

	\addplot[smooth,color=black!20!purple,thick]
	table[row sep=crcr]
	{
0	0.556393293	\\
0.001	0.554393293	\\
0.002	0.552393293	\\
0.003	0.550393293	\\
0.004	0.548393294	\\
0.005	0.546393294	\\
0.006	0.544393294	\\
0.007	0.542393294	\\
0.008	0.540393294	\\
0.009	0.538393295	\\
0.01	0.536393295	\\
0.011	0.534393295	\\
0.012	0.532393295	\\
0.013	0.530393295	\\
0.014	0.528393296	\\
0.015	0.526393296	\\
0.016	0.524393296	\\
0.017	0.522393296	\\
0.018	0.520393296	\\
0.019	0.518393297	\\
0.02	0.516393297	\\
0.021	0.514393297	\\
0.022	5.12E-01	\\
0.023	5.10E-01	\\
0.024	5.08E-01	\\
0.025	5.06E-01	\\
0.026	5.04E-01	\\
0.027	5.02E-01	\\
0.028	5.00E-01	\\
0.029	4.98E-01	\\
0.03	4.96E-01	\\
0.031	4.94E-01	\\
0.032	4.92E-01	\\
0.033	4.90E-01	\\
0.034	4.88E-01	\\
0.035	4.86E-01	\\
0.036	4.84E-01	\\
0.037	4.82E-01	\\
0.038	4.80E-01	\\
0.039	4.78E-01	\\
0.04	4.76E-01	\\
0.041	4.74E-01	\\
0.042	4.72E-01	\\
0.043	4.70E-01	\\
0.044	4.68E-01	\\
0.045	4.66E-01	\\
0.046	4.64E-01	\\
0.047	4.62E-01	\\
0.048	4.60E-01	\\
0.049	4.58E-01	\\
0.05	4.56E-01	\\
0.051	4.54E-01	\\
0.052	4.52E-01	\\
0.053	4.50E-01	\\
0.054	4.48E-01	\\
0.055	4.46E-01	\\
0.056	4.44E-01	\\
0.057	4.42E-01	\\
0.058	4.40E-01	\\
0.059	4.38E-01	\\
0.06	4.36E-01	\\
0.061	4.34E-01	\\
0.062	4.32E-01	\\
0.063	4.30E-01	\\
0.064	4.28E-01	\\
0.065	4.26E-01	\\
0.066	4.24E-01	\\
0.067	4.22E-01	\\
0.068	4.20E-01	\\
0.069	4.18E-01	\\
0.07	4.16E-01	\\
0.071	4.14E-01	\\
0.072	4.12E-01	\\
0.073	4.10E-01	\\
0.074	4.08E-01	\\
0.075	4.06E-01	\\
0.076	4.04E-01	\\
0.077	4.02E-01	\\
0.078	4.00E-01	\\
0.079	3.98E-01	\\
0.08	3.96E-01	\\
0.081	0.394393309	\\
0.082	0.392393309	\\
0.083	0.390393309	\\
0.084	0.38839331	\\
0.085	0.38639331	\\
0.086	0.38439331	\\
0.087	0.38239331	\\
0.088	0.379856152	\\
0.089	0.376856152	\\
0.09	0.373856152	\\
0.091	0.370856153	\\
0.092	0.367856153	\\
0.093	0.364856153	\\
0.094	0.361856154	\\
0.095	0.358856154	\\
0.096	0.355856154	\\
0.097	0.352856154	\\
0.098	0.349856155	\\
0.099	0.346856155	\\
0.1	0.343856155	\\
0.101	0.340856156	\\
0.102	0.337856156	\\
0.103	0.334856156	\\
0.104	0.331856157	\\
0.105	0.328856157	\\
0.106	0.325856157	\\
0.107	0.322856157	\\
0.108	0.319856158	\\
0.109	0.316856158	\\
0.11	0.313856158	\\
0.111	0.310856159	\\
0.112	0.307856159	\\
0.113	0.304856159	\\
0.114	0.30185616	\\
0.115	0.29885616	\\
0.116	0.29585616	\\
0.117	0.29285616	\\
0.118	0.289856161	\\
0.119	0.286856161	\\
0.12	0.283856161	\\
0.121	0.280856162	\\
0.122	0.277856162	\\
0.123	0.274856162	\\
0.124	0.271856163	\\
0.125	0.268856163	\\
0.126	0.265856163	\\
0.127	0.262856163	\\
0.128	0.259856164	\\
0.129	0.256856164	\\
0.13	0.253856164	\\
0.131	0.250856165	\\
0.132	0.247856165	\\
0.133	0.244856165	\\
0.134	0.241856166	\\
0.135	0.238856166	\\
0.136	0.235856166	\\
0.137	0.232856166	\\
0.138	0.229856167	\\
0.139	0.226856167	\\
0.14	0.223856167	\\
0.141	0.220856168	\\
0.142	0.217856168	\\
0.143	0.214856168	\\
0.144	0.211856169	\\
0.145	0.208856169	\\
0.146	0.205856169	\\
0.147	0.202856169	\\
0.148	0.19985617	\\
0.149	0.19685617	\\
0.15	0.19385617	\\
0.151	0.190856171	\\
0.152	0.187856171	\\
0.153	0.184856171	\\
0.154	0.181856172	\\
0.155	0.178856172	\\
0.156	0.175856172	\\
0.157	0.172856172	\\
0.158	0.169856173	\\
0.159	0.166856173	\\
0.16	0.163856173	\\
0.161	0.160856174	\\
0.162	0.157856174	\\
0.163	0.154856174	\\
0.164	0.151856175	\\
0.165	0.148856175	\\
0.166	0.145856175	\\
0.167	0.142856175	\\
0.168	0.139856176	\\
0.169	0.136856176	\\
0.17	0.133856176	\\
0.171	0.130856177	\\
0.172	0.127856177	\\
0.173	0.124856177	\\
0.174	0.121856178	\\
0.175	0.118856178	\\
0.176	0.115856178	\\
0.177	0.112856178	\\
0.178	0.109856179	\\
0.179	0.106856179	\\
0.18	0.103856179	\\
0.181	0.10085618	\\
0.182	0.09785618	\\
0.183	0.09485618	\\
0.184	0.091856181	\\
0.185	0.088856181	\\
0.186	0.085856181	\\
0.187	0.082856181	\\
0.188	0.079856182	\\
0.189	0.076856182	\\
0.19	0.073856182	\\
0.191	0.070856183	\\
0.192	0.067856183	\\
0.193	0.064856183	\\
0.194	0.061856184	\\
0.195	0.058856184	\\
0.196	0.055856184	\\
0.197	0.052856184	\\
0.198	0.049856185	\\
0.199	0.046856185	\\
0.2	0.043856185	\\
0.201	0.040856186	\\
0.202	0.037856186	\\
0.203	0.034856186	\\
0.204	0.031856187	\\
0.205	0.028856187	\\
0.206	0.025856187	\\
0.207	0.022856187	\\
0.208	0.019856188	\\
0.209	0.016856188	\\
0.21	0.013856188	\\
0.211	0.010856189	\\
0.212	0.007856189	\\
0.213	0.004856189	\\
0.214	0.00185619	\\
	};
	\addlegendentry{$E_{\mbox{\tiny r}}^{\mbox{\tiny opt}}(R),~p=0.1$}

	\end{axis}
	
	\end{tikzpicture}
	\caption{Random coding error exponents for the BSC under optimal decoding.}\label{fig:BSC-RC}
\end{figure}

In Figure \ref{fig:BSC-RC} we plot $E_{\mbox{\tiny r}}^{\mbox{\tiny opt}}(R)$ for two different values of $p$. As can be seen there, for $p < p^{*}$, the exponent function decreases with a slope of $-2$ at all coding rates (which is related to the error event of switching between two bees), but for $p > p^{*}$, it decreases with a slope of $-2$ as long as $R \leq R^{*}(p) \approx 0.087$, and with a slope of $-3$ otherwise (exchanging between three bees).

Similarly to ordinary channel coding, also in this scenario, the random coding error exponent can be improved at relatively low coding rates by expurgation. It should be pointed out, however, that the processes of expurgation in ordinary channel coding and in the bee identification problem slightly differ from one another. In ordinary channel coding, one draws $2M$ codewords, and expurgate the $M$ codewords with the highest conditional error probabilities, such that all remaining ones have error probabilities bounds above by $e^{-n E_{\mbox{\tiny ex}}(R)}$, where $E_{\mbox{\tiny ex}}(R)$ is the expurgated error exponent. In the bee identification problem, on the other hand, the specific performance of the individual codewords are no longer of interest, since all the codewords are being used together. Here, too, we draw $2M$ codewords, but prove the existence of a subset of $M$ codewords with a good collective behavior.

Define the following exponent function:
\begin{align} \label{OPTIMUM_Expurgated}
E_{\mbox{\tiny ex}}^{\mbox{\tiny opt}}(R) = \sup_{\sigma \geq 1} \left\{ \sigma \cdot \min \left[ -\log \Xi(\sigma)-2R, -2 \log \Omega(\sigma)-3R \right] \right\}.
\end{align}

Then, our second result is the following theorem, which is proved in Appendix F.
\begin{theorem} \label{Theorem_EX}
	Assume that $W$ is a symmetric channel and that $P_{X}$ is the uniform distribution. Then, under optimal decoding, there exists a sequence of i.i.d.\ codes, $\{\calC_{n},~n=1,2,\ldots\}$, such that 
	\begin{align}
	\liminf_{n \to \infty} -\frac{1}{n} \log \OEP \geq E_{\mbox{\tiny ex}}^{\mbox{\tiny opt}}(R).
	\end{align}
\end{theorem}

%\subsubsection*{Discussion}
The proof of Theorem \ref{Theorem_EX} relies on ideas and techniques from both \cite{Gal65} and \cite{BEES}. Most importantly, the proof in Appendix F uses the fact that every permutation of a set (e.g., of bees) is equivalent to a composition of disjoint cycles \cite{Her1975}. Since each cycle of incorrectly decoded bees can be analyzed relatively easily, we are able, exactly as in \cite{BEES}, to sum up the contributions of all possible permutations.    

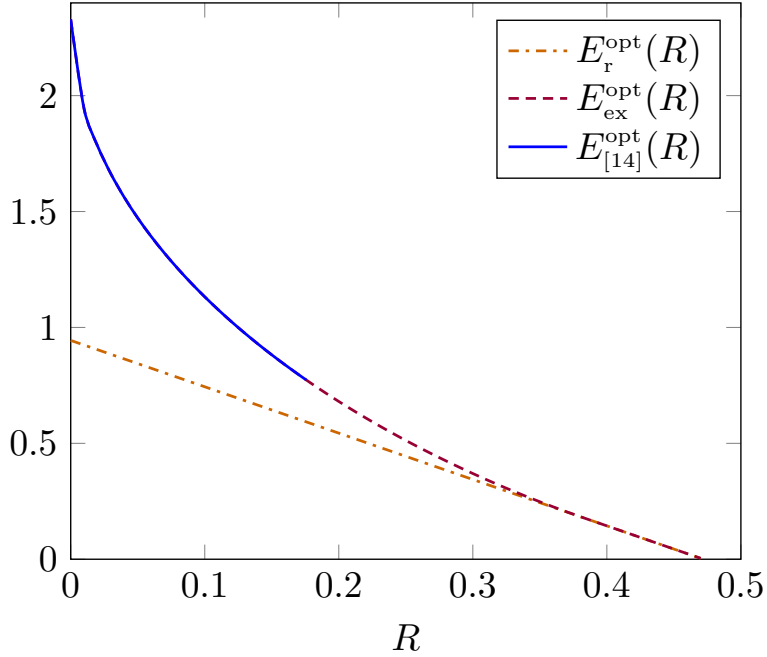
\begin{figure}[ht!]
	\centering
	\begin{tikzpicture}[scale=1.3]
	\begin{axis}[
	disabledatascaling,
	%x=25cm,
	%y=30cm,
	scaled x ticks=false,
	xticklabel style={/pgf/number format/fixed,
		/pgf/number format/precision=3},
	scaled y ticks=false,
	yticklabel style={/pgf/number format/fixed,
		/pgf/number format/precision=3},
	xlabel={$R$},
	xmin=0, xmax=0.5,
	ymin=0, ymax=2.4,
	xtick={0,0.1,0.2,0.3,0.4,0.5},
	legend pos=north east,
	%    ymajorgrids=true,
	%    grid style=dashed,
	]
	
	\addplot[smooth,color=black!20!orange,thick,dash pattern={on 3pt off 2pt on 1pt off 2pt}]
	table[row sep=crcr] 
	{
		0	0.943971461	\\
		0.01	0.923971461	\\
		0.02	0.903971461	\\
		0.03	0.883971461	\\
		0.04	0.863971461	\\
		0.05	0.843971461	\\
		0.06	0.823971461	\\
		0.07	0.803971461	\\
		0.08	0.783971461	\\
		0.09	0.763971461	\\
		0.1	0.743971461	\\
		0.11	0.723971461	\\
		0.12	0.703971461	\\
		0.13	0.683971461	\\
		0.14	0.663971461	\\
		0.15	0.643971461	\\
		0.16	0.623971461	\\
		0.17	0.603971461	\\
		0.18	0.583971461	\\
		0.19	0.563971461	\\
		0.2	0.543971461	\\
		0.21	0.523971461	\\
		0.22	0.503971461	\\
		0.23	0.483971461	\\
		0.24	0.463971461	\\
		0.25	0.443971461	\\
		0.26	0.423971461	\\
		0.27	0.403971461	\\
		0.28	0.383971461	\\
		0.29	0.363971461	\\
		0.3	0.343971461	\\
		0.31	0.323971461	\\
		0.32	0.303971461	\\
		0.33	0.283971461	\\
		0.34	0.263971461	\\
		0.35	0.243971461	\\
		0.36	0.223971461	\\
		0.37	0.203971461	\\
		0.38	0.183971461	\\
		0.39	0.163971461	\\
		0.4	0.143971461	\\
		0.41	0.123971461	\\
		0.42	0.103971461	\\
		0.43	0.083971461	\\
		0.44	0.063971461	\\
		0.45	0.043971461	\\
		0.46	0.023971461	\\
		0.47	0.003971461	\\
	};
	\legend{}
	\addlegendentry{$E_{\mbox{\tiny r}}^{\mbox{\tiny opt}}(R)$}

	\addplot[smooth,color=black!20!purple,thick,dash pattern={on 3pt off 2pt}]
	table[row sep=crcr]
	{
0	2.33E+00	\\
0.01	1.942243119	\\
0.02	1.783248372	\\
0.03	1.662129568	\\
0.04	1.560768231	\\
0.05	1.472130522	\\
0.06	1.392601807	\\
0.07	1.32003083	\\
0.08	1.253013607	\\
0.09	1.190573456	\\
0.1	1.131998387	\\
0.11	1.076750365	\\
0.12	1.024411019	\\
0.13	0.974647539	\\
0.14	0.927189988	\\
0.15	0.881815962	\\
0.16	0.838339983	\\
0.17	0.79660487	\\
0.18	0.756476879	\\
0.19	0.71784084	\\
0.2	0.680596961	\\
0.21	0.644657843	\\
0.22	0.609947083	\\
0.23	0.576397432	\\
0.24	0.543948986	\\
0.25	0.512548791	\\
0.26	0.482150065	\\
0.27	0.452710417	\\
0.28	0.424192539	\\
0.29	0.396563088	\\
0.3	0.369792388	\\
0.31	0.343854156	\\
0.32	0.318725254	\\
0.33	0.294385461	\\
0.34	0.270817346	\\
0.35	0.248006392	\\
0.36	0.225940484	\\
0.37	0.204610478	\\
0.38	0.184010114	\\
0.39	0.163971461	\\
0.4	0.143971461	\\
0.41	0.123971461	\\
0.42	0.103971461	\\
0.43	0.083971461	\\
0.44	0.063971461	\\
0.45	0.043971461	\\
0.46	0.023971461	\\
0.47	0.003971461	\\
	};
	\addlegendentry{$E_{\mbox{\tiny ex}}^{\mbox{\tiny opt}}(R)$}

	\addplot[smooth,color=blue,thick]
	table[row sep=crcr]
	{
0	2.329177915	\\
0.01	1.94224312	\\
0.02	1.783248372	\\
0.03	1.662129568	\\
0.04	1.560768234	\\
0.05	1.472130542	\\
0.06	1.392601825	\\
0.07	1.320030844	\\
0.08	1.253013608	\\
0.09	1.190573461	\\
0.1	1.131998402	\\
0.11	1.076750371	\\
0.12	1.024411029	\\
0.13	0.974647539	\\
0.14	0.927190005	\\
0.15	0.881816056	\\
0.16	0.838339988	\\
0.17	0.796604913	\\
0.1758	0.773142729	\\
	};
	\addlegendentry{$E_{\mbox{\tiny \cite{BEES}}}^{\mbox{\tiny opt}}(R)$}

	\end{axis}
	
	\end{tikzpicture}
	\caption{Error exponents for the BSC under optimal decoding ($p=0.01$).}\label{fig:BSC-Numeric-UT}
\end{figure}

In \cite{BEES}, two lower bounds on the reliability function of the bee identification problem are given. The first is a random coding bound, similarly to the bound in Theorem \ref{Theorem_OPT}. It can be easily shown that upon applying $E_{\mbox{\tiny r}}^{\mbox{\tiny opt}}(R)$ to the BSC, one arrives at the result in \cite[Theorem 2]{BEES}. The second bound in \cite{BEES} stems from characteristics of typical random binary codes \cite{BargForney} and is given by      
\begin{align} \label{OPTIMUM_Expurgated_TAN}
E_{\mbox{\tiny \cite{BEES}}}^{\mbox{\tiny opt}}(R) = -\delta_{\mbox{\tiny GV}}(2R) \cdot \log \left(\sqrt{4p(1-p)}\right),~~~R \in [0,R_{\mbox{\tiny TRC}}(p)),
\end{align}
where $\delta_{\mbox{\tiny GV}}(2R)$ is the Gilbert-Varshamov distance, defined as the value of $\delta \in [0,0.5]$ with $h_{2}(\delta)=1-2R$, $h_{2}(\cdot)$ being the binary entropy function, and where
\begin{align}
R_{\mbox{\tiny TRC}}(p) = \frac{1}{2} \left[1- h_{2}\left(\frac{\sqrt{4p(1-p)}}{1+\sqrt{4p(1-p)}}\right)\right].
\end{align}
Since \eqref{OPTIMUM_Expurgated} and \eqref{OPTIMUM_Expurgated_TAN} are given by relatively different optimization problems\footnote{Solving the non-linear equation $h_{2}(\delta)=1-2R$ can be recast as an optimization problem.}, it seems that comparing between $E_{\mbox{\tiny ex}}^{\mbox{\tiny opt}}(R)$ and $E_{\mbox{\tiny \cite{BEES}}}^{\mbox{\tiny opt}}(R)$ directly from their expressions may be rather difficult.
Hence, we compare between $E_{\mbox{\tiny ex}}^{\mbox{\tiny opt}}(R)$ and $E_{\mbox{\tiny \cite{BEES}}}^{\mbox{\tiny opt}}(R)$ numerically. As can be seen in Figure \ref{fig:BSC-Numeric-UT}, for $R \leq R_{\mbox{\tiny TRC}}(p) \approx 0.1758$, the two bounds are equal, but for $R \geq R_{\mbox{\tiny TRC}}(p)$, there exists an interval where $E_{\mbox{\tiny ex}}^{\mbox{\tiny opt}}(R)$ still improves upon $E_{\mbox{\tiny r}}^{\mbox{\tiny opt}}(R)$. The fact that $E_{\mbox{\tiny ex}}^{\mbox{\tiny opt}}(R) = E_{\mbox{\tiny \cite{BEES}}}^{\mbox{\tiny opt}}(R)$ at relatively low coding rates is quite surprising, at least to the authors of this work, since $E_{\mbox{\tiny \cite{BEES}}}^{\mbox{\tiny opt}}(R)$ is related to typical codes, while $E_{\mbox{\tiny ex}}^{\mbox{\tiny opt}}(R)$ is a byproduct of an expurgation process. As far as we know, the only scenario where TRCs and expurgated codes have similar performance is for linear codes \cite{BargForney}, while in any other case (e.g., \cite{MERHAV_TYPICAL} and \cite{TM}), the expurgated code performs strictly better than the TRC, at least at some interval of rates.

\section*{Appendix A}
\renewcommand{\theequation}{A.\arabic{equation}}
\setcounter{equation}{0}  
\subsection*{Proof of Eq.\ (\ref{THM1_UB}) of Theorem \ref{THEOREM_Naive}}

Assume that the codebook $\calC_{n}$ is given. Then, the enumerator $N_{\mbox{\tiny e}}(\calC_{n})$ is a sum of independent indicator random variables. Note that these indicators have different success probabilities. 
%Given $\calC_{n}$, 
The probability of erroneous decoding of the codeword $\bx_{m}$ is given by
\begin{align} \label{ErrorProb}
p_{m}(\calC_{n}) \dfn \sum_{\by \in \calY^{n}} W(\by|\bx_{m}) \cdot \frac{\sum_{m' \neq m} \exp\{n g(\hat{P}_{\bx_{m'}\by})\}}{\sum_{\tilde{m}=1}^{M} \exp\{n g(\hat{P}_{\bx_{\tilde{m}}\by})\}}.
\end{align} 
%which may be different for every codeword. 
%Hence, $N_{\mbox{\tiny e}}(\calC_{n})$ has a {\it Poisson binomial distribution}, since it is a sum of independent Bernoulli trials that are not necessarily identically distributed.
Denote the expectation of $N_{\mbox{\tiny e}}(\calC_{n})$ by
\begin{align}
\mu = \mu(\calC_{n}) \dfn \mathbb{E}\left[N_{\mbox{\tiny e}}(\calC_{n})\right] = \sum_{m=1}^{M} p_{m}(\calC_{n}).
\end{align}
Let $L \in \mathbb{N}$ be fixed and denote the indicator random variables $I_{m}
= \IND \left\{\text{Decoding of $\bx_{m}$ has failed} \right\}$, $m \in \{1,2, \ldots,M\}$. Then, for any $t \geq 0$, the Chernoff bound implies that
\begin{align}
P_{\mbox{\tiny e}}(\calC_{n}) 
&= \prob \{N_{\mbox{\tiny e}}(\calC_{n}) \geq L \} \\
&\leq e^{-t L} \cdot \mathbb{E} \left[ \exp\left\{t \cdot \sum_{m=1}^{M} I_{m} \right\} \right] \\
&= e^{-t L} \cdot \prod_{m=1}^{M} \mathbb{E} \left[ \exp\left\{t \cdot  I_{m} \right\} \right] \\
&= e^{-t L} \cdot \prod_{m=1}^{M}  \left( 1 - p_{m}(\calC_{n}) + p_{m}(\calC_{n}) e^{t}  \right) \\
&= e^{-t L} \cdot \exp \left\{M \cdot \frac{1}{M} \sum_{m=1}^{M} \log \left[ 1 + (e^{t}-1) p_{m}(\calC_{n})   \right] \right\} \\
\label{term10}
&\leq e^{-t L} \cdot \exp \left\{M \cdot \log \left[ 1 + \frac{1}{M} \sum_{m=1}^{M} (e^{t}-1) p_{m}(\calC_{n})   \right] \right\} \\
&= e^{-t L} \cdot \exp \left\{M \cdot \log \left[ 1 + \frac{\mu(\calC_{n})}{M} (e^{t}-1) \right] \right\} \\
\label{term8}
&= \exp \left\{M \cdot \log \left[ 1 + \frac{\mu(\calC_{n})}{M} (e^{t}-1) \right] -t L \right\} ,
\end{align}
where \eqref{term10} is due to Jensen's inequality and the concavity of the $\log(\cdot)$ function.
Next, we minimize with respect to $t$. Let us define the function
\begin{align}
f(t) = a \cdot \log \left[ 1 + b \cdot (e^{t}-1) \right] -c \cdot t ,
\end{align}
%Differentiating provides
whose derivative is given by
\begin{align}
f'(t) = a \cdot \frac{b \cdot e^{t}}{1 + b \cdot (e^{t}-1)} - c, 
\end{align}
and thus, solving $f'(t)=0$ provides
\begin{align}
&\frac{b \cdot e^{t}}{1 - b + b \cdot e^{t}} = \frac{c}{a} \DEF d \\
\Leftrightarrow~~~~& b \cdot e^{t} = d(1-b) + bd \cdot  e^{t} \\
\Leftrightarrow~~~~& b(1-d) \cdot e^{t} = d(1-b) \\
\Leftrightarrow~~~~& e^{t} = \frac{d(1-b)}{b(1-d)}.
\end{align}
Now, by substituting $b=\frac{\mu}{M}$ and $d=\frac{L}{M}$, we arrive at
\begin{align}
e^{t} 
= \frac{\frac{L}{M}(1-\frac{\mu}{M})}{\frac{\mu}{M}(1-\frac{L}{M})}
= \frac{\frac{L}{M}(\frac{M-\mu}{M})}{\frac{\mu}{M}(\frac{M-L}{M})}
\label{term11}
= \frac{L (M-\mu)}{\mu (M-L)}, 
\end{align}
where the right most expression of \eqref{term11} is greater or equal to one as long as $L \geq \mu(\calC_{n})$, 
and thus, the minimizer is given by
\begin{align}
t^{*} = \log \left[\frac{L (M-\mu)}{\mu (M-L)}\right],
\end{align}
for $L \geq \mu(\calC_{n})$, and $t^{*}=0$ otherwise.
In the former case, substituting $t^{*}$ back into \eqref{term8} provides
\begin{align}
P_{\mbox{\tiny e}}(\calC_{n}) 
&\leq \exp \left\{M \cdot \log \left[ 1 + \frac{\mu}{M} \left(\frac{L (M-\mu)}{\mu (M-L)} - 1\right) \right] - L \cdot \log \left[\frac{L (M-\mu)}{\mu (M-L)}\right] \right\} \\
&= \exp \left\{M \cdot \log \left[ 1 + \frac{\mu}{M} \cdot \frac{M(L - \mu)}{\mu (M-L)}  \right] - L \cdot \log \left[\frac{L (M-\mu)}{\mu (M-L)}\right] \right\} \\
&= \exp \left\{M \cdot \log \left(1 +  \frac{L - \mu}{M-L} \right) - L \cdot \log \left[\frac{L (M-\mu)}{\mu (M-L)}\right] \right\} \\
&= \exp \left\{M \cdot \log \left(\frac{M - \mu}{M-L} \right) 
- L \cdot \log \left(\frac{L}{\mu}\right)
- L \cdot \log \left(\frac{M-\mu}{M-L}\right) \right\} \\
&= \exp \left\{(M- L) \cdot \log \left(\frac{M - \mu}{M-L} \right) 
- L \cdot \log \left(\frac{L}{\mu}\right) \right\} \\
&\doteq \exp \left\{M \cdot \log \left(1- \frac{\mu}{M} \right) 
- L \cdot \log \left(\frac{L}{\mu}\right) \right\} ,
\end{align}
where the last passage is due to the assumption that $L$ is exponentially smaller than $M=e^{nR}$.
When $L < \mu(\calC_{n})$, substituting $t^{*}=0$ back into \eqref{term8} gives the trivial bound $P_{\mbox{\tiny e}}(\calC_{n}) \leq 1$. Hence, we have that 
\begin{align}
&P_{\mbox{\tiny e}}(\calC_{n}) \nn \\
&\leq \exp \left\{M \cdot \log \left(1- \frac{\mu(\calC_{n})}{M} \right) 
- L \cdot \log \left(\frac{L}{\mu(\calC_{n})}\right) \right\} \cdot \IND \left\{\mu(\calC_{n}) \leq L \right\} + \IND \left\{\mu(\calC_{n}) > L\right\} \\
&\doteq \exp \left\{M \cdot \log \left(1- \frac{\mu(\calC_{n})}{M} \right) 
+ L \cdot \log \left(\mu(\calC_{n})\right) \right\} \cdot \IND \left\{\mu(\calC_{n}) \leq L \right\} + \IND \left\{\mu(\calC_{n}) > L\right\} \\
&= \left(\mu(\calC_{n})\right)^{L} \cdot \left(1- \frac{\mu(\calC_{n})}{M} \right)^{M} \cdot \IND \left\{\mu(\calC_{n}) \leq L \right\} + \IND \left\{\mu(\calC_{n}) > L\right\} \\
&\doteq \left(\mu(\calC_{n})\right)^{L} \cdot \exp \left\{- \mu(\calC_{n}) \right\} \cdot \IND \left\{\mu(\calC_{n}) \leq L \right\} + \IND \left\{\mu(\calC_{n}) > L\right\} \\
&\doteq \left(\mu(\calC_{n})\right)^{L} \cdot \IND \left\{\mu(\calC_{n}) \leq L \right\} + \IND \left\{\mu(\calC_{n}) > L\right\} \\
\label{term1}
&\leq \min \left\{L^{L}, \left(\mu(\calC_{n})\right)^{L} \right\}  + \IND \left\{\mu(\calC_{n}) > L\right\} .
\end{align} 
Let us average \eqref{term1} over the ensemble of codebooks. It follows from Jensen's inequality and the concavity of the function $f(t) = \min \{A,t\}$ that
\begin{align}
\mathbb{E} \left[P_{\mbox{\tiny e}}(\calC_{n})\right] 
&\leq \mathbb{E} \left[ \min \left\{L^{L}, \mu(\calC_{n})^{L} \right\} + \IND \left\{\mu(\calC_{n}) > L\right\} \right] \\
\label{term3}
&\leq  \min \left\{L^{L}, \mathbb{E} \left[ \mu(\calC_{n})^{L} \right] \right\} + \prob \left\{\mu(\calC_{n}) > L\right\} .
\end{align}
Let
\begin{align}
\label{Z_DEF}
Z_{m}(\by) = \sum_{\tilde{m} \neq m} \exp\{n g(\hat{P}_{\bx_{\tilde{m}}\by}) \},
\end{align}
fix $\epsilon>0$ arbitrarily small, and for every $\by \in \calY^{n}$, define the set
\begin{align}
\label{B_DEF}
\calB_{\epsilon}(m,\by) = \left\{\calC_{n}:~ Z_{m}(\by) \leq \exp\{n \alpha(R-\epsilon, \hat{P}_{\by})\}  \right\}.
\end{align}
Following the result of \cite[Appendix B]{MERHAV2017}, we know that, considering the ensemble of randomly selected constant composition codes of type $Q_{X}$,
\begin{align}
\label{B_DE_UB}
\prob \{\calB_{\epsilon}(m,\by)\} \leq \exp\{-e^{n\epsilon} + n\epsilon + 1\},
\end{align} 
for every $m \in \{1,2,\dotsc,M\}$ and $\by \in \calY^{n}$, and so, by the union bound,
\begin{align}
\label{B_UNION_DEF}
\prob \left\{\bigcup_{m=1}^{M} \bigcup_{\by \in \calY^{n}}\calB_{\epsilon}(m,\by)\right\}
\DEF  \prob \left\{\calB_{\epsilon}\right\} 
&\leq \sum_{m=1}^{M} \sum_{\by \in \calY^{n}} \prob \left\{\calB_{\epsilon}(m,\by)\right\} \\
&\leq \sum_{m=1}^{M} \sum_{\by \in \calY^{n}} \exp\{-e^{n\epsilon} + n\epsilon + 1\} \\
&= e^{nR} \cdot |\calY|^{n} \cdot \exp\{-e^{n\epsilon} + n\epsilon + 1\},
\end{align}
which still decays double--exponentially fast.

Now, for the expectation inside the left expression of \eqref{term3}, we derive as follows:
\begin{align}
&\mathbb{E} \left[ \mu(\calC_{n})^{L} \right] \\
&= \mathbb{E} \left\{ \left[\sum_{m=1}^{M} \sum_{\by \in \calY^{n}} W(\by|\bx_{m}) \cdot \frac{\sum_{m' \neq m} \exp\{n  g(\hat{P}_{\bx_{m'}\by})\}}{\sum_{\tilde{m}=1}^{M} \exp\{n g(\hat{P}_{\bx_{\tilde{m}}\by})\}}\right]^{L} \right\} \\
&= \mathbb{E} \left\{ \left[\sum_{m=1}^{M} \sum_{m' \neq m} \sum_{\by \in \calY^{n}} W(\by|\bx_{m}) \cdot \frac{ \exp\{n  g(\hat{P}_{\bx_{m'}\by})\}}{\exp\{n g(\hat{P}_{\bx_{m}\by})\} + Z_{m}(\by)}\right]^{L} \right\} \\
&\lexe \mathbb{E} \left\{ \left[\sum_{m=1}^{M} \sum_{m' \neq m} \sum_{\by \in \calY^{n}} W(\by|\bx_{m}) \cdot \min \left\{1, \frac{ \exp\{n  g(\hat{P}_{\bx_{m'}\by})\}}{\exp\{n g(\hat{P}_{\bx_{m}\by})\} + \exp\{n \alpha(R-\epsilon,\hat{P}_{\by})\}} \right\} \right]^{L} \right\} \\
&\doteq \mathbb{E} \left\{ \left[\sum_{m=1}^{M} \sum_{m' \neq m} \exp \left\{-n \Gamma(\hat{P}_{\bx_{m}\bx_{m'}},R-\epsilon)\right\} \right]^{L} \right\} \\
&= \mathbb{E} \left\{ \left[\sum_{Q_{X'|X} \in \calQ(Q_{X})} N(Q_{XX'}) \cdot \exp \left\{-n \Gamma(Q_{XX'},R-\epsilon)\right\} \right]^{L} \right\} \\
\label{term2}
&\doteq \sum_{Q_{X'|X} \in \calQ(Q_{X})} \mathbb{E} \left\{ \left[ N(Q_{XX'}) \right]^{L} \right\} \cdot \exp \left\{-n \Gamma(Q_{XX'},R-\epsilon) \cdot L \right\}  .
\end{align}
Next, the $L$--th moment of $N(Q_{XX'})$ is given by \cite[Lemma 3]{TMWG}
\begin{align}
\mathbb{E} \left\{ \left[ N(Q_{XX'}) \right]^{L} \right\} 
&\lexe \exp \left\{n \cdot \left(L \cdot \left[2R - I_{Q}(X;X')\right]_{+} - \left[I_{Q}(X;X')-2R\right]_{+}\right) \right\}.
\end{align}
Substituting it back into \eqref{term2} and then into the left expression in \eqref{term3} provides
\begin{align}
&\min \left\{L^{L}, \mathbb{E} \left[ \mu(\calC_{n})^{L} \right] \right\} \nn \\
&\lexe \min \left\{L^{L}, \sum_{Q_{X'|X} \in \calQ(Q_{X})} e^{n \cdot \left(L \cdot \left[2R - I_{Q}(X;X')\right]_{+} - \left[I_{Q}(X;X')-2R\right]_{+}\right) } \cdot e^{-n \Gamma(Q_{XX'},R-\epsilon) \cdot L } \right\} \\
&\doteq \exp \left\{-n \cdot E_{\mbox{\tiny r}}^{\mbox{\tiny ub}}(R,L,\epsilon) \right\},
\end{align}
where,
\begin{align} 
&E_{\mbox{\tiny r}}^{\mbox{\tiny ub}}(R,L,\epsilon) \nn \\
&= \min_{Q_{X'|X} \in \calQ(Q_{X})}
\left[L \cdot \Gamma(Q_{XX'},R-\epsilon) - L \cdot [2R - I_{Q}(X;X')]_{+} + [I_{Q}(X;X')-2R]_{+}\right]_{+}.
\end{align}
For the right expression of \eqref{term3}, we derive in the following way:
\begin{align}
\prob \left\{\mu(\calC_{n}) > L\right\} 
&= \prob \left\{\sum_{m=1}^{M} \sum_{\by \in \calY^{n}} W(\by|\bx_{m}) \cdot \frac{\sum_{m' \neq m} \exp\{n  g(\hat{P}_{\bx_{m'}\by})\}}{\sum_{\tilde{m}=1}^{M} \exp\{n g(\hat{P}_{\bx_{\tilde{m}}\by})\}} > L \right\} \\
&\lexe \prob \left\{ \sum_{m=1}^{M} \sum_{m' \neq m} \exp \left\{-n \Gamma(\hat{P}_{\bx_{m}\bx_{m'}},R-\epsilon)\right\} > L \right\} \\
&\leq \prob \left\{\sum_{Q_{X'|X} \in \calQ(Q_{X})} N(Q_{XX'}) \cdot \exp \left\{-n \Gamma(Q_{XX'},R-\epsilon)\right\} > 1 \right\} \\
&\doteq \sum_{Q_{X'|X} \in \calQ(Q_{X})} \prob \left\{ N(Q_{XX'})  >  \exp \left\{n \Gamma(Q_{XX'},R-\epsilon)\right\} \right\} \\
&\doteq \max_{Q_{X'|X} \in \calQ(Q_{X})} \prob \left\{ N(Q_{XX'})  > \exp \left\{n \Gamma(Q_{XX'},R-\epsilon)\right\} \right\} \\
&\doteq \exp \left\{-n \cdot \tilde{E}_{\mbox{\tiny r}}(R,\epsilon) \right\},
\end{align}
where it follows from \cite[Theorem 3]{TMWG} that
\begin{align}
\tilde{E}_{\mbox{\tiny r}}(R,\epsilon) = \min_{\{Q_{X'|X} \in \calQ(Q_{X}):~ [2R-I_{Q}(X;X')]_{+} \geq \Gamma(Q_{XX'},R-\epsilon) \}}
\left[I_{Q}(X;X')-2R\right]_{+}.
\end{align}
As a last step, we prove that for any finite $L$, $E_{\mbox{\tiny r}}^{\mbox{\tiny ub}}(R,L,\epsilon)$ is lower or equal to $\tilde{E}_{\mbox{\tiny r}}(R,\epsilon)$. 
We first prove that $E_{\mbox{\tiny r}}^{\mbox{\tiny ub}}(R,L,\epsilon)$ is monotonically non--decreasing in $L$. We have that 
\begin{align}
& E_{\mbox{\tiny r}}^{\mbox{\tiny ub}}(R,L,\epsilon) \nn \\
&= \min_{Q_{X'|X} \in \calQ(Q_{X})}
\left[L \cdot \Gamma(Q_{XX'},R-\epsilon) - L \cdot \left[2R - I_{Q}(X;X')\right]_{+} + \left[I_{Q}(X;X')-2R\right]_{+}\right]_{+} \\
&= \min \left\{ \min_{\{Q_{X'|X} \in \calQ(Q_{X}):~ I_{Q}(X;X') \leq 2R\}}
L \cdot \left[\Gamma(Q_{XX'},R-\epsilon) + I_{Q}(X;X') - 2R \right]_{+}, \right. \nn \\
& ~~~~~~~~~~\left. \min_{\{Q_{X'|X} \in \calQ(Q_{X}):~ I_{Q}(X;X') > 2R\}}
\left[L \cdot \Gamma(Q_{XX'},R-\epsilon) + I_{Q}(X;X')-2R \right]_{+} \right\} \\
&\DEF \min \left\{A(L),B(L)\right\}. 
\end{align}
Now, the sequence $A(L)$ is trivially non--decreasing, and $B(L)$ is also non--decreasing, since $\Gamma(Q_{XX'},R)$ is non--negative. Hence, $E_{\mbox{\tiny r}}^{\mbox{\tiny ub}}(R,L,\epsilon)$ is non--decreasing as a minimum between two non--decreasing sequences. Letting $L$ grow without bound gives 
\begin{align}
&\lim_{L \to \infty} E_{\mbox{\tiny r}}^{\mbox{\tiny ub}}(R,L,\epsilon) \nn \\
&= \lim_{L \to \infty} \min_{Q_{X'|X} \in \calQ(Q_{X})}
\left[L \cdot \Gamma(Q_{XX'},R-\epsilon) - L \cdot \left[2R - I_{Q}(X;X')\right]_{+} + \left[I_{Q}(X;X')-2R\right]_{+}\right]_{+} \\
&=\min_{\{Q_{X'|X} \in \calQ(Q_{X}):~ [2R-I_{Q}(X;X')]_{+} \geq \Gamma(Q_{XX'},R-\epsilon) \}}
\left[I_{Q}(X;X')-2R\right]_{+} \\
&= \tilde{E}_{\mbox{\tiny r}}(R,\epsilon),
\end{align}
which proves that $E_{\mbox{\tiny r}}^{\mbox{\tiny ub}}(R,L,\epsilon) \leq \tilde{E}_{\mbox{\tiny r}}(R,\epsilon)$ for any finite $L$. Thus,
\begin{align}
\lim_{n \to \infty} -\frac{1}{n} \log \mathbb{E} \left[ P_{\mbox{\tiny e}}(\calC_{n}) \right] \geq \min \{ E_{\mbox{\tiny r}}^{\mbox{\tiny ub}}(R,L,\epsilon), \tilde{E}_{\mbox{\tiny r}}(R,\epsilon) \} = E_{\mbox{\tiny r}}^{\mbox{\tiny ub}}(R,L,\epsilon),
\end{align}
which complete the proof of the first part of Theorem \ref{THEOREM_Naive}, due to the arbitrariness of $\epsilon>0$.

\section*{Appendix B}
\renewcommand{\theequation}{B.\arabic{equation}}
\setcounter{equation}{0}  
\subsection*{Proof of Eq.\ (\ref{THM1_LB}) of Theorem \ref{THEOREM_Naive}}

Recall that the probability of error is given by
\begin{align}
P_{\mbox{\tiny e}}(\calC_{n}) 
%&= \prob \{N_{\mbox{\tiny e}}(\calC_{n}) \geq L \} \\
&= \prob \left\{ \sum_{m=1}^{M} I_{m} \geq L \right\} .
\end{align}
Let $\epsilon > 0$ be given. Define the sets
\begin{align}
\calA_{\epsilon}(\calC_{n},i) = \left\{m:~ e^{-ni\epsilon} \leq p_{m}(\calC_{n}) \leq 1 \right\},
\end{align}
and the enumerators
\begin{align}
N_{\epsilon}(\calC_{n},i) = \sum_{m=1}^{M} \IND \left\{e^{-ni\epsilon} \leq p_{m}(\calC_{n}) \leq 1 \right\},
\end{align}
where $p_{m}(\calC_{n})$ is the probability of error when message $m$ is transmitted, as given explicitly in \eqref{ErrorProb}. Now,
\begin{align}
P_{\mbox{\tiny e}}(\calC_{n}) 
&= \prob \left\{ \sum_{m=1}^{M} I_{m} \geq L \right\} \\
%&= \prob \left\{ \sum_{i=0}^{E_{\mbox{\tiny sp}}(R)/\epsilon} \sum_{m \in \calA_{\epsilon}(\calC_{n},i)} I_{m} \geq L \right\} \\
&= \prob \left\{ \bigcup_{i=1}^{\infty} \left\{ \sum_{m \in \calA_{\epsilon}(\calC_{n},i)} I_{m} \geq L \right\} \right\} \\
\label{Recall9}
&\geq \sup_{i \in \mathbb{N}} \prob \left\{\sum_{m \in \calA_{\epsilon}(\calC_{n},i)} I_{m} \geq L \right\} .
\end{align}
For any $i \in \mathbb{N}$ and a given codebook $\calC_{n}$, let $R(\calC_{n},i)$ be the exponential rate of the size of $\calA_{\epsilon}(\calC_{n},i)$, i.e., 
\begin{align}
R(\calC_{n},i) \dfn \frac{1}{n} \log N_{\epsilon}(\calC_{n},i).
\end{align} 
The probability in \eqref{Recall9} can be lower-bounded as follows:
\begin{align}
\prob \left\{\sum_{m \in \calA_{\epsilon}(\calC_{n},i)} I_{m} \geq L \right\}
&\geq \sum_{k=L}^{e^{n R(\calC_{n},i)}} \binom{e^{n R(\calC_{n},i)}}{k} \left(e^{-n i \epsilon}\right)^{k} \left(1- e^{-n i \epsilon}\right)^{e^{n R(\calC_{n},i)} - k} \\
&\geq \binom{e^{n R(\calC_{n},i)}}{L} \left(e^{-n i \epsilon}\right)^{L} \left(1- e^{-n i \epsilon}\right)^{e^{n R(\calC_{n},i)} - L} \\
&\DEXE \binom{e^{n R(\calC_{n},i)}}{L} \left(e^{-n i \epsilon}\right)^{L} \left(1- e^{-n i \epsilon}\right)^{e^{n R(\calC_{n},i)}} \\
\label{Recall10}
&\doteq e^{n R(\calC_{n},i) L} e^{-n i \epsilon L} \left(1- e^{-n i \epsilon}\right)^{e^{n R(\calC_{n},i)}} .
\end{align} 
As for the third factor in \eqref{Recall10}, we use the fact that $\log \left(1- e^{-n i \epsilon}\right) \doteq - e^{-n i \epsilon}$, and get
\begin{align}
\left(1- e^{-n i \epsilon}\right)^{e^{n R(\calC_{n},i)}}
&= \exp \left\{ e^{n R(\calC_{n},i)} \log \left(1- e^{-n i \epsilon}\right) \right\} \\
\label{DE1}
&\DEXE \exp \left\{- e^{n (R(\calC_{n},i)- i\epsilon)} \right\} \\
\label{DE2}
&\geq \left(1 - e^{n (R(\calC_{n},i)- i\epsilon)} \right) \IND \{R(\calC_{n},i) \leq  i\epsilon\},
\end{align}
where \eqref{DE2} is due to the fact that for any $t \in \mathbb{R}$, $e^{-t} \geq 1-t$.   
Substituting \eqref{DE2} back into \eqref{Recall10} yields
\begin{align}
P_{\mbox{\tiny e}}(\calC_{n}) 
%%%%%%%%%%%%%%%%%%%%%%%%%%%%%%%%%%%%%%%%%
&\geq \sup_{i \in \mathbb{N}}  \left[ e^{n (R(\calC_{n},i)- i\epsilon) L} \cdot \left(1 - e^{n (R(\calC_{n},i)- i\epsilon)} \right) \IND \{R(\calC_{n},i) \leq  i\epsilon\} \right] \\
%%%%%%%%%%%%%%%%%%%%%%%%%%%%%%%%%%%%%%%%%
&\geq \sup_{i \geq R/\epsilon}  \left[ e^{n (R(\calC_{n},i)- i\epsilon) L} \cdot \left(1 - e^{n (R(\calC_{n},i)- i\epsilon)} \right) \IND \{R(\calC_{n},i) \leq  i\epsilon\} \right] \\
%%%%%%%%%%%%%%%%%%%%%%%%%%%%%%%%%%%%%%%%%
\label{ToExp0}
&= \sup_{i \geq R/\epsilon}  \left[ e^{n (R(\calC_{n},i)- i\epsilon) L} \cdot \left(1 - e^{n (R(\calC_{n},i)- i\epsilon)} \right) \right] \\
%%%%%%%%%%%%%%%%%%%%%%%%%%%%%%%%%%%%%%%%%
\label{ToExp2}
&\doteq \sup_{i \geq R/\epsilon}  \left[ e^{n (R(\calC_{n},i)- i\epsilon) L} \right],
\end{align}
where \eqref{ToExp0} and \eqref{ToExp2} are due to the fact that $R(\calC_{n},i) \leq R$ for any $i$ with probability one. Taking the expectation provides
\begin{align}
\mathbb{E}\{P_{\mbox{\tiny e}}(\calC_{n})\} 
%%%%%%%%%%%%%%%%%%%%%%%%%%%%%%%%%%%%%%%%%%%%%%%%%%%%%%%%%%%%
&\geq \mathbb{E} \left\{ \sup_{i \geq R/\epsilon}  \left[ e^{n (R(\calC_{n},i)- i\epsilon) L} \right] \right\} \\
%%%%%%%%%%%%%%%%%%%%%%%%%%%%%%%%%%%%%%%%%%%%%%%%%%%%%%%%%%%%
&\geq \sup_{i \geq R/\epsilon} \mathbb{E} \left\{ e^{-n i \epsilon L} \cdot N_{\epsilon}(\calC_{n},i)^{L} \right\} \\
%%%%%%%%%%%%%%%%%%%%%%%%%%%%%%%%%%%%%%%%%%%%%%%%%%%%%%%%%%%%
&= \sup_{i \geq R/\epsilon} e^{-n i \epsilon L} \cdot \mathbb{E} \left\{ N_{\epsilon}(\calC_{n},i)^{L} \right\} \\
%%%%%%%%%%%%%%%%%%%%%%%%%%%%%%%%%%%%%%%%%%%%%%%%%%%%%%%%%%%%
\label{ToExp1}
&\geq \sup_{i \geq R/\epsilon} e^{-n i \epsilon L} \cdot \left( \mathbb{E} \left\{ N_{\epsilon}(\calC_{n},i) \right\} \right)^{L} ,
\end{align}
where \eqref{ToExp1} follows from Jensen's inequality and the convexity of the function $f(t)=t^{L}$, $L \in \mathbb{N}$.
As for the expectation in \eqref{ToExp1}, we have 
\begin{align}
\label{Recall11}
\mathbb{E} \left\{ N_{\epsilon}(\calC_{n},i) \right\}
%%%%%%%%%%%%%%%%%%%%%%%%%%%%%%%%%%%%%%%%%%%%%%%%%%%%%
=\sum_{m=1}^{M} \prob \left\{ p_{m}(\calC_{n}) \geq e^{-n i \epsilon} \right\}.
\end{align}
Next, we prove in Appendix C, that the probability in \eqref{Recall11}, which is given explicitly by 
\begin{align}
\prob \left\{ \sum_{\by \in \calY^{n}} W(\by|\bX_{m}) \cdot \frac{\sum_{m' \neq m} \exp\{n g(\hat{P}_{\bX_{m'}\by})\}}{\sum_{\tilde{m}=1}^{M} \exp\{n g(\hat{P}_{\bX_{\tilde{m}}\by})\}} \geq e^{-n i \epsilon} \right\},
\end{align}
is lower-bounded as
\begin{align}
\label{Result3}
\prob \left\{ p_{m}(\calC_{n}) \geq e^{-n i \epsilon} \right\} \gexe \exp\{-n E(R,i \epsilon)\},
\end{align}
where
\begin{align}
E(R,i \epsilon) 
= \min_{Q_{X'|X} \in \calJ(R,i\epsilon)} \left[I_{Q}(X;X')-R\right]_{+}  
\end{align}
and $\calJ(\cdot,\cdot)$ is defined by 
\begin{align}
\calJ(R,s) = \left\{Q_{X'|X} \in \calQ(Q_{X}):~ 
\left[R-I_{Q}(X;X')\right]_{+} \geq \Lambda(Q_{XX'},R) - s \right\}.
\end{align}
Substituting \eqref{Result3} back into \eqref{Recall11} and then into \eqref{ToExp1} yields 
\begin{align}
\mathbb{E}\{P_{\mbox{\tiny e}}(\calC_{n})\} 
%%%%%%%%%%%%%%%%%%%%%%%%%%%%%%%%%%%%%%%%%%%%%%%%%%%%%%%%%%%%
&\gexe \sup_{i \geq R/\epsilon} e^{-n i \epsilon L} \cdot \left( \sum_{m=1}^{M} \exp\{-n E(R,i\epsilon)\} \right)^{L} \\
%%%%%%%%%%%%%%%%%%%%%%%%%%%%%%%%%%%%%%%%%%%%%%%%%%%%%%%%%%%%
&= \sup_{i \geq R/\epsilon} e^{-n i \epsilon L} \cdot \exp\{-n [E(R,i\epsilon) - R] L\}  \\
%%%%%%%%%%%%%%%%%%%%%%%%%%%%%%%%%%%%%%%%%%%%%%%%%%%%%%%%%%%%
&= \sup_{i \geq R/\epsilon} \exp\{-n [E(R,i\epsilon) - R + i \epsilon] L\} .
\end{align}
Finally, since $\epsilon > 0$ is arbitrarily small, we conclude that
\begin{align}
\label{Ref1}
\lim_{n \to \infty} -\frac{1}{n} \log \mathbb{E}\{P_{\mbox{\tiny e}}(\calC_{n})\} 
%%%%%%%%%%%%%%%%%%%%%%%%%%%%%%%%%%%%%%%%%%%%%%%%%%%%%%%%%%%%
&\leq \inf_{s \geq R} \left\{ [E(R,s) - R + s] L \right\}.
\end{align}
It only remains to simplify the expression on the right-hand-side of \eqref{Ref1}. Let us define
\begin{align}
E_{\mbox{\tiny r}}^{\mbox{\tiny lb}}(R,L) = \inf_{s \geq R} \min_{Q_{X'|X} \in \calJ(R,s)} \left\{ \left([I_{Q}(X;X')-R]_{+} - R + s\right) \cdot L \right\},
\end{align} 
such that
\begin{align}
&E_{\mbox{\tiny r}}^{\mbox{\tiny lb}}(R,L) \nn \\
&= \inf_{s \geq R} \min_{\left\{\substack{Q_{X'|X} \in \calQ(Q_{X}), \\ [R-I_{Q}(X;X')]_{+} \geq \Lambda(Q_{XX'},R) - s}\right\}} \left\{ \left([I_{Q}(X;X')-R]_{+} - R + s\right) \cdot L \right\} \\
%%%%%%%%%%%%%%%%%%%%%%%%%%%%%%%%%%%%%%%%%%%%%%%%%%%%%%%%%%%%%%
&= \min_{Q_{X'|X} \in \calQ(Q_{X})} 
\inf_{s \geq \max\left\{R, \Lambda(Q_{XX'},R) - [R-I_{Q}(X;X')]_{+}\right\}}
\left\{ \left([I_{Q}(X;X')-R]_{+} - R + s\right) \cdot L \right\} \\
%%%%%%%%%%%%%%%%%%%%%%%%%%%%%%%%%%%%%%%%%%%%%%%%%%%%%%%%%%%%%%
&= \min_{Q_{X'|X} \in \calQ(Q_{X})} 
\left\{ \left([I_{Q}(X;X')-R]_{+} - R + \max\left\{R, \Lambda(Q_{XX'},R) - [R-I_{Q}(X;X')]_{+}\right\} \right) \cdot L \right\} \\
%%%%%%%%%%%%%%%%%%%%%%%%%%%%%%%%%%%%%%%%%%%%%%%%%%%%%%%%%%%%%%
&= \min_{Q_{X'|X} \in \calQ(Q_{X})} 
L \cdot \max\left\{[I_{Q}(X;X')-R]_{+}, \Lambda(Q_{XX'},R) + I_{Q}(X;X') - 2R \right\},
\end{align}
which complete the proof of the second part of Theorem \ref{THEOREM_Naive}.

\section*{Appendix C}
\renewcommand{\theequation}{C.\arabic{equation}}
\setcounter{equation}{0}  
\subsection*{Proof of Eq.\ (\ref{Result3})}

For a given $m$, $m' \neq m$, and $\by \in \calY^{n}$, define
\begin{align}
\label{Zmmtag_DEF}
Z_{mm'}(\by) = \sum_{\tilde{m}\in\{0,1,\ldots,M-1\} \setminus \{m,m'\}} \exp \{n g(\hat{P}_{\bx_{\tilde{m}}\by})\}.
\end{align}
Let $\delta>0$ and define the set
\begin{align}
\hat{\calB}_{n}(\delta,m,m',\by) = \left\{\calC_{n}:~ Z_{mm'}(\by) \geq \exp\{n \cdot (\beta(R,\hat{P}_{\by}) + \delta)\}  \right\},
\end{align}
and its complement $\hat{\calG}_{n}(\delta,m,m',\by)$,
where $\beta(R,Q_{Y})$ is defined as in \eqref{Beta_DEF}. Let
\begin{align}
\label{B_hat_DEF}
\hat{\calB}_{n}(\delta,m)
= \bigcup_{m' \neq m} \bigcup_{\by \in \calY^{n}} \hat{\calB}_{n}(\delta,m,m',\by),
\end{align}
and
\begin{align}
\label{G_hat_DEF}
\hat{\calG}_{n}(\delta,m) = \hat{\calB}_{n}^{\mbox{\tiny c}}(\delta,m).
\end{align}
Let us define the quantity
\begin{align}
\label{Lambda_Tilde_DEF}
\tilde{\Lambda}(Q_{XX'},R,\delta) &= \min_{Q_{Y|XX'}} \{ D(Q_{Y|X} \| W |Q_{X}) + I_{Q}(X';Y|X) \nonumber \\
&~~+ [\max\{g(Q_{XY}), \beta(R,Q_{Y})+\delta\} - g(Q_{X'Y})]_{+} \},
\end{align}
and the type class enumerator
\begin{align} \label{Def_Enum}
N_{m}(Q_{X'|X}|\bx_{m}) = \sum_{m' \neq m} \IND \left\{\bX_{m'} \in \calT(Q_{X'|X} | \bx_{m})\right\}.
\end{align}
%Denote the event $\bX_{m}=\bx_{m} = \{\bX_{m}=\bx_{m}\}$.
We get the following
\begin{align}
&\prob \left\{ p_{m}(\calC_{n}) \geq e^{-n i \epsilon} \middle| \bX_{m}=\bx_{m} \right\} \nonumber \\
%%%%%%%%%%%%%%%%%%%%%%%%%%%%%%%%%%%%%%%%%%%%%%%%%%%%%%%%%%%%%%%%%%%%%%%%%%%%%%%%%%%%%%%%%%%%%%
\label{aterminus1}
&= \prob \left\{\sum_{m' \neq m} \sum_{\by \in \calY^{n}} W(\by|\bx_{m}) \right. \nn \\ 
&\left. ~~~~\times \frac{\exp\{n g(\hat{P}_{\bX_{m'}\by}) \}}{\exp\{n g(\hat{P}_{\bx_{m}\by}) \} + \exp\{n g(\hat{P}_{\bX_{m'}\by}) \} + Z_{mm'}(\by)}  \geq e^{-n i \epsilon} \middle| \bX_{m}=\bx_{m} \right\} \\
%%%%%%%%%%%%%%%%%%%%%%%%%%%%%%%%%%%%%%%%%%%%%%%%%%%%%%%%%%%%%%%%%%%%%%%%%%%%%%%%%%%%%%%%%%%%%%
\label{aterminus2}
&\geq \prob \left\{\calC_{n} \in \hat{\calG}_{n}(\delta,m), \sum_{m' \neq m} \sum_{\by \in \calY^{n}} W(\by|\bx_{m}) \right. \nn \\ 
&\left. ~~~~\times \frac{\exp\{n g(\hat{P}_{\bX_{m'}\by}) \}}{\exp\{n g(\hat{P}_{\bx_{m}\by}) \} + \exp\{n g(\hat{P}_{\bX_{m'}\by}) \} + Z_{mm'}(\by)}  \geq e^{-n i \epsilon} \middle| \bX_{m}=\bx_{m} \right\} \\
%%%%%%%%%%%%%%%%%%%%%%%%%%%%%%%%%%%%%%%%%%%%%%%%%%%%%%%%%%%%%%%%%%%%%%%%%%%%%%%%%%%%%%%%%%%%%%
\label{aterminus3}
&\geq \prob \left\{\calC_{n} \in \hat{\calG}_{n}(\delta,m), \sum_{m' \neq m} \sum_{\by \in \calY^{n}} W(\by|\bx_{m}) \right. \nn \\
&\left. ~~~~\times \frac{\exp\{n g(\hat{P}_{\bX_{m'}\by}) \}}{\exp\{n g(\hat{P}_{\bx_{m}\by}) \} + \exp\{n g(\hat{P}_{\bX_{m'}\by}) \} + \exp\{n \cdot[\beta(R,\hat{P}_{\by}) + \delta]\} }  \geq e^{-n i \epsilon} \middle| \bX_{m}=\bx_{m} \right\} \\
%%%%%%%%%%%%%%%%%%%%%%%%%%%%%%%%%%%%%%%%%%%%%%%%%%%%%%%%%%%%%%%%%%%%%%%%%%%%%%%%%%%%%%%%%%%%%%
\label{aterminus4}
&\doteq \prob \left\{\calC_{n} \in \hat{\calG}_{n}(\delta,m), \sum_{m' \neq m} \sum_{\by \in \calY^{n}} W(\by|\bx_{m}) \right. \nn \\
&\left.~~~~ \times  \exp\{n \cdot [\max\{ g(\hat{P}_{\bx_{m}\by}), \beta(R,\hat{P}_{\by}) + \delta\} - g(\hat{P}_{\bX_{m'}\by})]_{+} \}   \geq e^{-n i \epsilon} \middle| \bX_{m}=\bx_{m} \right\} \\
%%%%%%%%%%%%%%%%%%%%%%%%%%%%%%%%%%%%%%%%%%%%%%%%%%%%%%%%%%%%%%%%%%%%%%%%%%%%%%%%%%%%%%%%%%%%%%
\label{aterminus5}
&\doteq \prob \left\{\calC_{n} \in \hat{\calG}_{n}(\delta,m), \sum_{m' \neq m} \exp \{-n \cdot \tilde{\Lambda}(\hat{P}_{\bx_{m}\bX_{m'}},R,\delta)\} \geq e^{-n i \epsilon} \middle| \bX_{m}=\bx_{m} \right\} \\
%%%%%%%%%%%%%%%%%%%%%%%%%%%%%%%%%%%%%%%%%%%%%%%%%%%%%%%%%%%%%%%%%%%%%%%%%%%%%%%%%%%%%%%%%%%%%%
\label{ToContinue0}
&= \prob \left\{\calC_{n} \in \hat{\calG}_{n}(\delta,m),  \sum_{Q_{X'|X} \in \calQ(Q_{X})} N_{m}(Q_{X'|X}|\bx_{m}) \cdot
\exp \{-n \cdot \tilde{\Lambda}(Q_{XX'},R,\delta)\}
\geq e^{-n i \epsilon} \middle| \bX_{m}=\bx_{m} \right\} ,
\end{align}
where \eqref{aterminus1} follows from the definitions of the probability of error and $Z_{mm'}(\by)$ in \eqref{ErrorProb} and \eqref{Zmmtag_DEF}, respectively. 
In \eqref{aterminus2}, we lower--bounded by intersecting with the event $\calC_{n} \in \hat{\calG}_{n}(\delta,m)$.
In \eqref{aterminus3}, the definition of the set $\hat{\calG}_{n}(\delta,m)$ in \eqref{G_hat_DEF} was used, 
in \eqref{aterminus4}, the exponential equivalence $e^{nB}/(e^{nA}+e^{nB}+e^{nC}) \doteq \exp\{-n \cdot [\max\{A,C\}-B]_{+}\}$,
in \eqref{aterminus5}, the method of types and the definition of $\tilde{\Lambda}(Q_{XX'},R,\delta)$ in \eqref{Lambda_Tilde_DEF}, 
and in \eqref{ToContinue0}, the definition of the type class enumerators $N_{m}(Q_{X'|X}|\bx_{m})$ in \eqref{Def_Enum}.

Next, we simplify the expression of $\tilde{\Lambda}(Q_{XX'},R,\delta)$.
First, note that for any $\hQ_{XY}$ with marginals $Q_{X}$ and $Q_{Y}$
\begin{align}
\beta(R,Q_{Y}) 
&= \max_{\{Q_{\tilde{X}|Y}:~ Q_{\tilde{X}}=Q_{X}\}} \{g(Q_{\tilde{X}Y}) + [R - I_{Q}(\tilde{X};Y)]_{+}\} \\
&\geq \max_{\{Q_{\tilde{X}|Y}:~ Q_{\tilde{X}}=Q_{X}\}} g(Q_{\tilde{X}Y}) \\
&\geq g(\hQ_{XY}).
\end{align}
Then,
\begin{align}
&\tilde{\Lambda}(Q_{XX'},R,\delta) \nn \\
&= \min_{Q_{Y|XX'}} \{ D(Q_{Y|X} \| W |Q_{X}) + I_{Q}(X';Y|X) \nonumber \\
&~~~+ [\max\{g(Q_{XY}), \beta(R,Q_{Y})+\delta\} - g(Q_{X'Y})]_{+} \} \\
%%%%%%%%%%%%%%%%%%%%%%%%%%%%%%%%%%%%%%%%%%%%%%%%%%%%%
\label{TERM2EXPa1}
&= \min_{Q_{Y|XX'}} \{ D(Q_{Y|X} \| W |Q_{X}) + I_{Q}(X';Y|X) + [\beta(R,Q_{Y}) + \delta - g(Q_{X'Y})]_{+} \} \\
%%%%%%%%%%%%%%%%%%%%%%%%%%%%%%%%%%%%%%%%%%%%%%%%%%%%%
\label{TERM2EXPa2}
&= \min_{Q_{Y|XX'}} \{ D(Q_{Y|X} \| W |Q_{X}) + I_{Q}(X';Y|X) + \beta(R,Q_{Y}) - g(Q_{X'Y}) + \delta \} \\
%%%%%%%%%%%%%%%%%%%%%%%%%%%%%%%%%%%%%%%%%%%%%%%%%%%%%
\label{TERM2EXPa3}
&= \Lambda(Q_{XX'},R) + \delta,
\end{align}
where \eqref{TERM2EXPa1} is due to $\beta(R,Q_{Y}) \geq g(Q_{XY})$, \eqref{TERM2EXPa2} is because $\beta(R,Q_{Y}) \geq g(Q_{X'Y})$, and \eqref{TERM2EXPa3} follows the definition in \eqref{LAMBDA_DEF}.
Let us now define 
\begin{align}
\Gset = \left\{\calC_{n}:~ \sum_{Q_{X'|X} \in \calQ(Q_{X})} N_{m}(Q_{X'|X}|\bx_{m}) \cdot
\exp \{-n \cdot \tilde{\Lambda}(Q_{XX'},R,\delta)\}
\geq e^{-n i \epsilon} \right\},
\end{align}
such that, continuing from \eqref{ToContinue0}:
\begin{align}
&\prob \left\{ p_{m}(\calC_{n}) \geq e^{-n i \epsilon} \middle| \bX_{m}=\bx_{m} \right\} \nn \\
%%%%%%%%%%%%%%%%%%%%%%%%%%%%%%%%%%%%%%%%%%%%%
\label{TERM2Call1}
&\gexe \prob \left\{\hat{\calG}_{n}(\delta,m) \cap \Gset \middle| \bX_{m}=\bx_{m} \right\} \\
%%%%%%%%%%%%%%%%%%%%%%%%%%%%%%%%%%%%%%%%%%%%%%%%%%%%%%%%%%%%%%%%%%%%%%%%%%%%%%%%%%%%%%%%%%%%%%
&= \prob \left\{ \bigcap_{m' \neq m}\bigcap_{\by\in\calY^n}\hat{\calG}_{n}(\delta,m,m',\by) \middle| \Gset, \bX_{m}=\bx_{m}  \right\} \cdot \prob \left\{ \Gset \middle| \bX_{m}=\bx_{m} \right\}  \\
%%%%%%%%%%%%%%%%%%%%%%%%%%%%%%%%%%%%%%%%%%%%%%%%%%%%%%%%%%%%%%%%%%%%%%%%%%%%%%%%%%%%%%%%%%%%%%
&= \left(1 - \prob \left\{ \bigcup_{m' \neq m}\bigcup_{\by\in\calY^n}\hat{\calB}_{n}(\delta,m,m',\by) \middle| \Gset,\bX_{m}=\bx_{m}  \right\} \right) \cdot \prob \left\{ \Gset \middle| \bX_{m}=\bx_{m} \right\}  \\
%%%%%%%%%%%%%%%%%%%%%%%%%%%%%%%%%%%%%%%%%%%%%%%%%%%%%%%%%%%%%%%%%%%%%%%%%%%%%%%%%%%%%%%%%%%%%%
&\geq \left(1 - \sum_{m' \neq m}\sum_{\by\in\calY^n} \prob \left\{ \hat{\calB}_{n}(\delta,m,m',\by) \middle| \Gset, \bX_{m}=\bx_{m}  \right\} \right) \cdot \prob \left\{ \Gset \middle| \bX_{m}=\bx_{m} \right\}  \\  
%%%%%%%%%%%%%%%%%%%%%%%%%%%%%%%%%%%%%%%%%%%%%%%%%%%%%%%%%%%%%%%%%%%%%%%%%%%%%%%%%%%%%%%%%%%%%%
\label{ToCall9}
&= \prob \left\{ \Gset \middle| \bX_{m}=\bx_{m}  \right\} - \sum_{m' \neq m}\sum_{\by\in\calY^n} \prob \left\{ \hat{\calB}_{n}(\delta,m,m',\by) \cap \Gset \middle| \bX_{m}=\bx_{m}  \right\} .
\end{align}

\subsubsection*{Assessing $\prob\{\Gset | \bX_{m}=\bx_{m} \}$ in \eqref{ToCall9}}
Now,
\begin{align}
&\prob\{\Gset | \bX_{m}=\bx_{m} \} \nn \\
%%%%%%%%%%%%%%%%%%%%%%%%%%%%%%%%%%%%%%%%%%%%%
&= \prob \left\{ \sum_{Q_{X'|X} \in \calQ(Q_{X})} N_{m}(Q_{X'|X}|\bx_{m}) \cdot
\exp \{-n \cdot (\Lambda(Q_{XX'},R) + \delta)\}
\geq e^{-n i \epsilon} \middle| \bX_{m}=\bx_{m} \right\} \\
%%%%%%%%%%%%%%%%%%%%%%%%%%%%%%%%%%%%%%%%%%%%%
\label{ENDofPROOF1}
&\doteq \prob \left\{ \max_{Q_{X'|X} \in \calQ(Q_{X})} N_{m}(Q_{X'|X}|\bx_{m}) \cdot
\exp \{-n \cdot (\Lambda(Q_{XX'},R) + \delta)\}
\geq e^{-n i \epsilon} \middle| \bX_{m}=\bx_{m} \right\} \\
%%%%%%%%%%%%%%%%%%%%%%%%%%%%%%%%%%%%%%%%%%%%%%%%
&= \prob \left\{ \bigcup_{Q_{X'|X} \in \calQ(Q_{X})} \left\{ N_{m}(Q_{X'|X}|\bx_{m})
\geq \exp \{n \cdot(\Lambda(Q_{XX'},R) - i \epsilon + \delta)\} \right\} \middle| \bX_{m}=\bx_{m} \right\} \\
%%%%%%%%%%%%%%%%%%%%%%%%%%%%%%%%%%%%%%%%%%%%%%%%
&\doteq \sum_{Q_{X'|X} \in \calQ(Q_{X})} \prob \left\{ N_{m}(Q_{X'|X}|\bx_{m})
\geq \exp \{n \cdot(\Lambda(Q_{XX'},R) - i \epsilon + \delta)\} \middle| \bX_{m}=\bx_{m} \right\} \\
%%%%%%%%%%%%%%%%%%%%%%%%%%%%%%%%%%%%%%%%%%%%%%%
\label{ENDofPROOF2}
&\doteq \max_{Q_{X'|X} \in \calQ(Q_{X})} \prob \left\{N_{m}(Q_{X'|X}|\bx_{m}) \geq \exp\left\{n \cdot (\Lambda(Q_{XX'},R) - i \epsilon + \delta) \right\} \middle| \bX_{m}=\bx_{m} \right\},
\end{align}
where \eqref{ENDofPROOF1} and \eqref{ENDofPROOF2} follow by the SME. 
%and are similar to the steps between \eqref{WHYMAX0}--\eqref{WHYMAX2}. 
Since $N_{m}(Q_{X'|X}|\bx_{m})$ is a binomial sum of $e^{nR}-1$ trials and probability of success $e^{-n I_{Q}(X;X')}$, the last expression decays exponentially with the following rate function 
\begin{align}
&\min_{Q_{X'|X} \in \calQ(Q_{X})} \left\{   
\begin{array}{l l}
\left[I_{Q}(X;X')-R\right]_{+}    & \quad \text{  $\left[R - I_{Q}(X;X')\right]_{+} \geq \Lambda(Q_{XX'},R) - i \epsilon + \delta$  }   \\
\infty    & \quad \text{  $\left[R - I_{Q}(X;X')\right]_{+} < \Lambda(Q_{XX'},R) - i \epsilon + \delta$  }
\end{array} \right. \\
&= \min_{\{Q_{X'|X} \in \calQ(Q_{X}):~ [R - I_{Q}(X;X')]_{+} \geq \Lambda(Q_{XX'},R) - i \epsilon + \delta\}} \left[I_{Q}(X;X')-R\right]_{+} \\
&\equiv E(R,i\epsilon-\delta), 
\end{align}
and thus
\begin{align}
\label{ToCall13}
\prob\{\Gset | \bX_{m}=\bx_{m} \} \doteq \exp\{-n \cdot E(R,i\epsilon-\delta)\}.
\end{align}

%In order to make sure that $E_{\mbox{\tiny lt}}^{\mbox{\tiny lb}}(R,\expVAR) < \infty$, observe that when 
%\begin{align} 
%\expVAR > \min_{Q_{XX'} \in \calQ(Q_{X})} \{\Lambda(Q_{XX'},R) - [2R - I_{Q}(X;X')]_{+}\} + R, 
%\end{align}
%the set $\calM(R,\expVAR)$ is nonempty.
%The proof of the lower bound is complete.

\subsubsection*{Upper--bounding $\prob \{ \hat{\calB}_{n}(\delta,m,m',\by) \cap \Gset | \bX_{m}=\bx_{m} \}$ in \eqref{ToCall9}}

Define the type class enumerator 
\begin{align}
\label{NQ_def2}
N_{m}(Q_{X|Y}|\by) = \sum_{\tilde{m} \neq m} \IND \left\{ \bX_{\tilde{m}} \in \calT(Q_{X|Y}|\by) \right\}. 
\end{align} 
Then, we have the following
\begin{align}
&\prob \{ \hat{\calB}_{n}(\delta,m,m',\by) \cap \Gset | \bX_{m}=\bx_{m} \} \nn \\
%%%%%%%%%%%%%%%%%%%%%%%%%%%%%%%%%%%%%%%%%%%%%%%%%%%%%%%%%%%
%%%%%%%%%%%%%%%%%%%%%%%%%%%%%%%%%%%%%%%%%%%%%%%%%%%%%%%%%%%
&=\prob \left\{ \sum_{\tilde{m}\in\{1,2,\ldots,M\} \setminus \{m,m'\}} \exp \{n g(\hat{P}_{\bX_{\tilde{m}}\by})\} \geq \exp\{n \cdot(\beta(R,\hat{P}_{\by}) + \delta)\}, \nn \right. \\
&\left. ~~~~~~~~~~~~~
\sum_{m'' \neq m} \exp \{-n \cdot (\Lambda(\hat{P}_{\bx_{m}\bX_{m''}},R) + \delta)\} \geq e^{-n i \epsilon} \middle| \bX_{m}=\bx_{m} \right\} \\
%%%%%%%%%%%%%%%%%%%%%%%%%%%%%%%%%%%%%%%%%%%%%%%%%%%%%%%%%%%
%%%%%%%%%%%%%%%%%%%%%%%%%%%%%%%%%%%%%%%%%%%%%%%%%%%%%%%%%%%
&\leq \prob \left\{ \sum_{\tilde{m} \neq m} \exp \{n g(\hat{P}_{\bX_{\tilde{m}}\by})\} \geq \exp\{n \cdot(\beta(R,\hat{P}_{\by}) + \delta)\}, \nn \right. \\
&\left. ~~~~~~~~~~~~~
\sum_{m'' \neq m} \exp \{-n \cdot (\Lambda(\hat{P}_{\bx_{m}\bX_{m''}},R) + \delta)\} \geq e^{-n i \epsilon} \middle| \bX_{m}=\bx_{m} \right\} \\
%%%%%%%%%%%%%%%%%%%%%%%%%%%%%%%%%%%%%%%%%%%%%%%%%%%%%%%%%%%
%%%%%%%%%%%%%%%%%%%%%%%%%%%%%%%%%%%%%%%%%%%%%%%%%%%%%%%%%%%
&=\prob \left\{ \sum_{Q_{X|Y}} N_{m}(Q_{X|Y}|\by) \exp \{n g(Q_{XY})\} \geq \exp\{n \cdot(\beta(R,\hat{P}_{\by}) + \delta)\}, \nn \right. \\
&\left. ~~~~~~~~~~~~~
\sum_{Q_{X'|X}} N_{m}(Q_{X'|X}|\bx_{m}) \exp \{-n \cdot (\Lambda(Q_{XX'},R) + \delta)\} \geq e^{-n i \epsilon} \middle| \bX_{m}=\bx_{m} \right\} \\
%%%%%%%%%%%%%%%%%%%%%%%%%%%%%%%%%%%%%%%%%%%%%%%%%%%%%%%%%%%
%%%%%%%%%%%%%%%%%%%%%%%%%%%%%%%%%%%%%%%%%%%%%%%%%%%%%%%%%%%
\label{abterminus0}
&\doteq \prob \left\{ \bigcup_{Q_{X|Y}} \left\{N_{m}(Q_{X|Y}|\by) \geq e^{n \cdot(\beta(R,\hat{P}_{\by}) - g(Q_{XY}) + \delta)} \right\}, 
\nn \right. \\ &\left. ~~~~~~~~~~~~~
\bigcup_{Q_{X'|X}} \left\{ N_{m}(Q_{X'|X}|\bx_{m}) \geq e^{n \cdot (\Lambda(Q_{XX'},R) - i \epsilon + \delta)} \right\} \middle| \bX_{m}=\bx_{m} \right\} \\
%%%%%%%%%%%%%%%%%%%%%%%%%%%%%%%%%%%%%%%%%%%%%%%%%%%%%%%%%%%
%%%%%%%%%%%%%%%%%%%%%%%%%%%%%%%%%%%%%%%%%%%%%%%%%%%%%%%%%%%
%&= \prob \left\{ \bigcup_{Q_{\tilde{X}Y}} \left\{[N_{\by}(Q_{\tilde{X}Y})]^{k} \geq e^{n \cdot[\beta(R,\hat{P}_{\by}) - g(Q_{\tilde{X}Y}) + \epsilon] \cdot k} \right\}, 
%%\nn \right. \\ &\left. ~~~~~~~~~~~~~
%\bigcup_{Q_{XX'}} \left\{ [N(Q_{XX'})]^{l} \geq e^{n \cdot [\tilde{\Lambda}(Q_{XX'},R) + R - \expVAR] \cdot l} \right\} \right\} \\
%%%%%%%%%%%%%%%%%%%%%%%%%%%%%%%%%%%%%%%%%%%%%%%%%%%%%%%%%%%
%%%%%%%%%%%%%%%%%%%%%%%%%%%%%%%%%%%%%%%%%%%%%%%%%%%%%%%%%%%
&\doteq \sum_{Q_{X|Y}} \sum_{Q_{X'|X}} \prob \left\{N_{m}(Q_{X|Y}|\by)^{\ell} \geq e^{n \cdot(\beta(R,\hat{P}_{\by}) - g(Q_{XY}) + \delta) \cdot \ell}, 
\nn \right. \\ &\left. ~~~~~~~~~~~~~~~~~~~~~~~~~
N_{m}(Q_{X'|X}|\bx_{m})^{k} \geq e^{n \cdot (\Lambda(Q_{XX'},R) - i \epsilon + \delta) \cdot k} \middle| \bX_{m}=\bx_{m} \right\} \\
%%%%%%%%%%%%%%%%%%%%%%%%%%%%%%%%%%%%%%%%%%%%%%%%%%%%%%%%%%%
%%%%%%%%%%%%%%%%%%%%%%%%%%%%%%%%%%%%%%%%%%%%%%%%%%%%%%%%%%%
&\doteq \max_{Q_{X|Y}} \max_{Q_{X'|X}} \prob \left\{N_{m}(Q_{X|Y}|\by)^{\ell} \geq e^{n \cdot(\beta(R,\hat{P}_{\by}) - g(Q_{XY}) + \delta) \cdot \ell}, 
\nn \right. \\ &\left. ~~~~~~~~~~~~~~~~~~~~~~~~~
N_{m}(Q_{X'|X}|\bx_{m})^{k} \geq e^{n \cdot (\Lambda(Q_{XX'},R) - i \epsilon + \delta) \cdot k} \middle| \bX_{m}=\bx_{m} \right\} \\
%%%%%%%%%%%%%%%%%%%%%%%%%%%%%%%%%%%%%%%%%%%%%%%%%%%%%%%%%%%
%%%%%%%%%%%%%%%%%%%%%%%%%%%%%%%%%%%%%%%%%%%%%%%%%%%%%%%%%%%
\label{abterminus1}
&\leq \max_{Q_{X|Y}} \max_{Q_{X'|X}} \prob \left\{N_{m}(Q_{X|Y}|\by)^{\ell} \cdot
N_{m}(Q_{X'|X}|\bx_{m})^{k} 
\nn \right. \\ &\left. ~~~~~~~~~~~~~~~~~~~~~~~~~
\geq e^{n \cdot(\beta(R,\hat{P}_{\by}) - g(Q_{XY}) + \delta) \cdot \ell} \cdot e^{n \cdot (\Lambda(Q_{XX'},R) - i \epsilon + \delta) \cdot k} \middle| \bX_{m}=\bx_{m} \right\} \\
%%%%%%%%%%%%%%%%%%%%%%%%%%%%%%%%%%%%%%%%%%%%%%%%%%%%%%%%%%%
%%%%%%%%%%%%%%%%%%%%%%%%%%%%%%%%%%%%%%%%%%%%%%%%%%%%%%%%%%%
\label{abterminus2}
&\leq \max_{Q_{X|Y}} \max_{Q_{X'|X}} \prob \left\{N_{m}(Q_{X|Y}|\by)^{\ell} \cdot
N_{m}(Q_{X'|X}|\bx_{m})^{k} 
\nn \right. \\ &\left. ~~~~~~~~~~~~~~~~~~~~~~~~~
\geq e^{n \cdot ([R - I_{Q}(X;Y)]_{+} + \delta) \cdot \ell} \cdot e^{n \cdot (\Lambda(Q_{XX'},R) - i \epsilon + \delta) \cdot k} \middle| \bX_{m}=\bx_{m} \right\} ,
\end{align}
where $k$ and $\ell$ are arbitrary positive integers. 
Step \eqref{abterminus1} is due to the fact that $\prob\{X\geq a, Y\geq b\} \leq \prob\{X\cdot Y \geq a \cdot b\}$, under the assumption that $a,b$ are positive.
In \eqref{abterminus2}, we use the definition of $\beta(R,Q_{Y})$ in \eqref{Beta_DEF}, which implies that $\beta(R,Q_{Y}) \geq g(Q_{XY}) + \left[R - I_{Q}(X;Y)\right]_{+}$. 

It follows from Markov's inequality that
\begin{align}
&\prob \left\{N_{m}(Q_{X|Y}|\by)^{\ell} \cdot
N_{m}(Q_{X'|X}|\bx_{m})^{k} \geq e^{n \cdot ([R - I_{Q}(X;Y)]_{+} + \delta) \cdot \ell} \cdot e^{n \cdot (\Lambda(Q_{XX'},R) - i \epsilon + \delta) \cdot k} \middle| \bX_{m}=\bx_{m} \right\} \nn \\
%%%%%%%%%%%%%%%%%%%%%%%%%%%%%%%%%%%%%%%%%%%%%%%%%%%%%%%%%%%
%%%%%%%%%%%%%%%%%%%%%%%%%%%%%%%%%%%%%%%%%%%%%%%%%%%%%%%%%%%
&~~~~\leq \inf_{\ell \in \mathbb{N}} \inf_{k \in \mathbb{N}}  
\frac{\mathbb{E} \left[ N_{m}(Q_{X|Y}|\by)^{\ell} \cdot
	N_{m}(Q_{X'|X}|\bx_{m})^{k} \middle| \bX_{m}=\bx_{m}  \right]}{e^{n \cdot([R - I_{Q}(X;Y)]_{+} + \delta) \cdot \ell} \cdot e^{n \cdot (\Lambda(Q_{XX'},R) - i \epsilon + \delta) \cdot k}},
\end{align}
and substituting it back into \eqref{abterminus2} yields 
\begin{align}
\label{ToCall10}
&\prob \{ \hat{\calB}_{n}(\epsilon,m,m',\by) \cap \Gset | \bX_{m}=\bx_{m} \} \nn \\
&~~~~\lexe \max_{Q_{X|Y}} \max_{Q_{X'|X}} \inf_{\ell \in \mathbb{N}} \inf_{k \in \mathbb{N}} 
\frac{\mathbb{E} \left[ N_{m}(Q_{X|Y}|\by)^{\ell} \cdot
	N_{m}(Q_{X'|X}|\bx_{m})^{k} \middle| \bX_{m}=\bx_{m}  \right]}{e^{n \cdot ([R - I_{Q}(X;Y)]_{+} + \delta) \cdot \ell} \cdot e^{n \cdot (\Lambda(Q_{XX'},R) - i \epsilon + \delta) \cdot k}} .
\end{align}
For $S \geq 0$, a joint distribution $Q_{UV}$, and an integer $j \in \mathbb{N}$, define the following quantity
\begin{align}
\label{DEF_F}
F(S,Q_{UV},j)
=\left\{   
\begin{array}{l l}
\exp \{n j \left(S - I_{Q}(U;V)\right) \}    & \quad \text{  $I_{Q}(U;V) < S$  }   \\
\exp \{n \left(S - I_{Q}(U;V)\right) \}    & \quad \text{  $I_{Q}(U;V) > S$  }
\end{array} \right. .
\end{align}
%with the understanding that the mutual information is between the two random variables involved in the joint distribution.
We use the following proposition:
\begin{proposition} \label{Prop_Moments}
	Let $N_{m}(Q_{X'|X}|\bx_{m})$ and $N_{m}(Q_{X|Y}|\by)$ be as in \eqref{Def_Enum} and \eqref{NQ_def2}, respectively. 
	Then, for any $\ell,k \in \mathbb{N}$,
	\begin{align}
	\mathbb{E} \left[ N_{m}(Q_{X|Y}|\by)^{\ell}
	N_{m}(Q_{X'|X}|\bx_{m})^{k} \middle| \bX_{m}=\bx_{m}  \right]
	&\lexe F(R,Q_{XY},\ell) \cdot F(R,Q_{XX'},k).
	\end{align}
\end{proposition}
Since Proposition \ref{Prop_Moments} is very close in spirit to \cite[Proposition 4]{TMWG}, we omit the proof.
Substituting the result of Proposition \ref{Prop_Moments} back into \eqref{ToCall10} provides
\begin{align}
&\prob \{ \hat{\calB}_{n}(\delta,m,m',\by) \cap \Gset | \bX_{m}=\bx_{m}  \} \nn \\
\label{ToCall11}
&~~~~\lexe \max_{Q_{X|Y}}  \inf_{\ell \in \mathbb{N}} 
\frac{\exp\{n \cdot (\ell \cdot [R-I_{Q}(X;Y)]_{+} - [I_{Q}(X;Y)-R]_{+})\}}{\exp\{n \cdot ([R - I_{Q}(X;Y)]_{+} + \delta) \cdot \ell\}} \nn \\
&~~~~\times \max_{Q_{X'|X}} \inf_{k \in \mathbb{N}} 
\frac{\exp\{n \cdot (k \cdot [R-I_{Q}(X;X')]_{+} - [I_{Q}(X;X')-R]_{+})\}}{\exp\{n \cdot (\Lambda(Q_{XX'},R) - i \epsilon + \delta) \cdot k\}}.
\end{align} 
As for the left--hand term in \eqref{ToCall11}, we have that
\begin{align}
&-\frac{1}{n} \log\max_{Q_{X|Y}}  \inf_{\ell \in \mathbb{N}} 
\frac{\exp\{n \cdot (\ell \cdot [R-I_{Q}(X;Y)]_{+} - [I_{Q}(X;Y)-R]_{+})\}}{\exp\{n \cdot ([R - I_{Q}(X;Y)]_{+} + \delta) \cdot \ell\}} \nn \\
%%%%%%%%%%%%%%%%%%%%%%%%%%%%%%%%%%%%%%%%%%%%%%%%%%%%%%%%%%
&=-\frac{1}{n} \log\max_{Q_{X|Y}}  \inf_{\ell \in \mathbb{N}} 
\exp\{-n \cdot \left([I_{Q}(X;Y)-R]_{+} + \ell \delta \right)\}\\
%%%%%%%%%%%%%%%%%%%%%%%%%%%%%%%%%%%%%%%%%%%%%%%%%%%%%%%%%%
&=\min_{Q_{X|Y}}  \sup_{\ell \in \mathbb{N}} 
\left([I_{Q}(X;Y)-R]_{+} + \ell \delta \right)\\
%%%%%%%%%%%%%%%%%%%%%%%%%%%%%%%%%%%%%%%%%%%%%%%%%%%%%%%%%%
&= \infty .
\end{align}
For the right--hand term in \eqref{ToCall11}, we get the following
\begin{align}
& -\frac{1}{n} \log \max_{Q_{X'|X}} \inf_{k \in \mathbb{N}} 
\frac{\exp\{n \cdot (k \cdot [R-I_{Q}(X;X')]_{+} - [I_{Q}(X;X')-R]_{+})\}}{\exp\{n \cdot (\Lambda(Q_{XX'},R) - i \epsilon + \delta) \cdot k\}} \nn \\
%%%%%%%%%%%%%%%%%%%%%%%%%%%%%%%%%%%%%%%%%%%%%%%%%%%%%%%%%%%%
%%%%%%%%%%%%%%%%%%%%%%%%%%%%%%%%%%%%%%%%%%%%%%%%%%%%%%%%%%%%
&= \min_{Q_{X'|X}} \sup_{k \in \mathbb{N}} \left( k \cdot \left(\Lambda(Q_{XX'},R) - i \epsilon + \delta  - [R-I_{Q}(X;X')]_{+} \right) + [I_{Q}(X;X')-R]_{+} \right) \\
%%%%%%%%%%%%%%%%%%%%%%%%%%%%%%%%%%%%%%%%%%%%%%%%%%%%%%%%%%%%
%%%%%%%%%%%%%%%%%%%%%%%%%%%%%%%%%%%%%%%%%%%%%%%%%%%%%%%%%%%%
&= \min_{\{Q_{X'|X} \in \calQ(Q_{X}):~ [R - I_{Q}(X;X')]_{+} \geq \Lambda(Q_{XX'},R) - i \epsilon + \delta\}} \left[I_{Q}(X;X')-R\right]_{+}  \\
%%%%%%%%%%%%%%%%%%%%%%%%%%%%%%%%%%%%%%%%%%%%%%%%%%%%%%%%%%%%
%%%%%%%%%%%%%%%%%%%%%%%%%%%%%%%%%%%%%%%%%%%%%%%%%%%%%%%%%%%%
&= E(R,i\epsilon-\delta) .
\end{align}
Thus,
\begin{align}
\label{ToCall12}
\prob \{ \hat{\calB}_{n}(\delta,m,m',\by) \cap \Gset | \bX_{m}=\bx_{m} \}
&\lexe e^{-n \infty} \cdot \exp\{-n \cdot E(R,i\epsilon-\delta)\} .
\end{align}

\subsubsection*{Final Steps}
Finally, we continue from \eqref{ToCall9} and use the results of \eqref{ToCall13} and \eqref{ToCall12} to provide 
\begin{align}
&\prob \left\{ p_{m}(\calC_{n}) \geq e^{-n i \epsilon} \middle| \bX_{m}=\bx_{m} \right\}  \nn \\
%%%%%%%%%%%%%%%%%%%%%%%%%%%%%%%%%%%%%%%%%%%%%%%%%%%%%%%%%%%%%%%%%%%%%%%%%%%%%%%%%%%%%%%%%%%%%%
&\gexe \prob \left\{ \Gset \middle| \bX_{m}=\bx_{m} \right\} - \sum_{m' \neq m}\sum_{\by\in\calY^n} \prob \left\{ \hat{\calB}_{n}(\delta,m,m',\by) \cap \Gset \middle| \bX_{m}=\bx_{m} \right\} \\
%%%%%%%%%%%%%%%%%%%%%%%%%%%%%%%%%%%%%%%%%%%%%%%%%%%%%%%%%%
%%%%%%%%%%%%%%%%%%%%%%%%%%%%%%%%%%%%%%%%%%%%%%%%%%%%%%%%%%
&\gexe \exp\{-n \cdot E(R,i\epsilon-\delta)\} - \sum_{m' \neq m}\sum_{\by\in\calY^n} e^{-n \infty} \cdot \exp\{-n \cdot E(R,i\epsilon-\delta)\} \\
%%%%%%%%%%%%%%%%%%%%%%%%%%%%%%%%%%%%%%%%%%%%%%%%%%%%%%%%%%
%%%%%%%%%%%%%%%%%%%%%%%%%%%%%%%%%%%%%%%%%%%%%%%%%%%%%%%%%%
&\doteq \left(1 - e^{nR} \cdot |\calY|^n \cdot e^{-n \infty} \right) \cdot \exp\{-n \cdot E(R,i\epsilon-\delta)\} \\
%%%%%%%%%%%%%%%%%%%%%%%%%%%%%%%%%%%%%%%%%%%%%%%%%%%%%%%%%%
%%%%%%%%%%%%%%%%%%%%%%%%%%%%%%%%%%%%%%%%%%%%%%%%%%%%%%%%%%
\label{Recall1}
&\doteq \exp\{-n \cdot E(R,i\epsilon-\delta)\}.
\end{align}
Since \eqref{Recall1} is independent of the specific realization of $\bX_{m}$, it immediately follows that 
\begin{align}
\prob \left\{ p_{m}(\calC_{n}) \geq e^{-n i \epsilon}  \right\}
%%%%%%%%%%%%%%%%%%%%%%%%%%%%%%%%%%%%%%%%%%%%%%%%%%%%%%%%%%
%%%%%%%%%%%%%%%%%%%%%%%%%%%%%%%%%%%%%%%%%%%%%%%%%%%%%%%%%%
\gexe \exp\{-n \cdot E(R,i\epsilon-\delta)\},
\end{align}
and due to the arbitrariness of $\delta > 0$, we conclude that
\begin{align}
\prob \left\{ p_{m}(\calC_{n}) \geq e^{-n i \epsilon} \right\} 
%%%%%%%%%%%%%%%%%%%%%%%%%%%%%%%%%%%%%%%%%%%%%%%%%%%%%%%%%%%%%%%%%%%%%%%%%%%%%%%%%%%%%%%%%%%%%%
\gexe \exp\{-n \cdot E(R,i\epsilon)\} ,
\end{align}
which is exactly \eqref{Result3}.

\section*{Appendix D}
\renewcommand{\theequation}{D.\arabic{equation}}
\setcounter{equation}{0}  
\subsection*{Proof of Theorem \ref{THEOREM_Naive_TRC}}

We have proved in \eqref{term1} that
\begin{align}
\label{term4}
P_{\mbox{\tiny e}}(\calC_{n}) 
\leq \min \left\{L^{L}, \left(\mu(\calC_{n})\right)^{L} \right\}  + \IND \left\{\mu(\calC_{n}) > L\right\} .
\end{align}
Note that for every codebook, 
the first term on the right hand side of \eqref{term4} is at least as large as the second term, and hence, the right hand side of \eqref{term4} can be further upper--bounded by 
\begin{align}
\label{term7}
P_{\mbox{\tiny e}}(\calC_{n}) 
\leq 2 \min \left\{L^{L}, \left(\mu(\calC_{n})\right)^{L} \right\} .
\end{align}
It follows that
\begin{align}
\mathbb{E} \left[ \log P_{\mbox{\tiny e}}(\calC_{n}) \right]  
&\lexe \mathbb{E} \left[ \min \left\{L\log(L), L\log \left(\mu(\calC_{n})\right) \right\} \right]\\
\label{term6}
&\leq \min \left\{L\log(L), L \cdot \mathbb{E} \left[\log \left(\mu(\calC_{n})\right) \right] \right\}  .
\end{align}
In order to derive $\mathbb{E} \left[ \log(\mu(\calC_{n})) \right]$, we note that $\mu(\calC_{n})$ is very similar to the probability of error in ordinary channel coding, which is given by 
\begin{align}
\frac{1}{M}\sum_{m=1}^{M} \sum_{m' \neq m} \sum_{\by \in \calY^{n}} W(\by|\bx_{m}) \cdot \frac{\exp\{n  g(\hat{P}_{\bx_{m'}\by})\}}{\sum_{\tilde{m}=1}^{M} \exp\{n g(\hat{P}_{\bx_{\tilde{m}}\by})\}},
\end{align}  
and hence, we rely on the derivation in \cite[Subsection 5.1]{MERHAV_TYPICAL} and only provide a proof sketch.
Assessing the $1/\rho$--th moment of $\mu(\calC_{n})$, for any $\rho>1$, we get that
\begin{align}
\label{term5}
\mathbb{E} \left\{ \left[ \mu(\calC_{n}) \right]^{1/\rho} \right\} 
\lexe \sum_{Q_{X'|X} \in \calQ(Q_{X})} \mathbb{E} \left\{ \left[ N(Q_{XX'}) \right]^{1/\rho} \right\} \cdot \exp \left\{-n \Gamma(Q_{XX'},R-\epsilon) / \rho \right\} .
\end{align}
The $1/\rho$--th moment of $N(Q_{XX'})$ is upper-bounded by \cite{MERHAV_TYPICAL} 
\begin{align}
\mathbb{E} \left\{ \left[ N(Q_{XX'}) \right]^{1/\rho} \right\} 
\leq \exp \left\{n \cdot \left( [2R - I_{Q}(X;X')]_{+} / \rho - [I_{Q}(X;X')-2R]_{+}\right) \right\},
\end{align}
and then
\begin{align}
\lim_{\rho \to \infty} \left(\mathbb{E} \left\{ \left[ N(Q_{XX'}) \right]^{1/\rho} \right\} \right)^{\rho} 
\leq
\left\{   
\begin{array}{l l}
\exp\{n \cdot[2R-I_{Q}(X;X')]\}    & \quad \text{  $2R \geq I_{Q}(X;X')$  }   \\
0    & \quad \text{  $2R < I_{Q}(X;X')$  }
\end{array} \right. .
\end{align}
Substituting it back into \eqref{term5} gives 
\begin{align}
&\lim_{\rho \to \infty} \left(\mathbb{E} \left\{ \left[ \mu(\calC_{n}) \right]^{1/\rho} \right\} \right)^{\rho} \nn \\
&\lexe \sum_{\{Q_{X'|X} \in \calQ(Q_{X}):~ I_{Q}(X;X') \leq 2R\}} e^{n \cdot[2R-I_{Q}(X;X')]} \cdot \exp \left\{-n \Gamma(Q_{XX'},R-\epsilon)  \right\} \\
&\doteq \exp \left\{-n \cdot \min_{\{Q_{X'|X} \in \calQ(Q_{X}):~ I_{Q}(X;X') \leq 2R\}}
\left[\Gamma(Q_{XX'},R-\epsilon) + I_{Q}(X;X') - 2R \right] \right\} ,
\end{align}
and hence, it follows from the identity
\begin{align} \label{StartingPoint}
\mathbb{E} [\log \mu(\calC_{n}) ]
= \lim_{\rho \to \infty} \log \left( \mathbb{E} [\mu(\calC_{n})]^{1/\rho}   \right)^{\rho}
\end{align}
that
\begin{align}
\mathbb{E} \left[ \log(\mu(\calC_{n})) \right]
\lexe -n \cdot \min_{\{Q_{X'|X} \in \calQ(Q_{X}):~ I_{Q}(X;X') \leq 2R\}}
\left[\Gamma(Q_{XX'},R-\epsilon) + I_{Q}(X;X') - 2R \right] .
\end{align}
According to \eqref{term6}, 
\begin{align}
&\lim_{n \to \infty} -\frac{1}{n} \mathbb{E} \left[\log P_{\mbox{\tiny e}}(\calC_{n}) \right] \nn \\
%%%%%%%%%%%%%%%%%%%%%%%%%%%%%%%%%%%%%%%%%%%%%%%%%%%% 
&\geq \lim_{n \to \infty} -\frac{1}{n}
\min \left\{L\log(L), L \cdot \mathbb{E} \left[\log \left(\mu(\calC_{n})\right) \right] \right\} \\
%%%%%%%%%%%%%%%%%%%%%%%%%%%%%%%%%%%%%%%%%%%%%%%%%%%% 
&= \max \left\{\lim_{n \to \infty} -\frac{1}{n} L\log(L), \lim_{n \to \infty} -\frac{1}{n} L \cdot \mathbb{E} \left[\log \left(\mu(\calC_{n})\right) \right] \right\} \\
%%%%%%%%%%%%%%%%%%%%%%%%%%%%%%%%%%%%%%%%%%%%%%%%%%%
&\geq \max \left\{0, \min_{\{Q_{X'|X} \in \calQ(Q_{X}):~ I_{Q}(X;X') \leq 2R\}}
L \cdot \left[\Gamma(Q_{XX'},R-\epsilon) + I_{Q}(X;X') - 2R \right] \right\} \\
%%%%%%%%%%%%%%%%%%%%%%%%%%%%%%%%%%%%%%%%%%%%%%%%%%%
&= \min_{\{Q_{X'|X} \in \calQ(Q_{X}):~ I_{Q}(X;X') \leq 2R\}}
L \cdot \left[\Gamma(Q_{XX'},R-\epsilon) + I_{Q}(X;X') - 2R \right]_{+},
\end{align}
and it follows from the arbitrariness of $\epsilon>0$ that
\begin{align}
\lim_{n \to \infty} -\frac{1}{n} \mathbb{E} \left[\log P_{\mbox{\tiny e}}(\calC_{n}) \right] \geq E_{\mbox{\tiny trc}}(R,L),
\end{align}
which proves Theorem \ref{THEOREM_Naive_TRC}.

\section*{Appendix E}
\renewcommand{\theequation}{E.\arabic{equation}}
\setcounter{equation}{0}  
\subsection*{Proof of Theorem \ref{THEOREM_Naive_EX}}

Let us first recall the following result from \cite{MERHAV2017}, which provides an expurgated error exponent in the settings of ordinary channel coding.
\begin{theorem}[Theorem 2 in \cite{MERHAV2017}]
	There exists a sequence of constant composition codes, $\{\calC_{n},~n=1,2,\dotsc\}$, with composition $Q_{X}$, such that
	\begin{align}
	\liminf_{n \to \infty} \left[- \frac{1}{n} \log \max_{m} p_{m}(\calC_{n})\right] \geq E_{\mbox{\tiny ex}}(R,Q_{X}),
	\end{align}
	where,
	\begin{align}
	E_{\mbox{\tiny ex}}(R,Q_{X}) = \min_{\{Q_{X'|X} \in \calQ(Q_{X}):~ I_{Q}(X;X') \leq R\}}
	\left[\Gamma(Q_{XX'},R) + I_{Q}(X;X') - R \right].
	\end{align}
\end{theorem}  
Assume that we use this sequence of good constant composition codes. Then, we continue from \eqref{term7} and arrive at 
\begin{align}
P_{\mbox{\tiny e}}(\calC_{n}) 
&\lexe \min \left\{L^{L}, \left(\mu(\calC_{n})\right)^{L} \right\} \\
&= \min \left\{L^{L}, \left(\sum_{m=1}^{M} p_{m}(\calC_{n}) \right)^{L} \right\} \\
&\leq \min \left\{L^{L}, \left(\sum_{m=1}^{M} \exp\left\{-n \cdot E_{\mbox{\tiny ex}}(R,Q_{X}) \right\} \right)^{L} \right\} \\
&= \min \left\{L^{L}, \exp\left\{-n \cdot L \cdot \left[E_{\mbox{\tiny ex}}(R,Q_{X}) - R \right] \right\}  \right\} \\
&= \exp\left\{-n \cdot L \cdot \left[E_{\mbox{\tiny ex}}(R,Q_{X}) - R \right]_{+} \right\},
\end{align}
which proves Theorem \ref{THEOREM_Naive_EX}.

\section*{Appendix F}
\renewcommand{\theequation}{F.\arabic{equation}}
\setcounter{equation}{0}  
\subsection*{Proof of Theorem \ref{Theorem_EX}}

Assume that we draw a codebook $\calC_{0}=\{\bx_{1}, \bx_{2}, \ldots, \bx_{M_{0}}\}$, where $\bx_{i}$, $i \in \{1,2,\ldots,M_{0}\}$, is drawn i.i.d.\ according to $P_{X}$, and $M_{0} = 2M = e^{nR}$. 
Let $\mathbb{C}(M,\calC_{0})$ be the set of all subsets (codebooks) of $\calC_{0}$ with size $M$.
%Let $\IND(\calC_{n})$ be the index set of the codewords in $\calC_{n}$.
Denote $\xi_{n} \dfn |\mathbb{C}(M,\calC_{0})| = \binom{2M}{M}$ and let us enumerate the codebooks in $\mathbb{C}(M,\calC_{0})$ 
by $m \in \{1,2,\ldots,\xi_{n}\}$ and denote them by $\calC_{n}^{m}$.

We assume, without loss of generality, that the permutation induced by the channel is the identity permutation, denoted by $\pi_{0}$. 
The probability of error, associated with $\calC_{n}^{m} \in \mathbb{C}(M,\calC_{0})$ is given by
\begin{align}
P_{\mbox{\tiny e}}(\calC_{n}^{m}) 
&= \sum_{\by_{1} \in \calY^{n}} \cdots \sum_{\by_{M} \in \calY^{n}} \prod_{m=1}^{M} W(\by_{m}|\bx_{m}) \IND \{\hat{\pi}(\by_{1},\ldots,\by_{M}) \neq \pi_{0} \} \\
&= \sum_{\by_{1} \in \calY^{n}} \cdots \sum_{\by_{M} \in \calY^{n}} \prod_{m=1}^{M} W(\by_{m}|\bx_{m}) 
\IND \left\{ \bigcup_{\substack{\pi \in \Pi(M) \\ \pi \neq \pi_{0}}} \left\{ \prod_{m=1}^{M} W(\by_{m}|\bx_{\pi(m)}) \geq  \prod_{m=1}^{M} W(\by_{m}|\bx_{m}) \right\} \right\} \\
&\leq \sum_{\by_{1} \in \calY^{n}} \cdots \sum_{\by_{M} \in \calY^{n}} \prod_{m=1}^{M} W(\by_{m}|\bx_{m}) 
\sum_{\substack{\pi \in \Pi(M) \\ \pi \neq \pi_{0}}}
\frac{\sqrt{\prod_{m=1}^{M} W(\by_{m}|\bx_{\pi(m)})}}{\sqrt{\prod_{m=1}^{M} W(\by_{m}|\bx_{m})}} \\
&= \sum_{\by_{1} \in \calY^{n}} \cdots \sum_{\by_{M} \in \calY^{n}} 
\sum_{\substack{\pi \in \Pi(M) \\ \pi \neq \pi_{0}}}
\sqrt{\prod_{m=1}^{M} W(\by_{m}|\bx_{m})} \sqrt{\prod_{m=1}^{M} W(\by_{m}|\bx_{\pi(m)})} \\
&= \sum_{\by_{1} \in \calY^{n}} \cdots \sum_{\by_{M} \in \calY^{n}} 
\sum_{\substack{\pi \in \Pi(M) \\ \pi \neq \pi_{0}}}
\prod_{m=1}^{M} 
\sqrt{W(\by_{m}|\bx_{m}) W(\by_{m}|\bx_{\pi(m)})} \\
&= \sum_{\substack{\pi \in \Pi(M) \\ \pi \neq \pi_{0}}}
\prod_{m=1}^{M} \sum_{\by_{m} \in \calY^{n}} 
\sqrt{W(\by_{m}|\bx_{m}) W(\by_{m}|\bx_{\pi(m)})} .
\end{align}
Now, raising it to the $1/\sigma$-th power for some $\sigma \geq 1$ and averaging over the codebook yields
\begin{align}
\mathbb{E} \left[P_{\mbox{\tiny e}}(\calC_{n}^{m})^{1/\sigma}\right] 
%%%%%%%%%%%%%%%%%%%%%%%%%%%%%%%%%%%%%%%%%%%%%%%%%%%%%%%%%%%
&\leq 
\mathbb{E} \left[\left(\sum_{\substack{\pi \in \Pi(M) \\ \pi \neq \pi_{0}}} 
\prod_{m=1}^{M} \sum_{\by_{m} \in \calY^{n}} 
\sqrt{W(\by_{m}|\bX_{m}) W(\by_{m}|\bX_{\pi(m)})}\right)^{1/\sigma}\right] \\
%%%%%%%%%%%%%%%%%%%%%%%%%%%%%%%%%%%%%%%%%%%%%%%%%%%%%%%%%%%
&\leq 
\mathbb{E} \left[\sum_{\substack{\pi \in \Pi(M) \\ \pi \neq \pi_{0}}} \left( 
\prod_{m=1}^{M} \sum_{\by_{m} \in \calY^{n}} 
\sqrt{W(\by_{m}|\bX_{m}) W(\by_{m}|\bX_{\pi(m)})}\right)^{1/\sigma}\right] \\
%%%%%%%%%%%%%%%%%%%%%%%%%%%%%%%%%%%%%%%%%%%%%%%%%%%%%%%%%%%
&= 
\sum_{\substack{\pi \in \Pi(M) \\ \pi \neq \pi_{0}}} \mathbb{E} \left[\left( 
\prod_{m=1}^{M} \sum_{\by_{m} \in \calY^{n}} 
\sqrt{W(\by_{m}|\bX_{m}) W(\by_{m}|\bX_{\pi(m)})}\right)^{1/\sigma}\right] \\
%%%%%%%%%%%%%%%%%%%%%%%%%%%%%%%%%%%%%%%%%%%%%%%%%%%%%%%%%%
\label{Recall5}
&\dfn \sum_{\substack{\pi \in \Pi(M) \\ \pi \neq \pi_{0}}} G(\pi,\sigma)  .
\end{align}

\subsubsection*{Step 1: The Permutation is a Transposition}

Assume, without loss of generality, a permutation with $\pi(1)=2, \pi(2)=1$, and $\pi(m)=m$, $\forall m \geq 3$. Then, we get that
\begin{align}
G(\pi,\sigma)
&= \left[\sum_{x_{1} \in \calX} \sum_{x_{2} \in \calX} P_{X}(x_{1}) P_{X}(x_{2}) \left(\sum_{y_{1} \in \calY} 
\sqrt{W(y_{1}|x_{1}) W(y_{1}|x_{2})} \right)^{1/\sigma} \right. \nn \\
&~~~~~~~~~~~~~~~~~~~~~~~~~~~~~~~~~~~~~~\left. \times \left(\sum_{y_{2} \in \calY} 
\sqrt{W(y_{2}|x_{2}) W(y_{2}|x_{1})} \right)^{1/\sigma} \right]^{n} \\
%%%%%%%%%%%%%%%%%%%%%%%%%%%%%%%%%%%%%%%%%%%%%%%%%%%%%
&= \left[\sum_{x_{1} \in \calX} \sum_{x_{2} \in \calX} P_{X}(x_{1}) P_{X}(x_{2}) \left[B(x_{1},x_{2})\right]^{2/\sigma} \right]^{n} \\
%%%%%%%%%%%%%%%%%%%%%%%%%%%%%%%%%%%%
&= \left(\mathbb{E} \left[ B(X_{1},X_{2})^{2/\sigma} \right]\right)^{n} \\
&= [\Xi(\sigma)]^{n}.
\end{align}

\subsubsection*{Step 2: The Permutation is a Cycle}

In this case, assume, without loss of generality, that $\pi(i)=i+1$ for $1 \leq i \leq k-1$, $\pi(k)=1$ and $\pi(m)=m$, $\forall m \geq k+1$. We have that
\begin{align}
G(\pi,\sigma)
%%%%%%%%%%%%%%%%%%%%%%%%%%%%%%%%%%%%%%%%%%%%%%%%%%
&= \left[\sum_{x_{1} \in \calX} \cdots \sum_{x_{k} \in \calX} \left( \prod_{i=1}^{k} P_{X}(x_{i}) \right) 
\left( B(x_{1},x_{2}) B(x_{2},x_{3}) \cdots
B(x_{k-1},x_{k}) B(x_{k},x_{1}) \right)^{1/\sigma} \right]^{n} \\
%%%%%%%%%%%%%%%%%%%%%%%%%%%%%%%%%%%%%%%%%%%%%%%%%%
\label{Recall4}
&= \left(\mathbb{E} \left[ \left( B(X_{1},X_{2}) B(X_{2},X_{3}) \cdots B(X_{k-1},X_{k}) B(X_{k},X_{1}) \right)^{1/\sigma} \right]\right)^{n} .
%&\dfn [\Phi_{k}(\rho)]^{n}.
\end{align}
In order to proceed, observe the following. First, we have that for any $x,x' \in \calX$, $B(x,x') \leq 1$, which follows immediately by the Cauchy--Schwarz inequality.  
We also have the following result, which is proved in Appendix G.
\begin{lemma} \label{Lemma1}
	For a symmetric channel and a uniform input distribution,
	\begin{align}
	\mathbb{E} \left[ \left( B(X_{1},X_{2}) B(X_{2},X_{3}) \cdots B(X_{k-2},X_{k-1}) B(X_{k-1},X_{k}) \right)^{1/\sigma} \right] =  \left[\Omega(\sigma)\right]^{k-1} .
	\end{align} 
\end{lemma}
Let us continue from \eqref{Recall4} and conclude that 
\begin{align}
G(\pi,\sigma)
%%%%%%%%%%%%%%%%%%%%%%%%%%%%%%%%%%%%%%%%%%%%%%%%%%
&= \left(\mathbb{E} \left[ \left( B(X_{1},X_{2}) B(X_{2},X_{3}) \cdots B(X_{k-1},X_{k}) B(X_{k},X_{1}) \right)^{1/\sigma} \right]\right)^{n} \\
%%%%%%%%%%%%%%%%%%%%%%%%%%%%%%%%%%%%%%%%%%%%%%%%%%
&\leq \left(\mathbb{E} \left[ \left( B(X_{1},X_{2}) B(X_{2},X_{3}) \cdots B(X_{k-1},X_{k}) \right)^{1/\sigma} \right]\right)^{n} \\
&= \left[\Omega(\sigma)\right]^{(k-1)n} \\ 
&\leq \left[\Omega(\sigma)\right]^{\frac{2}{3}kn} ,
\end{align}
where the last step is due to the fact that $\Omega(\sigma) \leq 1$.

\subsubsection*{Step 3: A Unified Upper Bound for a Transposition and a Cycle}
Let us now define 
\begin{align}
\Upsilon(\sigma) \dfn \min \left\{-\frac{1}{2} \log \Xi(\sigma), -\frac{2}{3} \log \Omega(\sigma) \right\}.
\end{align}
Now, for a transposition:
\begin{align}
G(\pi,\sigma)
&= [\Xi(\sigma)]^{n} \\
&= \exp \left\{-n \left[-\log \Xi(\sigma) \right] \right\} \\
&= \exp \left\{-2n \left[- \frac{1}{2}\log \Xi(\sigma) \right] \right\} \\
&\leq \exp \left\{-2n \Upsilon(\sigma) \right\} ,
\end{align}
and for a $k$-cycle:
\begin{align}
G(\pi,\sigma)
&\leq \left[\Omega(\sigma)\right]^{\frac{2}{3}kn} \\
&= \exp \left\{-kn \left[-\frac{2}{3} \log \Omega(\sigma) \right] \right\} \\
&\leq \exp \left\{-kn \Upsilon(\sigma) \right\} .
\end{align}

\subsubsection*{Step 4: A Composition of Disjoint Cycles}

Let $\bi = \{i_1,i_2, \ldots, i_k\}$ and $\bj = \{j_1,j_2, \ldots, j_\ell\}$ be two arbitrary disjoint sets of indices of arbitrary lengths $k$ and $\ell$. Assume a permutation $\pi$ composed by two disjoint cycles defined over the sets $\bi$ and $\bj$. Then, it follows from the independence of codewords that
\begin{align}
&G(\pi,\sigma) \nn \\
%%%%%%%%%%%%%%%%%%%%%%%%%%%%%%%%%%%%%%%%%%%%%%%%%%
&= \left(\mathbb{E} \left[ \left( B(X_{i_1},X_{i_2}) \cdots B(X_{i_k},X_{i_1}) \cdot B(X_{j_1},X_{j_2}) \cdots B(X_{j_\ell},X_{j_1}) \right)^{1/\sigma} \right]\right)^{n} \\
%%%%%%%%%%%%%%%%%%%%%%%%%%%%%%%%%%%%%%%%%%%%%%%%%%
&= \left(\mathbb{E} \left[ \left( B(X_{i_1},X_{i_2}) \cdots B(X_{i_k},X_{i_1})\right)^{1/\sigma} \cdot \left( B(X_{j_1},X_{j_2}) \cdots B(X_{j_\ell},X_{j_1}) \right)^{1/\sigma} \right]\right)^{n} \\
%%%%%%%%%%%%%%%%%%%%%%%%%%%%%%%%%%%%%%%%%%%%%%%%%%
&= \left(\mathbb{E} \left[ \left( B(X_{i_1},X_{i_2}) \cdots B(X_{i_k},X_{i_1})\right)^{1/\sigma} \right] \cdot \mathbb{E} \left[ \left( B(X_{j_1},X_{j_2}) \cdots B(X_{j_\ell},X_{j_1}) \right)^{1/\sigma} \right]\right)^{n} \\
%%%%%%%%%%%%%%%%%%%%%%%%%%%%%%%%%%%%%%%%%%%%%%%%%%
&\leq \exp \left\{-kn \Upsilon(\sigma) \right\} \cdot \exp \left\{- \ell n \Upsilon(\sigma) \right\} \\
%%%%%%%%%%%%%%%%%%%%%%%%%%%%%%%%%%%%%%%%%%%%%%%%%%
&= \exp \left\{-(k + \ell) n \Upsilon(\sigma) \right\}.
\end{align}
This result can be easily extended by induction to permutations composed by an arbitrary number of disjoint cycles. Assume such a permutation with $c$ disjoint cycles of arbitrary lengths $\{\ell_1,\ell_2,\ldots,\ell_c\}$. Denote $L = \ell_1+\ell_2+\ldots+\ell_c$. Then, for such a permutation, one arrives at
\begin{align}
\label{GreatBound}
G(\pi,\sigma) 
\leq \exp \left\{-L n \Upsilon(\sigma) \right\}.
\end{align}

\subsubsection*{Step 5: Wrapping Up}
Let us recall the fact that every permutation is equivalent to a composition of disjoint cycles \cite{Her1975}. Let $\Pi_{j}(M)$, $j \in \{2,3, \ldots,M\}$, be the set of all permutations where exactly $j$ bees changed their places.
At this point, it is important to notice that the bound in \eqref{GreatBound} holds for any permutation for which the sum of lengths of all cycles is the same one.   
Continuing from \eqref{Recall5},   
\begin{align}
\mathbb{E} \left[P_{\mbox{\tiny e}}(\calC_{n}^{m})^{1/\sigma}\right] 
%%%%%%%%%%%%%%%%%%%%%%%%%%%%%%%%%%%%%%%%%%%%%%%%%%
&\leq \sum_{\substack{\pi \in \Pi(M) \\ \pi \neq \pi_{0}}} G(\pi,\sigma)  \\
%%%%%%%%%%%%%%%%%%%%%%%%%%%%%%%%%%%%%%%%%%%%%%%%%%
&= \sum_{\pi \in \Pi_{2}(M)} G(\pi,\sigma) + \sum_{j=3}^{M}\sum_{\pi \in \Pi_{j}(M)} G(\pi,\sigma) \\
%%%%%%%%%%%%%%%%%%%%%%%%%%%%%%%%%%%%%%%%%%%%%%%%%%
\label{ToExp3}
&\leq \sum_{\pi \in \Pi_{2}(M)} \exp \left\{-n [-\log \Xi(\sigma)] \right\} + \sum_{j=3}^{M}\sum_{\pi \in \Pi_{j}(M)} \exp \left\{-jn \Upsilon(\sigma) \right\}  \\
%%%%%%%%%%%%%%%%%%%%%%%%%%%%%%%%%%%%%%%%%%%%%%%%%%
&\leq M^{2} \exp \left\{-n [-\log \Xi(\sigma)] \right\} + \sum_{j=3}^{M} M^{j} \exp \left\{-jn \Upsilon(\sigma) \right\}  \\
%%%%%%%%%%%%%%%%%%%%%%%%%%%%%%%%%%%%%%%%%%%%%%%%%%
&= \exp \left\{-n [-\log \Xi(\sigma)-2R] \right\} + \sum_{j=3}^{M} \exp \left\{-jn [\Upsilon(\sigma)-R] \right\}  \\
%%%%%%%%%%%%%%%%%%%%%%%%%%%%%%%%%%%%%%%%%%%%%%%%%%
&\leq \exp \left\{-n [-\log \Xi(\sigma)-2R] \right\} + \sum_{j=3}^{\infty} \exp \left\{-jn [\Upsilon(\sigma)-R] \right\}   \\
%%%%%%%%%%%%%%%%%%%%%%%%%%%%%%%%%%%%%%%%%%%%%%%%%%
&= \exp \left\{-n [-\log \Xi(\sigma)-2R] \right\} 
+ \frac{\exp \left\{-3n [\Upsilon(\sigma)-R] \right\}}{1 - \exp \left\{-n [\Upsilon(\sigma)-R] \right\}}  \\
%%%%%%%%%%%%%%%%%%%%%%%%%%%%%%%%%%%%%%%%%%%%%%%%%%
\label{Recall6}
&\doteq  \exp \left\{-n [-\log \Xi(\sigma)-2R] \right\} 
+ \exp \left\{-3n [\Upsilon(\sigma)-R] \right\},
\end{align}
Where \eqref{ToExp3} follows from \eqref{GreatBound}.
Note that 
\begin{align}
&\exp \left\{-3n [\Upsilon(\sigma)-R] \right\} \nn \\
%%%%%%%%%%%%%%%%%%%%%%%%%%%%%%%%%%%%%%%%%%%%%
&~~~= \exp \left\{-3n \min \left\{-\frac{1}{2} \log \Xi(\sigma)-R, -\frac{2}{3} \log \Omega(\sigma)-R \right\} \right\} \\
%%%%%%%%%%%%%%%%%%%%%%%%%%%%%%%%%%%%%%%%%%%%%
&~~~\doteq \exp \left\{-3n \left[-\frac{1}{2} \log \Xi(\sigma)-R \right] \right\} + \exp \left\{-3n \left[-\frac{2}{3} \log \Omega(\sigma)-R \right] \right\} \\
%%%%%%%%%%%%%%%%%%%%%%%%%%%%%%%%%%%%%%%%%%%%%
&~~~= \exp \left\{-\frac{3}{2}n \left[-\log \Xi(\sigma)-2R \right] \right\} + \exp \left\{-n \left[-2 \log \Omega(\sigma)-3R \right] \right\} .
\end{align}
Substituting it back into \eqref{Recall6} yields
\begin{align}
\mathbb{E} \left[P_{\mbox{\tiny e}}(\calC_{n}^{m})^{1/\sigma}\right] 
%%%%%%%%%%%%%%%%%%%%%%%%%%%%%%%%%%%%%%%%%%%%%%%%%%
&\lexe \exp \left\{-n [-\log \Xi(\sigma)-2R] \right\} 
+ \exp \left\{-\frac{3}{2}n \left[-\log \Xi(\sigma)-2R \right] \right\}  \nn \\
&~~~~~~~~~~~~~ + \exp \left\{-n \left[-2 \log \Omega(\sigma)-3R \right] \right\}  \\
%%%%%%%%%%%%%%%%%%%%%%%%%%%%%%%%%%%%%%%%%%%%%%%%%%
&\doteq \exp \left\{-n [-\log \Xi(\sigma)-2R] \right\} 
+ \exp \left\{-n \left[-2 \log \Omega(\sigma)-3R \right] \right\} \\
%%%%%%%%%%%%%%%%%%%%%%%%%%%%%%%%%%%%%%%%%%%%%%%%%%
&\doteq \exp \left[-n \cdot  \min \left\{-\log \Xi(\sigma)-2R, -2 \log \Omega(\sigma)-3R \right\} \right].
\end{align}
Let us denote 
\begin{align}
E(R,\sigma) \dfn \min \left\{-\log \Xi(\sigma)-2R, -2 \log \Omega(\sigma)-3R \right\}, 
\end{align}
such that, for every $m \in \{1,2,\ldots,\xi_{n}\}$, 
\begin{align}
\mathbb{E} \left[P_{\mbox{\tiny e}}(\calC_{n}^{m})^{1/\sigma}\right] 
%%%%%%%%%%%%%%%%%%%%%%%%%%%%%%%%%%%%%%%%%%%%%%%%%%
&\lexe \exp \left\{-n \cdot E(R,\sigma) \right\}.
\end{align}
Now, according to Markov's inequality, it follows that
\begin{align}
\prob \left\{ \frac{1}{\xi_{n}} \sum_{m=1}^{\xi_{n}} P_{\mbox{\tiny e}}(\calC_{n}^{m})^{1/\sigma} > 2 \exp \left\{-n \cdot E(R,\sigma) \right\} \right\} \leq \frac{1}{2},
\end{align}
which means that there exists a code with
\begin{align}
\frac{1}{\xi_{n}} \sum_{m=1}^{\xi_{n}} P_{\mbox{\tiny e}}(\calC_{n}^{m})^{1/\sigma} \leq 2 \exp \left\{-n \cdot E(R,\sigma) \right\}.
\end{align}
We conclude that there exists a code $\calC_{n}$ with $M$ codewords for which 
\begin{align}
P_{\mbox{\tiny e}}(\calC_{n})^{1/\sigma} \leq 2 \exp \left\{-n \cdot E(R,\sigma) \right\},
\end{align}
and so
\begin{align}
P_{\mbox{\tiny e}}(\calC_{n}) 
\lexe  \exp \left\{-n \cdot \sigma \cdot E(R,\sigma) \right\},
\end{align}
thus,
\begin{align}
\liminf_{n \to \infty} - \frac{1}{n} \log P_{\mbox{\tiny e}}(\calC_{n}) 
\geq \sigma \cdot E(R,\sigma).
\end{align}
Since it holds for every $\sigma \geq 1$, the negative exponential rate of the error probability can be bounded as
\begin{align}
\liminf_{n \to \infty} - \frac{1}{n} \log P_{\mbox{\tiny e}}(\calC_{n}) 
&\geq \sup_{\sigma \geq 1} \left\{ \sigma \cdot E(R,\sigma) \right\},
\end{align}
and the proof of Theorem \ref{Theorem_EX} is now complete.

\section*{Appendix G}
\renewcommand{\theequation}{G.\arabic{equation}}
\setcounter{equation}{0}  
\subsection*{Proof of Lemma \ref{Lemma1}}

First, note that
\begin{align}
\mathbb{E} \left[ B(X,X')^{1/\sigma} \middle| X \right]
%%%%%%%%%%%%%%%%%%%%%%%%%%%%%%%%%%%%%%%%%%%%%%%%%%%%%%%%%
\label{Recall8}
&= \sum_{x' \in \calX} P_{X}(x') \left( \sum_{y \in \calY} 
\sqrt{ W(y|X) W(y|x') } \right)^{1/\sigma}
\end{align}
has the same value for every realization of $X$, thanks to the symmetry of the channel and the fact that $P_{X}$ is uniform across $\calX$. Averaging the right-hand-side of \eqref{Recall8} yields
\begin{align}
\mathbb{E} \left[ B(X,X')^{1/\sigma} \middle| X \right]
%%%%%%%%%%%%%%%%%%%%%%%%%%%%%%%%%%%%%%%%%%%%%%%%%%%%%%%%%
%%%%%%%%%%%%%%%%%%%%%%%%%%%%%%%%%%%%%%%%%%%%%%%%%%%%%%%%%
&= \sum_{x \in \calX}  \sum_{x' \in \calX} P_{X}(x) P_{X}(x') \left( \sum_{y \in \calY} 
\sqrt{ W(y|x) W(y|x') } \right)^{1/\sigma} \\
%%%%%%%%%%%%%%%%%%%%%%%%%%%%%%%%%%%%%%%%%%%%%%%%%%%%%%%%%
&= \sum_{x \in \calX}  \sum_{x' \in \calX} P_{X}(x) P_{X}(x') B(x,x')^{1/\sigma} \\
%%%%%%%%%%%%%%%%%%%%%%%%%%%%%%%%%%%%%%%%%%%%%%%%%%%%%%%%
&= \Omega(\sigma),
\end{align}
hence, it follows that $\mathbb{E} \left[B(X,X')^{1/\sigma} \right] = \Omega(\sigma)$ as well.
Now,
\begin{align}
&\mathbb{E} \left[ \left( B(X_{1},X_{2}) B(X_{2},X_{3}) \cdots B(X_{k-2},X_{k-1}) B(X_{k-1},X_{k}) \right)^{1/\sigma} \right] \nn \\
%%%%%%%%%%%%%%%%%%%%%%%%%%%%%%%%%%%%%%%%%%%%%%%%%%%%
&= \mathbb{E} \left[ \mathbb{E} \left[ \left( B(X_{1},X_{2}) B(X_{2},X_{3}) \cdots B(X_{k-2},X_{k-1}) B(X_{k-1},X_{k}) \right)^{1/\sigma} \middle| X_{1}, \ldots, X_{k-1} \right] \right] \\
%%%%%%%%%%%%%%%%%%%%%%%%%%%%%%%%%%%%%%%%%%%%%%%%%%%%
&= \mathbb{E} \left[ \left( B(X_{1},X_{2}) B(X_{2},X_{3}) \cdots B(X_{k-2},X_{k-1})  \right)^{1/\sigma} \cdot \mathbb{E} \left[  B(X_{k-1},X_{k})^{1/\sigma} \middle| X_{1}, \ldots, X_{k-1} \right] \right] \\
%%%%%%%%%%%%%%%%%%%%%%%%%%%%%%%%%%%%%%%%%%%%%%%%%%%%
&= \mathbb{E} \left[ \left( B(X_{1},X_{2}) B(X_{2},X_{3}) \cdots B(X_{k-2},X_{k-1})  \right)^{1/\sigma} \cdot \mathbb{E} \left[  B(X_{k-1},X_{k})^{1/\sigma} \middle| X_{k-1} \right] \right] \\
%%%%%%%%%%%%%%%%%%%%%%%%%%%%%%%%%%%%%%%%%%%%%%%%%%%%
&= \mathbb{E} \left[ \left( B(X_{1},X_{2}) B(X_{2},X_{3}) \cdots B(X_{k-2},X_{k-1})  \right)^{1/\sigma} \cdot \Omega(\sigma) \right] \\
%%%%%%%%%%%%%%%%%%%%%%%%%%%%%%%%%%%%%%%%%%%%%%%%%%%%
&= \mathbb{E} \left[ \left( B(X_{1},X_{2}) B(X_{2},X_{3}) \cdots B(X_{k-2},X_{k-1})  \right)^{1/\sigma} \right] \cdot \Omega(\sigma),
\end{align}
which proves the lemma upon repeating this process $k-1$ times.

\end{document}